\documentstyle[12pt]{article}

\catcode`\@=11
\long\def\@makefntext#1{ %\parindent 1em
\protect\noindent \hbox to 3.2pt {\hskip-.9pt
$^{{\ninerm\@thefnmark}}$\hfil}#1\hfill} %can be used

\def\thefootnote{\fnsymbol{footnote}}
 \def\@makefnmark{\hbox to 0pt{$^{\@thefnmark}$\hss}}  %original

\def\ps@myheadings{\let\@mkboth\@gobbletwo
\def\@oddhead{\hbox{} %\sl
\rightmark\hfil\ninerm\thepage}
\def\@oddfoot{}\def\@evenhead{\ninerm\thepage\hfil %\sl
\leftmark\hbox{}}\def\@evenfoot{}
\def\sectionmark##1{}\def\subsectionmark##1{}}

\evensidemargin
 0.30truein
\renewcommand{\thefootnote}{\fnsymbol{footnote}}

\newcounter{appendixc}
\newcounter{subappendixc}[appendixc]
\newcounter{subsubappendixc}[subappendixc]

\renewcommand{\appendix}[1] {\vspace{0.6cm}
        \refstepcounter{appendixc}
        \setcounter{figure}{0}
        \setcounter{table}{0}
        \setcounter{equation}{0}
        \renewcommand{\thefigure}{\Alph{appendixc}.\arabic{figure}}
        \renewcommand{\thetable}{\Alph{appendixc}.\arabic{table}}
        \renewcommand{\theappendixc}{\Alph{appendixc}}
        \renewcommand{\theequation}{\Alph{appendixc}.\arabic{equation}}
        \noindent{\bf Appendix \theappendixc #1}\par\vspace{0.4cm}}

\renewenvironment{thebibliography}[1]
        {\begin{list}{\arabic{enumi}.}
        {\usecounter{enumi}\setlength{\parsep}{0pt}
\setlength{\leftmargin 1.25cm}{\rightmargin 0pt}
         \setlength{\itemsep}{0pt} \settowidth
        {\labelwidth}{#1.}\sloppy}}{\end{list}}

\newcounter{itemlistc}
\newcounter{romanlistc}
\newcounter{alphlistc}
\newcounter{arabiclistc}

\newcommand{\fcaption}[1]{
        \refstepcounter{figure}
        \setbox\@tempboxa = \hbox{\tenrm Fig.~\thefigure. #1}
        \ifdim \wd\@tempboxa > 6in
           {\begin{center}
        \parbox{6in}{\tenrm\baselineskip=12pt Fig.~\thefigure. #1 }
            \end{center}}
        \else
             {\begin{center}
             {\tenrm Fig.~\thefigure. #1}
              \end{center}}
        \fi}

\newcommand{\tcaption}[1]{
        \refstepcounter{table}
        \setbox\@tempboxa = \hbox{\tenrm Table~\thetable. #1}
        \ifdim \wd\@tempboxa > 6in
           {\begin{center}
        \parbox{6in}{\tenrm\baselineskip=12pt Table~\thetable. #1 }
            \end{center}}
        \else
             {\begin{center}
             {\tenrm Table~\thetable. #1}
              \end{center}}
        \fi}

\def\@citex[#1]#2{\if@filesw\immediate\write\@auxout
        {\string\citation{#2}}\fi
\def\@citea{}\@cite{\@for\@citeb:=#2\do
        {\@citea\def\@citea{,}\@ifundefined
        {b@\@citeb}{{\bf ?}\@warning
        {Citation `\@citeb' on page \thepage \space undefined}}
        {\csname b@\@citeb\endcsname}}}{#1}}

\newif\if@cghi
\def\cite{\@cghitrue\@ifnextchar [{\@tempswatrue
        \@citex}{\@tempswafalse\@citex[]}}
\def\citelow{\@cghifalse\@ifnextchar [{\@tempswatrue
        \@citex}{\@tempswafalse\@citex[]}}
\def\@cite#1#2{{$\null^{#1}$\if@tempswa\typeout
        {IJCGA warning: optional citation argument
        ignored: `#2'} \fi}}

\newcommand{\be}{\begin{equation}}
\newcommand{\ee}{\end{equation}}
\newcommand{\beq}{\begin{equation}}
\newcommand{\eeq}{\end{equation}}
\newcommand{\ba}{\begin{array}}
\newcommand{\ea}{\end{array}}
\newcommand{\beqn}{\begin{eqnarray}}
\newcommand{\eeqn}{\end{eqnarray}}
\newcommand{\bea}{\begin{eqnarray}}
\newcommand{\eea}{\end{eqnarray}}
\newcommand{\beqan}{\begin{eqnarray*}}
\newcommand{\eeqan}{\end{eqnarray*}}
\newcommand{\cL}{{\cal L}}
\newcommand{\cD}{{\cal D}}
\newcommand{\cM}{{\cal M}}
\newcommand{\cO}{{\cal O}}

\newcommand{\cT}{{\cal T}}

\newcommand{\cH}{{\cal H}}
\newcommand{\dg}{\dagger}
\newcommand{\dfrac}{\displaystyle \frac}
\newcommand{\no}{\nonumber}
\newcommand{\rms}{\rm\scriptsize}

\newcommand{\vp}{\varphi}

\newcommand{\tr}{\rm tr}
\newcommand{\ra}{\rightarrow}
\newcommand{\Ra}{\Rightarrow}

\newcommand{\noi}{\noindent}

\def\fnt#1#2{\footnotetext{\kern-.3em
        {$^{\mbox{\sevenrm #1}}$}{#2}}}

 1
\font\twelverm=cmr10 scaled\magstep 1
 1

\font\tenrm=cmr10

\font\ninerm=cmr9

\font\fifteen=cmbx10 at 15pt
\font\twelve=cmbx10 at 12pt

\def\th{$^{\mbox{\scriptsize th}}$}

\def\tableofcontents{\section*{Table of Contents\@mkboth{CONTENTS}{CONTENTS}}
 \@starttoc{toc}}
\def\l@part#1#2{\addpenalty{\@secpenalty}
 \addvspace{2.25em plus 1pt} \begingroup
 \@tempdima 3em \parindent \z@ \rightskip \@pnumwidth \parfillskip
-\@pnumwidth
 {\large \bf \leavevmode #1\hfil \hbox to\@pnumwidth{\hss #2}}\par
 \nobreak \endgroup}
\def\l@section#1#2{\addpenalty{\@secpenalty} \addvspace{0.5em plus 1pt}
\@tempdima 1.5em \begingroup
 \parindent \z@ \rightskip \@pnumwidth
 \parfillskip -\@pnumwidth
 \bf \leavevmode \advance\leftskip\@tempdima \hskip -\leftskip #1\nobreak\hfil
\nobreak\hbox to\@pnumwidth{\hss #2}\par
 \endgroup}

\begin{document}

%----------------------START OF DATA FILE------------------------------

\begin{titlepage}

\begin{center}

{\twelve Centre de Physique Th\'eorique\footnote{
Unit\'e Propre de Recherche 7061} - CNRS - Luminy, Case 907}

{\twelve F-13288 Marseille Cedex 9 - France }

\vspace{1 cm}

\setcounter{footnote}{0}
\renewcommand{\thefootnote}{\arabic{footnote}}

{\fifteen CHIRAL LAGRANGIANS} \\
{\bf AND} \\[1mm]
{\fifteen KAON CP-VIOLATION\footnote{
Lectures given at the 1994\th\ TASI-School on CP--Violation
and the limits of the Standard Model. University of Colorado at Boulder.
}
}

\vspace{1 cm}

{\bf Eduardo de RAFAEL}

\vspace{2,7 cm}

{\bf Abstract}

\end{center}

These lectures are an introduction to the subject of chiral
effective Lagrangians of the Standard Model and their applications, mostly in
the sector of non--leptonic kaon decays, with special emphasis on
CP--violation. The first lecture gives an introduction to the phenomenological
description of the
$K^0-\bar{K}^0$ system and $K\ra \pi\pi$ decays. In the second lecture I give
an overview of the basic ideas behind the chiral perturbation theory
($\chi$PT) --approach to hadron dynamics at low energies. The study of the
weak interactions of $K$--particles within the framework of $\chi$PT is the
subject of the third lecture. The fourth lecture is an overview of various
models of the QCD low--energy effective action which have been developed
during the last few years. The fifth lecture is dedicated to a discussion of
the CP--violation $\epsilon$ and
$\epsilon'$ parameters, and to the study of the decay mode: $K_{L}\ra
\pi^{0}e^{+}e^{-}$.

\vspace{2,7 cm}


\bigskip

\noindent January 1995

\noindent CPT-95/P.3161

\bigskip

\noindent anonymous ftp or gopher: cpt.univ-mrs.fr

\end{titlepage}

\setcounter{footnote}{0}
\renewcommand{\thefootnote}{\fnsymbol{footnote}}

\newpage

\pagenumbering{roman}

\tableofcontents

\newpage

\pagenumbering{arabic}

\vfil
%\vspace{0.8cm}
\twelverm   %modified by CLee 23/07/93
\baselineskip=14pt
%%%%%%%%%%%%%%%%%%%%%%%%%%%%%%%%%%%%%%%%%%%%%%%%%%%%%%%%%%%%
%%%%%%%%%%%%%%%%%%%%%%%%%%%%%%%%%%%%%%%%%%%%%%%%%%%%%%%%%%%%
\section{Phenomenology of $K^{0}-\bar{K^{0}}$ Mixing and
  CP--Violation in $K\ra 2\pi$ Decays}
\label{sec:phenomenology}

The purpose of this first lecture is to introduce the basics of the
phenomenological description of the $K^0$--$\bar{K}^0$ system. There is
practically no theory behind this description. It is only based on first
principles: the superposition principle, Lorentz invariance, and general
invariance properties under the P, C and T symmetries. The basic idea is to
reduce the description of this system to a minimum of phenomenological
parameters which, eventually, an underlying theory  --like the Standard Model--
should be able to predict. There are many reviews on this subject. I recommend
the reader to consult Ref.~\cite{S90} for an excellent historical overview and
complementary information.

%%%%%%%%%%%%%%%%%%%%%%%%%%%%%%%%%%%%%%%%%%%%%%%%%%%%%%%%%%%%%
%%%%%%%%%%%%%%%%%%%%%%%%%%%%%%%%%%%%%%%%%%%%%%%%%%%%%%%%%%%%%
\subsection{Phenomenology of $K^{0}$--$\bar{K^{0}}$ Mixing}
\label{subsec:mixing}

In the absence of the weak interactions, the $K^{0}$ and
$\bar{K^{0}}$ particles produced by the strong interactions are stable
eigenstates of strangeness with eigenvalues $\pm 1$. In the presence of the
weak interaction they become unstable. The states with an exponential time
dependence law ($\tau$ is the proper time)
\be\label{eq:Kls}
\mid K_{L}\rangle \ra e^{-iM_{L}\tau} \mid K_{L}\rangle
\qquad {\rm and} \qquad
\mid K_{S}\rangle \ra e^{-iM_{S}\tau} \mid K_{S}\rangle\,,
\ee

\noi are linear superpositions of the eigenstates of strangeness:
\nopagebreak
\beqn\label{eq:KL}
\mid K_{L}\rangle & = & \frac{1}{\sqrt{{\mid p \mid }^2 + {\mid q \mid }^2}}
\left(p\mid K^0 \rangle + q\mid \bar{K^0} \rangle\right) \\
\mid K_{S}\rangle & = & \frac{1}{\sqrt{{\mid p \mid }^2 + {\mid q \mid }^2}}
\left(p\mid K^0 \rangle - q\mid \bar{K^0}\rangle\right)\,,
\eeqn

\noi where $p$ and $q$ are complex numbers and CPT--invariance, which is a
property of the Standard Model in any case, has been assumed. The parameters
$M_{L,S}$ in Eq.~(\ref{eq:Kls})  are also complex
\be\label{eq:masses} M_{L,S}= m_{L,S}-\frac{i}{2}\, \Gamma_{L,S}\, ,
\ee

\noi with $m_{L,S}$ the masses and $\Gamma_{L,S}$ the decay widths of the
long--lived and short--lived neutral kaon states.

As we shall see, experimentally, the $\mid K_{S}\rangle$ and $\mid
K_{L}\rangle$ states are very close to the CP-eigenstates
\be\label{eq:K12}
\mid K_{1}^{0}\rangle =\frac{1}{\sqrt{2}}\left(\mid K^0\rangle - \mid
\bar{K^0}\rangle\right) \ {\rm and}\
\mid K_{2}^{0}\rangle =\frac{1}{\sqrt{2}}\left(\mid K^0\rangle + \mid
\bar{K^0}\rangle \right)
\ee

\noi with
\be\label{eq:CPK12}  {\rm CP}\mid K_{1}^{0}\rangle =+\mid K_{1}^{0}\rangle
\qquad {\rm and}\qquad  {\rm CP}\mid K_{2}^{0}\rangle = -\mid K_{2}^{0}\rangle
\, .
\ee

\noi This is characterized by the small complex parameter
$\tilde{\epsilon}$,
\be\label{eq:epstilde}
\tilde{\epsilon}=\frac{p-q}{p+q}\,;
\ee

\noi in terms of which,
\be\label{eq:KLS12}
\mid K_{L,S}\rangle =\frac{1}{\sqrt{1+\mid \tilde{\epsilon} \mid ^2}} (\mid
K_{2,1}^0\rangle + \tilde{\epsilon} \mid K_{1,2}^0\rangle)\,.
\ee

According to Eqs.~(\ref{eq:Kls}) and (\ref{eq:KL}), a state
initially pure $\mid K^0 \rangle$ evolves, in a period of time
$\tau$ to a state which is a superposition of $\mid K^0 \rangle$ and
$\mid \bar{K^0} \rangle$\,:
\be\label{eq:Kevol}
\mid K^0 \rangle \ra \frac{1}{2}[e^{-iM_{L}\tau }+e^{-iM_{S}\tau }]
\mid K^0 \rangle+\frac{1}{2}\frac{p}{q} [e^{-iM_{L}\tau }-e^{-iM_{S}\tau }]
\mid \bar{K^0} \rangle\,;
\ee

\noi and, likewise
\be\label{eq:Kbarevol}
\mid \bar{K^0} \rangle \ra \frac{1}{2}[e^{-iM_{L}\tau } +e^{-iM_{S}\tau }] \mid
\bar{K^0}\rangle+\frac{1}{2}\frac{q}{p} [e^{-iM_{L}\tau}-e^{-iM_{S}\tau }]\mid
K^0 \rangle\,.
\ee

\noi For a small period of time $\delta \tau$ we then have
\beqn\label{eq:Kevolinf}
\mid K^0 \rangle & \ra & \mid K^0 \rangle-i\delta \tau (\cM_{11}\mid K^0
\rangle+\cM_{12}\mid \bar{K^0} \rangle)\, ;\\
\label{eq:Kbarevolinf}
\mid \bar{K^0} \rangle & \ra & \mid \bar{K^0}
\rangle-i\delta \tau  (\cM_{21}\mid K^0 \rangle+\cM_{22}\mid \bar{K^0}
\rangle),
\eeqn

\noi where
\be\label{eq:mmatrix}
\cM_{ij}\, = \, \frac{1}{2}\left(
\begin{array}{cc} M_{L}+M_{S} & \frac{p}{q}(M_{L}-M_{S}) \\
                  \frac{q}{p}(M_{L}-M_{S}) & M_{L}+M_{S}
\end{array} \right)\, .
\ee

\noi This is the complex mass--matrix of the $K^{0}-\bar{K^0}$ system.

In full generality, the mass--matrix $\cM_{ij}$ admits a decomposition, similar
to the one of the complex parameters $M_{L,S}$ in Eq.~(\ref{eq:masses}), in
terms of an absorptive part
$\Gamma_{ij}$ and a dispersive part $M_{ij}$ :
\be\label{eq:mmatrixad}
\cM_{ij}=M_{ij}-\frac{i}{2}\Gamma_{ij}\,.
\ee

\noi In a given quantum field theory, like e.g., the Standard Electroweak
Model, the complex $K^{0}-\bar{K^0}$ mass--matrix is defined via the transition
matrix $T$ which characterizes $S$--matrix elements. More precisely, the
off--diagonal absorptive matrix element
$\Gamma_{12}$ for example, is given by the sum of products of on--shell matrix
elements:
\be\label{eq:abs12}
\Gamma_{12}=\sum_{\Gamma} \int d\Gamma  (\langle\Gamma \mid T \mid
\bar{K^0}\rangle)^{*}\, \langle\Gamma \mid T \mid K^0 \rangle\, ,
\ee

\noi where the sum is extended to all possible states $\mid\Gamma\rangle$ to
which the states
$\mid K^0 \rangle$ and $\mid \bar{K^0}\rangle$ can decay.  The symbol $d\Gamma$
denotes the phase space measure appropriate to the particle content of the
state $\Gamma$. The corresponding matrix element
$M_{12}$ is defined by the dispersive principal part integral
\be\label{eq:dis12} M_{12}=\frac{1}{\pi}\wp \int ds
\frac{1}{m_{K}^{2}-s}\Gamma_{12}(s)+ {\rm ``local-terms"}\,.
\ee

The fact that ${\cM}_{11}={\cM}_{22}$ in Eq.~(\ref{eq:mmatrix}) is a
consequence of CPT--invariance. In general, if we have a transition between an
initial state $\mid {\it IN}\rangle$ and a final state
$\mid {\it FN}\rangle$, CPT--invariance relates the matrix elements of this
transition to the one between the corresponding CPT--transformed states
$\mid {\it \overline{FN}^{'}}\rangle$ and $\mid {\it
\overline{IN}^{'}}\rangle$, where $\mid {\it \overline{IN}^{'}}\rangle$
denotes the state obtained from
$\mid {\it IN}\rangle$ by interchanging all particles into antiparticles (this
is the meaning of the bar symbol in ${\it
\overline{IN}}$), and taking the mirror image of the kinematic variables:
$[(E,\vec{p})\ra (E,-\vec{p})\,; (\sigma^{0},\vec{\sigma})\ra
(-\sigma^{0},\vec{\sigma}) ]$, as well as their motion reversal image:
$[(E,\vec{p})\ra (E,-\vec{p})\,; (\sigma^{0},\vec{\sigma})\ra
(\sigma^{0},-\vec{\sigma}) ]$. (These kinematic changes are the meaning of the
prime symbol in  ${\it IN^{'}}\rangle$.) Altogether, CPT--invariance implies
then:
\be\label{eq:CPT} \langle{\it FN}\mid T \mid {\it IN}\rangle
 = \langle{\it \overline{IN}^{'}}\mid T \mid {\it
\overline{FN}^{'}}\rangle\,.
\ee

\noi Since, for the $K^{0}$-states: $\mid (\bar{K^{0}})^{'}\rangle =\mid
K^{0}\rangle$, the CPT--invariance relation implies
\be\label{eq:CPT1122}
\cM_{11}=\cM_{22}\,.
\ee

The off--diagonal matrix elements in Eq.~(\ref{eq:mmatrix}) are also related by
CPT--invariance, plus the hermiticity property of the
$T$-matrix in the absence of strong final state interactions; certainly the
case when the $\mid IN\rangle$ and $\mid FN\rangle$ states are
$\mid K^{0}\rangle$ and $\mid\bar{K^{0}}\rangle$. In general, in the absence of
strong final state interactions, we have:
\be\label{eq:hermit} \langle{\it \overline{IN}^{'}}\mid T \mid {\it
\overline{FN}^{'}}\rangle = \left(\langle{\it \overline{FN}^{'}}\mid T \mid
{\it
\overline{IN}^{'}}\rangle\right)^{*}\,.
\ee

\noi This relation, together with the CPT--invariance relation in
(\ref{eq:CPT}) implies then:
\be\label{eq:CPT1221}
\cM_{12}=(\cM_{21})^{*}\,.
\ee

There are a number of interesting constraints between the various
phenomenological parameters we have introduced. With $M_{12}$ and
$\Gamma_{12}$ defined in Eqs.~(\ref{eq:dis12}) and (\ref{eq:abs12}) and using
Eqs.~(\ref{eq:mmatrix}, \ref{eq:masses} and
\ref{eq:epstilde}), we have
\be\label{eq:relations}
\frac{q}{p}\, =\, \frac{1-\tilde{\epsilon}}{1+\tilde{\epsilon}}\, =
\, \frac{1}{2}\frac{\Delta m +i\frac{1}{2} \Delta \Gamma} {M_{12}-\frac{i}{2}
\Gamma_{12}}\, =
\, \frac{M_{21}-\frac{i}{2} \Gamma_{21}} {\frac{1}{2} (\Delta m +\frac{i}{2}
\Delta \Gamma)},
\ee

\noi where
\be\label{eq:dif}
\Delta m \equiv m_{L}-m_{S} \qquad {\rm and} \qquad
\Delta \Gamma \equiv \Gamma_{S}-\Gamma_{L}.
\ee

\noi
 As already discussed, CPT--invariance implies
\be\label{eq:cpt1221} M_{21}=(M_{12})^{*} \qquad {\rm and} \qquad
\Gamma_{21}=(\Gamma_{12})^{*}.
\ee

Experimentally, the masses $m_{L,S}$ and widths $\Gamma_{L,S}$ are well
measured, and in what follows they will be used as known parameters. (There is
no way for theory at present to do better than experiments in the determination
of these parameters...) The precise values for the masses and widths can be
found in the Particle Data Booklet~\cite{PDB94} (PDB). Nevertheless, it is
important to keep in mind some orders of magnitude:
\beqn\label{eq:widthshort}
\Gamma_{S}^{-1} & \simeq & 0.9\times 10^{-10}{\rm sec.}\,; \\
\label{eq:widthlong}
\Gamma_{L} & \simeq & 1.7\times 10^{-3}\Gamma_{S}; \\
\label{eq:difmass}
\Delta m & \simeq & 0.5 \Gamma_{S}.
\eeqn

%%%%%%%%%%%%%%%%%%%%%%%%%%%%%%%%%%%%%%%%%%%%%%%%%%%%%%%%%%%%
\subsection{The Bell--Steinberger Unitarity Constraint}
\label{subsec:Bell_Steinberger}

Let us consider a state $\mid \Psi\rangle$ to be an arbitrary  superposition of
the short--lived and long--lived kaon states:
\be\label{eq:psi}
\mid \Psi\rangle\, = \, \alpha \mid K_{S}\rangle +
\beta \mid K_{L}\rangle\, .
\ee

\noi The total decay rate of this state must be compensated by a decrease of
its norm:
\be\label{eq:norm}
\sum_{\Gamma}\mid \langle\Gamma \mid T \mid \Psi\rangle\mid^{2}=
-\frac{d}{d\tau}\mid \Psi \mid^{2}\,.
\ee

\noi The change in rate is governed by the mass matrix defined by
Eq.~(\ref{eq:Kevolinf}). Equating terms proportional to $\mid \alpha
\mid^{2}$ and $\mid \beta \mid^{2}$ in both sides of Eq.~(\ref{eq:norm})
results in the trivial relations:
\beqn\label{eq:GL}
\Gamma_{L} & = & \sum_{\Gamma} \int d\Gamma
\mid \langle\Gamma \mid T \mid K_{L}\rangle\mid^{2}\,, \\
\label{eq:GS}
\Gamma_{S} & = & \sum_{\Gamma} \int d\Gamma
\mid \langle\Gamma \mid T \mid K_{S}\rangle\mid^{2}.
\eeqn

\noi The mixed terms, proportional to $\alpha \beta^{*}$ and
$\alpha^{*}\beta$, lead however to a highly non--trivial relation, first
derived by Bell and Steinberger~\cite{BS65}:
\be\label{eq:BSconstr}  -i(M_{L}^{*}-M_{S})\langle K_{L}\mid K_{S}\rangle =
\sum_{\Gamma} \int d\Gamma \left(\langle\Gamma \mid T \mid K_{L}
\rangle\right)^{*} \langle\Gamma
\mid T \mid K_{S}\rangle\,.
\ee

\noi Notice that
\be\label{eq:KLKSNO}
\langle K_{L} \mid K_{S}\rangle =
\frac{\mid p \mid^{2}-\mid q \mid^{2}} {\mid p \mid^{2}+\mid q
\mid^{2}}=
\frac{2{\rm Re} \tilde{\epsilon}}{1+\mid \tilde{\epsilon}
\mid^{2}}\,.
\ee

\noi The l.h.s. of Eq.~(\ref{eq:BSconstr}) can be expressed in terms of
measurable physical parameters with the result
\be\label{eq:BSphys} \left(\frac{\Gamma_{S}+\Gamma_{L}}{2}-i\Delta m\right)
\frac{2{\rm Re}\tilde{\epsilon}}{1+\mid \tilde{\epsilon} \mid^{2}}=
\sum_{\Gamma} \int d\Gamma \left(\langle\Gamma \mid T
\mid K_{L}\rangle\right)^{*} \langle\Gamma\mid T \mid K_{S}\rangle\, .
\ee

\noi The r.h.s. of this equation can be bounded, using the Schwartz inequality,
with the result
\be\label{eq:BSSi}
\Big| \frac{\Gamma_{S}+\Gamma_{L}}{2}-i\Delta m \Big|
\frac{2{\rm Re}\tilde{\epsilon}}{1+\mid \tilde{\epsilon} \mid^{2}}
\leq \sqrt{\Gamma_{L}\Gamma_{S}}\,.
\ee

\noi Inserting the experimental values for $\Gamma_{S,L}$ and $\Delta m$,
results in an interesting bound for the non--orthogonality of the
$K_{L}$ and $K_{S}$ states [see eq~(\ref{eq:KLKSNO})]:
\be\label{eq:boundKLKSNO}
\frac{2{\rm Re} \tilde{\epsilon}}{1+\mid \tilde{\epsilon} \mid^{2}}
\leq 2.9\times 10^{-2}\,,
\ee

\noi indicating also that the admixture of $K_{1}^{0}(K_{2}^{0})$ in
$K_{L}(K_{S})$ has to be rather small.

It is possible to obtain further information from the unitarity constraint in
(\ref{eq:BSphys}), if one uses the experimental fact that the $2\pi$ states are
by far the dominant terms in the sum over hadronic states $\Gamma$. One can
then write the r.h.s. of Eq.~(\ref{eq:BSphys}) in the form
\be\label{eq:BS2pi}
\sum_{\pi \pi} \int d(\pi \pi) \left(\langle\pi \pi \mid T \mid
K_{L}\rangle\right)^{*} \langle\pi
\pi \mid T \mid K_{S}\rangle+ \gamma \Gamma_{S}\,.
\ee

\noi It is possible to obtain a bound for $\gamma$, by considering other states
than $2\pi $ in the sum of the r.h.s. in Eq.~(\ref{eq:BSphys}) and applying the
Schwartz inequality to individual sets of states separated by selection rules.
The contribution from the various semileptonic modes, for example, is known to
be smaller than
\be\label{eq:semilep}
\mid \sum_{\rm lep. modes} \int \cdots \mid \ll 10^{-3}\Gamma_{S}\,;
\ee

\noi and the contribution from the $3\pi$--states
\be\label{eq:3pi}
\mid \sum_{3\pi } \int \cdots \mid \ll 10^{-3}\Gamma_{S}\,.
\ee

\noi We conclude that to a good approximation we can restrict the
Bell--Steinberger relation to $2\pi$--states. We shall later come back to this
inequality, but first we have to discuss the phenomenology of the dominant
$K\ra \pi \pi$ transitions.

%%%%%%%%%%%%%%%%%%%%%%%%%%%%%%%%%%%%%%%%%%%%%%%%%%%%%%%%%%%%
\subsection{$K\ra \pi \pi$ Amplitudes}
\label{subsec:K2pi}

In the limit where CP is conserved the states $K_{S}(K_{L})$ become eigenstates
of CP; i.e., the states $K_{1}^{0}(K_{2}^{0})$ introduced in (\ref{eq:K12})
with eigenvalues
${\rm CP}=+1({\rm CP}=-1)$. On the other hand a state of two--pions with total
angular momentum
$J=0$ has ${\rm CP}=+1$. Therefore, the observation of a transition from the
long--lived component of the neutral kaon system to a two--pion final state is
evidence for  CP--violation. The first observation of such a transition to the
$\pi^{+} \pi^{-}$ mode was made by Christenson, Cronin, Fitch, and
Turlay~\cite{ChCFT64} in 1964, with the result
\be\label{eq:ChCFT}
\frac{\Gamma_{L}(+,-)}{\Gamma_{L}({\rm all})}=(2\pm 0.4)\times 10^{-3}\,.
\ee

\noi Since then the transition to the $\pi^{0} \pi^{0}$ mode has also been
observed, as well as  the phases of the amplitude ratios
\be\label{eq:etapm}
\eta_{+-}=\frac{\langle\pi^{+} \pi^{-} \mid T \mid K_{L}\rangle}
{\langle\pi^{+}\pi^{-} \mid T \mid K_{S}\rangle}\,\qquad {\rm and}\qquad
\eta_{00}=\frac{\langle\pi^{0} \pi^{0} \mid T \mid K_{L}\rangle}
{\langle\pi^{0}\pi^{0} \mid T \mid K_{S}\rangle}\,,
\ee

\noi with the results~\cite{PDB94}:
\beqn
\label{eq:etapmexp}
\eta_{+-} & = & (2.269\pm 0.023)\times 10^{-3} e^{i(44.3\pm0.8)^{\circ}}\,;  \\
\label{eq:etazzexp}
\eta_{00} & = & (2.259\pm 0.023)\times 10^{-3} e^{i(43.3\pm1.3)^{\circ}}\,.
\eeqn

In order to make a phenomenological analysis of $K\ra \pi
\pi$ transitions, it is convenient to express the states $\mid \pi^{+}
\pi^{-}\rangle$ and $\mid \pi^{0} \pi^{0}\rangle$ in terms of well  defined
isospin $I=0$, and $I=2$ states. (The $I=1$ state in this case is forbidden by
Bose statistics.):
\beqn
\label{eq:pmisospin}
\mid +-\rangle & = & \sqrt{\frac{2}{3}} \mid 0\rangle +
\sqrt{\frac{1}{3}} \mid 2\rangle\,;
\\
\label{eq:zzisospin}
\mid 00\rangle & = & \sqrt{\frac{2}{3}} \mid 2\rangle -\sqrt{\frac{1}{3}}\mid
0\rangle \, .
\eeqn

\noi The reason for introducing pure isospin states, is that the matrix
elements of transitions from $K^{0}$ and the $\bar{K^{0}}$ states to the same
$(\pi \pi)_{I}$--state can be related by
$CPT$--invariance plus Watson's theorem on final state interactions. The
relation in question is the following
\be
\label{eq:Watson}  e^{-2i\delta_{I}}\langle I \mid T \mid K^{0}\rangle =
(\langle I \mid T \mid \bar{K^{0}}\rangle)^{*}\,,
\ee

\noi where $\delta_{I}$ denotes the appropriate $J=0$, isospin $I$ $\pi
\pi$ phase--shift at the energy of the neutral kaon mass.

The proof of this relation is rather simple. With $S=1+iT$, the unitarity of
the $S$--matrix, $SS^{\dagger}=1$, implies
\be\label{eq:Tunitarity} T^{\dagger}T=i(T^{\dagger}-T)\,.
\ee

\noi If one takes matrix elements of this operator relation between an initial
state $K^{0}$, and a final $2\pi$--state with isospin $I$, we then have
\be\label{eq:Watson1}
\sum_{F}\langle I \mid T^{\dagger} \mid F\rangle\langle F \mid T
\mid K^{0}\rangle = i\langle I \mid T^{\dagger} \mid K^{0}\rangle-i\langle I
\mid T \mid K^{0}\rangle,
\ee

\noi where we have inserted a complete set of states $\sum\mid F\rangle\langle
F\mid =1$ between
$T$ and
$T^{\dagger}$. The crucial observation is that, in the strong interaction
sector of the
$S$--matrix, only the state
$F=I$ can contribute to the $T^{\dagger}$--matrix element. All the other states
are suppressed by selection rules; e.g., the
$3\pi$--states have opposite $G$--parity than the $2\pi$--states; the
$\pi l\nu$-- states are not related to $2\pi$--states by the strong
interactions alone; etc. Then, introducing the $\pi\pi$ phase--shift
definition:
\be
\label{eq:phaseshift} \langle I \mid S \mid I\rangle  =e^{2i\delta_{I}}\,,
\ee

\noi results in the relation
\beqn
\label{eq:Watson2} i(e^{-2i\delta_{I}}-1)\langle I \mid T \mid K^{0}\rangle & =
&  i\langle I \mid T^{\dagger} \mid K^{0}\rangle-i\langle I
\mid T \mid K^{0}\rangle, \no \\ & = & i(\langle K^{0} \mid T \mid
I\rangle)^{*}-i\langle I
\mid T \mid K^{0}\rangle\,.
\eeqn

\noi We can next use CPT--invariance [recall Eq.~(\ref{eq:CPT}), which in our
case implies the relation: $\langle K^{0} \mid T \mid I\rangle^{*}=  (\langle I
\mid T \mid \bar{K^{0}} \rangle)^{*}$.] The result in  Eq.~(\ref{eq:Watson})
then follows.

As a consequence of the relation we have proved, we can use in full generality
the following parametrization for
$K^{0}(\bar{K^{0}})\ra (\pi \pi)_{I}$ amplitudes:
\beqn
\label{eq:KI} \langle I \mid T \mid K^{0}\rangle & = & iA_{I}e^{i\delta_{I}}\,;
\\
\label{eq:KBI} \langle I \mid T \mid \bar{K^{0}}
\rangle &  = & -iA_{I}^{*}e^{i\delta_{I}}\, .
\eeqn

One possible quantity we can introduce to characterize the amount of
CP--violation in
$K\ra 2\pi$ transitions is the parameter
\be
\label{eq:eps}
\epsilon =
\frac{A[K_{L}\ra (\pi\pi)_{I=0}]}{A[K_{S}\ra (\pi\pi)_{I=0}]}\,.
\ee

\noi This parameter is related to the $\tilde{\epsilon}$--parameter introduced
in Eq.~(\ref{eq:epstilde})); as well as to the complex
$A_{0}$-amplitude defined in~(\ref{eq:KI}) and~(\ref{eq:KBI}), in the following
way
\be
\label{eq:eps1}
\epsilon =
\frac{(1+\tilde{\epsilon})A_{0}-(1-\tilde{\epsilon})A_{0}^{*}}
{(1+\tilde{\epsilon})A_{0}+(1-\tilde{\epsilon})A_{0}^{*}}\,.
\ee

\noi i.e.,
\be
\label{eq:eps2}
\epsilon =
\frac{\tilde{\epsilon}+i\frac{{\rm Im}A_{0}}{{\rm Re}A_{0}}}
{1+i\tilde{\epsilon}\frac{{\rm Im}A_{0}}{{\rm Re}A_{0}}}.
\ee

This is a good place to comment on the history of phase conventions in neutral
$K$--decays. In their pioneering paper on the phenomenology of the $K-\bar{K}$
system, Wu and Yang~\cite{WY64} chose to freeze the arbitrary relative phase
between the $ K^{0}$ and
$\bar{K^{0}}$ states, with the choice Im$A_{0}=0$. With this convention,
$\epsilon = \tilde{\epsilon}$. In fact, the parameter
$\epsilon$ is phase--convention independent; while neither
$\tilde{\epsilon}$, nor $A_{I}$ are. Indeed, under a small arbitrary phase
change of the $K^{0}$--state:
\be
\label{eq:phaserot}
\mid K^{0}\rangle\ra e^{-i\vp}\mid K^{0}\rangle,
\ee

\noi the parameters $A_{I}$, $M_{12}$, and $\tilde{\epsilon}$ change
as follows:
\beqn
\label{eq:ImArot} {\rm Im}A_{I} & \ra & {\rm Im}A_{I}-\vp {\rm Re}A_{I};
\\
\label{eq:M12rot} {\rm Im}M_{12} & \ra & {\rm Im}M_{12}+\vp \Delta m;
\\
\label{eq:epstilderot}
\tilde{\epsilon} & \ra & \tilde{\epsilon} +i\vp ;
\eeqn

\noi while $\epsilon$ remains invariant. The Wu--Yang phase convention was made
prior to the development of the electroweak theory. In the standard model, the
conventional way by which the freedom in the choice of relative phases of the
quark--fields has been frozen, is not compatible with the Wu--Yang convention.
Since $\epsilon$ is convention independent, we shall keep it as one of the
fundamental parameters. Then, however, we need a second parameter which
characterizes the amount of {\it intrinsic} CP--violation specific to the $K\ra
2\pi$ decay, by contrast to the CP--violation in the $K^{0}-\bar{K^{0}}$
mass--matrix. The parameter we are looking for has to be sensitive then to the
lack of relative reality of the the two isospin amplitudes $A_{0}$ and
$A_{2}$. This is the origin of the famous $\epsilon^{'}$--parameter, which we
shall next discuss.

In general, we can define three independent ratios of the
$K_{L,S}\ra (2\pi)_{I=0,2}$ transition amplitudes. One is the
$\epsilon$--parameter in (\ref{eq:eps}). Two other natural ratios are
\be
\frac{A[K_{L}\ra (\pi\pi)_{I=2}]}{A[K_{S}\ra (\pi\pi)_{I=0}]} \qquad {\rm
and}\qquad
\omega \equiv \frac{A[K_{S}\ra (\pi\pi)_{I=2}]}{A[K_{S}\ra (\pi\pi)_{I=0}]}\,.
\ee

\noi Both ratios can be expressed in terms of the
$\tilde{\epsilon}$--parameter introduced in Eq.~(\ref{eq:epstilde}), and the
complex
$A_{I}$--amplitudes defined in~(\ref{eq:KI}) and~(\ref{eq:KBI}):
\beqn
\label{eq:epsilonprime1}
\frac{A[K_{L}\ra (\pi\pi)_{I=2}]}{A[K_{S}\ra (\pi\pi)_{I=0}]} & = &
\frac{(1+\tilde{\epsilon})A_{2}-(1-\tilde{\epsilon})A_{2}^{*}}
{(1+\tilde{\epsilon})A_{0}+(1-\tilde{\epsilon})A_{0}^{*}}
e^{i(\delta_{2}-\delta_{0})}\no \\ & = &
\frac{i\frac{{\rm Im}A_{2}}{{\rm Re}A_{0}}+\tilde{\epsilon}
\frac{{\rm Re}A_{2}}{{\rm Re}A_{0}}}{1+i\tilde{\epsilon}\frac{{\rm
Im}A_{0}}{{\rm Re}A_{0}}} e^{i(\delta_{2}-\delta_{0})}\, ;
\eeqn

\noi and
\beqn
\label{eq:omega}
\omega \equiv \frac{A[K_{S}\ra (\pi\pi)_{I=2}]}{A[K_{S}\ra (\pi\pi)_{I=0}]}
& = & \frac{(1+\tilde{\epsilon})A_{2}+(1-\tilde{\epsilon})A_{2}^{*}}
{(1+\tilde{\epsilon})A_{0}+(1-\tilde{\epsilon})A_{0}^{*}}
e^{i(\delta_{2}-\delta_{0})} \no \\
& = & \frac{\frac{{\rm Re}A_{2}}{{\rm Re}A_{0}}+\tilde{\epsilon}
\frac{{\rm Im}A_{2}}{{\rm Re}A_{0}}}{1+i\tilde{\epsilon}\frac{{\rm Im}A_{0}}
{{\rm Re}A_{0}}}e^{i(\delta_{2}-\delta_{0})}\,.
\eeqn

The $\epsilon'$--parameter is then defined as the following combination of
these ratios :
\be
\label{eq:epsilonprime}
\epsilon'=\frac{1}{\sqrt{2}} \left(\frac{A[K_{L}\ra (\pi\pi)_{I=2}]}{A[K_{S}\ra
(\pi\pi)_{I=0}]}-
\epsilon\times\omega\right)\,.
\ee

\noi From these results, and using the expression for $\epsilon$ we obtained in
Eq.~(\ref{eq:eps2}), we finally get
\be
\label{eq:epsilonprime2}
\epsilon'=\frac{i}{\sqrt{2}}
\frac{(1-\tilde{\epsilon}^2 )e^{i(\delta_{2}-\delta_{0})}} {({\rm
Re}A_{0}+i\tilde{\epsilon}{\rm Im}A_{0})^2} ({\rm Im}A_{2}{\rm Re}A_{0}-{\rm
Im}A_{0}{\rm Re}A_{2})\,,
\ee

\noi an expression which clearly shows the proportionality to the lack of
relative reality between the $A_{0}$ and $A_{2}$ amplitudes.

We shall next establish contact with the parameters $\eta_{+-}$ and
$\eta_{00}$, which were introduced in Eq.~(\ref{eq:etapm}, and which are
directly accessible to experiment. Using Eqs.~(\ref{eq:pmisospin}),
(\ref{eq:zzisospin}), as well as the definitions of
$\epsilon$,
$\epsilon^{'}$, and $\omega$ above, one finds
\beqn
\label{eq:etapm1}
\eta_{+-} & = & \epsilon+\epsilon^{'}\frac{1}{1+\frac{1}{\sqrt{2}}\omega}\,;\\
\label{eq:etazz1}
\eta_{00} & = & \epsilon-2\epsilon^{'}\frac{1}{1-\sqrt{2}\omega}\,.
\eeqn

So far, we have made no approximations in our phenomenological analysis of the
$K^0 -\bar{K^0}$ mass--matrix and $K\ra 2\pi$ decays. It is however useful to
try to thin down in some way the exact expressions we have derived, by taking
into account the relative size of the various phenomenological parameters which
appear in the expressions above. The strategy will be to neglect first, terms
which are products of  CP--violation parameters. For example, in
Eq.~(\ref{eq:omega}), we have introduced the parameter $\omega$, which a priori
we can reasonably expect to be dominated by the term
\be
\label{eq:omega1}
\omega \simeq \frac{{\rm Re}A_{2}}{{\rm Re}A_{0}}e^{i(\delta_{2}-\delta_{0})}.
\ee

\noi We can justify this approximation by the fact that non--leptonic
$\Delta I=\frac{3}{2}$ \, transitions, although suppressed with respect to the
$\Delta I=\frac{1}{2}$\, transitions, are nevertheless larger than the
observed  CP--violation effects. Notice that the amplitude
$A_{2}$ is responsible for the deviation from an exact $\Delta I=
\frac{1}{2}$\, rule. The ratio $\frac{{\rm Re}A_{2}}{{\rm Re}A_{0}}$ can be
obtained from the experimentally known branching ratios
$\Gamma (K_{S}\ra \pi^{+}\pi^{-})$ and
$\Gamma (K_{S}\ra \pi^{0}\pi^{0})$. More precisely, correcting for the
phase--space effects, one must compare the normalized decay rates:
\be
\label{eq:NPSBR}
\gamma (1,2)\equiv \frac{\Gamma (K\ra \pi_{1}\pi_{2})} {\frac{1}{16\pi
M}\sqrt{1-\frac{(m_{1}+m_{2})^2 }{M^2 }}
\sqrt{1-\frac{(m_{1}-m_{2})^2 }{M^2 }}}\, ,
\ee

\noi where the denominator here is the two--body phase space factor for the
mode $K\ra \pi_{1}\pi_{2}$, ($M$ is the mass of the $K$-particle and
$m_{1,2}$ the pion masses.) Then, we have
\be
\label{eq:A2vA0}
\frac{\gamma_{S}(+-)}{2\gamma_{S}(00)}= 1+3\sqrt{2}\frac{{\rm Re}A_{2}}{{\rm
Re}A_{0}}\cos (\delta_{2}-\delta_{1})+
\cO (\frac{\alpha}{\pi}).
\ee

\noi Experimentally, from the PDB~\cite{PDB94}, one finds
\be
\label{eq:A2vA0exp}
\frac{\gamma_{S}(+-)}{2\gamma_{S}(00)}=1.109\pm 0.012,
\ee

\noi and using the present experimental information on
$(\delta_{2}-\delta_{1})$, [see the discussion in {\bf Sec.\,5.},] we find,
with neglect of radiative corrections
\be
\label{eq:A2vA0exp1}
\frac{{\rm Re}A_{2}}{{\rm Re}A_{0}}= (+22.2)^{-1}.
\ee

\noi I shall later discuss some of the qualitative dynamical explanations,
within the standard model, of how this small number appears. It is fair to say
however, that a reliable calculation of this ratio is still lacking at present.

Using the approximations
\be
\label{eq:approx}
\tilde{\epsilon}\, {\rm Im}A_{0}\ll {\rm Re}A_{0}\qquad {\rm and}
\qquad
\tilde{\epsilon}^2 \ll 1\, ,
\ee

\noi we can rewrite $\epsilon'$ in a simpler form
\be
\label{eq:epsilonprime3}
\epsilon' \simeq \frac{1}{\sqrt{2}}
e^{i(\delta_{2}-\delta_{0}+\frac{\pi}{2})}\,
\frac{{\rm Re}A_{2}}{{\rm Re}A_{0}}
\left(\frac{{\rm Im}A_{2}}{{\rm Re}A_{2}}-\frac{{\rm Im}A_{0}} {{\rm
Re}A_{0}}\right)\, ,
\ee

\noi clearly showing the fact that {\it $\epsilon'$ is proportional to direct
CP--violation in $K\ra 2\pi$ transitions and is also suppressed by the $\Delta
I=\frac{1}{2}$ selection rule.}

The same approximations in Eq.~(\ref{eq:approx}), when applied to
$\epsilon$, lead to
\be
\label{eq:eps3}
\epsilon \simeq \tilde{\epsilon}+i\frac{{\rm Im}A_{0}} {{\rm Re}A_{0}}\,.
\ee

Let us next go back to the mass matrix equations in (\ref{eq:relations}) which,
expanding in powers of
$\tilde{\epsilon}$, we can rewrite as follows
\be\label{eq:relations1} 1-2\tilde{\epsilon}\simeq
\frac{{\rm Re}M_{12}-\frac{i}{2}{\rm Re}\Gamma_{12}} {\frac{1}{2}(\Delta
m+\frac{i}{2}\Delta \Gamma )}-i
\frac{{\rm Im}M_{12}-\frac{i}{2}{\rm Im}\Gamma_{12}} {\frac{1}{2}(\Delta
m+\frac{i}{2}\Delta \Gamma )}\,.
\ee

\noi To a first approximation, neglecting CP--violation effects altogether, we
find that
\be\label{eq:mdiff}
\label{eq:relations2} {\rm Re}M_{12}\simeq \frac{\Delta m}{2}\qquad
{\rm and} \qquad  {\rm Re}\Gamma_{12}\simeq -\frac{\Delta\Gamma}{2}.
\ee

\noi
If furthermore, we restrict the sum over intermediate states in
$\Gamma_{12}$ [see Eq.~(\ref{eq:abs12})] to $2\pi$--states, an
approximation which we have already seen to be rather good [see
Eqs.~(\ref{eq:semilep}) and (\ref{eq:3pi})] we can write
\be
\label{eq:relations3}
\Gamma_{12}\simeq (-iA_{0}^{*}e^{i\delta_{0}})^{*}
iA_{0}e^{i\delta_{0}}=-({\rm Re}A_{0}+i{\rm Im}A_{0})^{2}\, ,
\ee

\noi
{}from where it follows that
\be
\label{eq:relations4}
\frac{{\rm Im}\Gamma_{12}}{{\rm Re}\Gamma_{12}}\simeq
\frac{2{\rm Re}A_{0}{\rm Im}A_{0}}{{\rm Re}A_{0}^{2}+{\rm Im}A_{0}^{2}}\simeq
2\frac{{\rm Im}A_{0}}{{\rm Re}A_{0}}\, .
\ee

\noi Then, using the empirical fact that
$\Delta m\simeq \frac{\Gamma_{S}}{2}$, and
$\Gamma_{L}\ll \Gamma_{S}$, we finally arrive at the simplified
expression
\be
\label{eq:epstilde1}
\tilde{\epsilon}\simeq
\frac{1}{1+i}\left(i\frac{{\rm Im}M_{12}}{\Delta m}
+\frac{{\rm Im}A_{0}}{{\rm Re}A_{0}}\right)\,.
\ee

\noi
and, using Eq.~(\ref{eq:eps3}),
\be
\label{eq:eps4}
\epsilon \simeq \frac{1}{\sqrt{2}}e^{i\frac{\pi}{4}}
\left(\frac{{\rm Im}M_{12}}{\Delta m}+\frac{{\rm Im}A_{0}}
{{\rm Re}A_{0}}\right)\, .
\ee

This is as much as one can do, within a strict phenomenological
analysis of the CP--violation in $K$--decays. We have reduced the
problem to the knowledge of two parameters: $\epsilon$ in
Eq.(\ref{eq:eps4}), and $\epsilon'$ in Eq.(\ref{eq:epsilonprime3}).
We shall come back to these parameters in {\bf Sec.\,5.} There, we
shall discuss what predictions for these fundamental parameters can be
made at present within the framework of the Standard Model. As we
shall see, the main difficulty comes from the lack of
quantitative understanding of the low--energy sector of the strong
interactions. In terms of QCD, the sector in question is the one of
the interactions between the states with lowest masses: the octet of
the pseudoscalar particles $(\pi, K, \eta)$. It seems therefore
appropriate to examine the possibility of describing the interactions of
these particles within the framework of an effective Lagrangian of the
Standard Model at very low--energies. This will be the subject of the next
three lectures.

%%%%%%%%%%%%%%%%%%%%%%%%%%%%%%%%%%%%%%%%%%%%%%%%%%%%%%%%%%%%
\section{Introduction to Chiral Perturbation Theory}
\label{sec:ICPT}

Chiral perturbation theory ($\chi$PT) is the effective field theory  of quantum
chromodynamics (QCD) at low energies. In this lecture I shall give an overview
of the basic ideas behind the
$\chi$PT--approach to hadron dynamics at low energies. This will give us the
basis for a study of possible applications of the
$\chi$PT--approach to the non--leptonic weak interactions of
$K$--particles, which will then be the subject of the next lecture.

There have already been several sets of TASI--lectures in previous years
dedicated to
$\chi$PT. The reader should consult them for complementary
information~\cite{L91}$^{,}$~\cite{D93}. Some recent review articles can be
found in~\cite{L92}$^{,}$~\cite{E92}$^{,}$~\cite{P93}$^{,~{\rm
and}}$~\cite{L94}.

\subsection{An Overview of the Basic Ideas in the $\chi$PT-Approach.}
\label{subsec:OBICPT}

In the limit where the heavy quark fields $t$, $b$, and $c$ are integrated out,
and the masses of the light quarks
$u$, $d$ and
$s$ are set to zero, the QCD Lagrangian is invariant under global
$SU(3)_{L}\times SU(3)_{R}$ rotations
$(V_L,V_R)$ of the left--handed and right--handed quark triplets
\be
q_L\equiv {1-\gamma_5\over 2}q \qquad {\rm and} \qquad q_R\equiv
{1+\gamma_5\over 2}q \qquad ;\: q=u,\ d,\ s.
\ee
\be
q_{L}\ra V_{L}\,q_{L} \qquad {\rm and} \qquad  q_{R}\ra V_{R}\,q_{R}\,.
\ee

\noi Formally, the global symmetry of the QCD--Lagrangian is in fact larger.
The full symmetry group is
$SU(3)_{L}\times SU(3)_{R}\times U(1)_{V}\times U(1)_{A}$. The
$U(1)_{A}$ symmetry is broken however at the quantum level by the abelian axial
anomaly; the Adler~\cite{A69}--Bell--Jackiw~\cite{BJ69} anomaly. The
$U(1)_{V}$ quark--number symmetry is trivially realized in the mesonic sector.
At the level of the hadronic spectrum, this chiral--$SU(3)$ symmetry of the
QCD--Lagrangian is however not apparent. Although the low--lying hadronic
states can indeed be neatly classified in irreducible representations of the
famous
$SU(3)$ symmetry of the Eightfold Way~\cite{GN64}, there do not appear
degenerate multiplets with opposite parity. In QCD it is therefore expected,
(and there are some good theoretical reasons for it~\cite{CW80}; as well as
numerical evidence from lattice QCD simulations~\cite{Sharpe}) that the
chiral--$SU(3)$ global symmetry is {\it spontaneously broken} down to the
diagonal
$SU(3)_{V=L+R}$ group of the Eightfold Way.  This pattern of spontaneously
broken symmetry implies specific constraints on the dynamics of the strong
interactions between the low--lying pseudoscalar states $(\pi,\ K,\
\eta)$, which are the massless  Nambu--Goldstone bosons associated to the
``broken'' chiral generators. As a result of the spontaneous symmetry breaking,
there appears a mass--gap in the hadronic spectrum between the ground state of
the octet of
 $0^-$--pseudoscalars and the lowest hadronic states which become massive in
the chiral limit $m_u=m_d=m_s=0$~; i.e., the octet of
$1^-$ vector--meson states and the octet of $1^+$ axial--vector--meson states.

The basic idea of the
$\chi$PT--approach is that in order to describe the physics at energies within
this gap region, it may be more useful to formulate the strong interactions of
the low--lying pseudoscalar particles in terms of an effective low--energy
Lagrangian of QCD, with the octet of Nambu--Goldstone fields
($\overrightarrow{\lambda}$ are the eight
$3\times 3$ Gell-Mann matrices, with ${\rm tr}\lambda^{a}\lambda^{b}=
2\delta_{ab}$)
\be \label{eq:phi}
\Phi(x)={\overrightarrow{\lambda}\over
\sqrt{2}}\cdot\overrightarrow{\varphi}(x)=
\normalbaselineskip=18pt\pmatrix{\pi^{\circ}/\sqrt{2}+
\eta/\sqrt{6}&\hfill\pi^+\hfill &\hfill K^+\hfill\cr
\hfill\pi^-\hfill &-\pi^{\circ}/\sqrt{2}+\eta/\sqrt{6}&
\hfill K^{\circ}\hfill \cr
\hfill K^- \hfill &\hfill\overline K^{\circ} \hfill &-2\eta/\sqrt{6}},
\ee

\noi as explicit degrees of freedom, rather than in terms of the quark and
gluon fields of the usual QCD Lagrangian. In the conventional formulation, the
Nambu--Goldstone fields are collected in a unitary
$3\times 3$ matrix $U(x)$ with $\det U=1$, which under chiral--$SU(3)$
transformations $(V_L,V_R)$ is chosen to transform linearly:
\be  U\rightarrow V_R U V_L^{\dag}.
\ee

\noi The effective Lagrangian we look for has to be then a sum of chirally
invariant terms with increasing number of derivatives of $U$. For example, to
lowest order in the number of derivatives, only one independent term can be
constructed which is invariant under
$(V_L,\ V_R)$ transformations~:
\be \label{eq:LO2} {\cal L}_{\rm eff}^{(2)}={1\over 4}f^2_{\pi}tr
\partial_{\mu}
U(x)\partial^{\mu}U^{\dagger}(x),
\ee

\noi where, as I shall soon explain in more detail, the normalization
$f_{\pi}$ is fixed in such a way that the corresponding axial--current with the
quantum numbers of the pion, obtained from this Lagrangian, is  the one which
induces the
$\pi\to\mu\nu$ transition. An explicit representation of $U$ is
\be\label{eq:U}
U(x)=\exp\left(-i{1\over f_{\pi}}\overrightarrow{\lambda}\cdot
\overrightarrow{\varphi}(x)\right)\ ;
\ee

\noi and, from experiment,
\be f_{\pi}\simeq 92.4{\rm MeV}.
\ee

\noi Because of the non-linearity in
${\varphi}$, processes with different numbers of pseudoscalar mesons are then
related. These are the successful current--algebra relations~\cite{AD68} of the
60's which the effective Lagrangian formulation above~\cite{W67} incorporates
in a compact way.

The low--energy effective Lagrangian ${\cal L}_{\rm eff}^{(2)}$ describes
physical amplitudes by means of the lowest order term in a Taylor expansion in
powers of momenta:
\be {\cal A}(p_1,p_2,...)=\sum a_{ij}^{(2)}p_{i}\cdot p_{j}  + {\cal O}(p^4).
\ee

\noi The expansion has no constant ${\cal O}(p^0)$ term, due to the fact that
the lowest order Lagrangian in (\ref{eq:LO2}) has two derivatives. As an
exercise you should check some examples, like elastic $\pi^+ \pi^0$ scattering,
where the induced amplitude one finds is:
\be {\cal A}(p_{+},p_{0},p_{+}^{'},p_{0}^{'})=
\frac{(p_{+}^{'}-p_{+})^2}{f_{\pi}^2}.
\ee

We expect this effective Lagrangian description to be useful for values of
invariants of momenta sufficiently small as compared to the scale
$\Lambda_{\chi}$ where spontaneous chiral symmetry breaking (S$\chi$SB) occurs
in QCD:
\be p^2/\Lambda_{\chi}^2\ll1.
\ee

\noi It seems also reasonable to expect $\Lambda_{\chi}$ to be of the same
order of magnitude as the masses of the lowest states which become massive due
to S$\chi$SB, i.e.,
\be\label{eq:Lscale} M_{\rho}(770 MeV)\leq \Lambda_{\chi}\leq M_A(1260 MeV).
\ee

\noi As we shall see, these expected features are justified phenomenologically.

It is convenient to promote the global chiral--$SU(3)$ symmetry to a local
$SU(3)_L\times SU(3)_R\ $ gauge symmetry. This can be accomplished by adding
appropriate quark bilinear couplings with external field sources to ${\cal
L}^{(0)}_{QCD}$, the usual QCD Lagrangian with massless quarks:
\be \label{eq:LQCD} {\cal L}_{QCD}(x)={\cal L}^{(0)}_{QCD}(x)+\overline
q(x)\gamma^{\mu}[v_{\mu}(x)+\gamma_5 a_{\mu}(x)]q(x)-\overline
q(x)[s(x)-i\gamma_5 p(x)]q(x).
\ee

\noi The external field sources
$v_{\mu},\ a_{\mu},\ s$ and $p$ are Hermitian $3\times 3$ matrices in flavour,
and colour singlets. Under chiral--$SU(3)$ gauge transformations ($V_{L}(x)$,
$V_{R}(x)$), they are required to transform as follows:
\bea \label{eq:lrsources} l_{\mu}\equiv v_{\mu}-a_{\mu} & \rightarrow &
V_{L}l_{\mu}V_{L}^{\dagger} + iV_{L}\partial_{\mu}V_{L}^{\dagger}, \\
r_{\mu}\equiv v_{\mu}+a_{\mu} & \rightarrow &  V_{R}r_{\mu}V_{R}^{\dagger}+
iV_{R}\partial_{\mu}V_{R}^{\dagger};
\eea

\noi and
\be s+ip \rightarrow V_R(s+ip)V_L^{\dagger}.
\ee
\noi In the presence of these external field sources, the possible terms in
${\cal L}_{\rm eff}^{(2)}$ with the lowest chiral dimension,  i.e.,
${\cO}(p^2)$ are
\be \label{eq:LOp2} {\cal L}_{\rm eff}^{(2)}={1\over 4}f^2_{\pi}
\left\{{\rm tr} D_{\mu}UD^{\mu}U^{\dagger} +{\rm \tr}\left(\chi U^{\dagger} +
U\chi^{\dagger}\right)\right\},
\ee

\noi where $D_{\mu}$ denotes the covariant derivative
\be D_{\mu}U=\partial_{\mu}U-ir_{\mu}U+iUl_{\mu},
\ee
and
\be\label{eq:chies}
\chi=2B[s(x)+i p(x)],
\ee

\noi with $B$ a constant, which like
$f_{\pi}$, is not fixed by symmetry requirements alone. Once special directions
in flavour space (like the ones selected by the electroweak Standard Model
couplings) are fixed for the external fields, the chiral symmetry is then
explicitly broken. In particular, the choice
\be \label{eq:spmasses} s+i p={\cal M}=\hbox{\rm diag}\, (m_u,\ m_d,\ m_s)
\ee
\noi takes into account the explicit breaking due to the quark masses in the
underlying QCD Lagrangian.

Notice that the quark bilinear $\bar{q}_L^{j}(x)q_R^{i}(x)$ has the same
transformation properties, under chiral--$SU(3)$, as the
$U(x)$--matrix field. In fact, $U(x)$ can be viewed as the low--energy
parametrization of the  Nambu--Goldstone field excitations of the vacuum. To
lowest order in the chiral expansion, the constant $B$ appears then as the
parameter which fixes the relative normalization to the light quark condensate:
\be
\left<0|\overline q^jq^i|0\right>=-f^2_{\pi}B\delta_{ij}.
\ee

The relation between the physical pseudoscalar masses and the quark masses, to
lowest order in the chiral expansion, is fixed by identifying quadratic terms
in ${\varphi}$ in the expansion of the second term in (\ref{eq:LOp2}), with the
result
\be
\chi=\normalbaselineskip=18pt\pmatrix{ m^2_{\pi^+}+M^2_{K^+}-M^2_{K^0} &\hfill
0\hfill &\hfill 0\hfill\cr
\hfill 0\hfill & m^2_{\pi^+}-M^2_{K^+}+M^2_{K^0} &\hfill 0\hfill\cr
\hfill 0\hfill &\hfill 0\hfill & -m^2_{\pi^+}+M^2_{K^+}+M^2_{K^0} }.
\ee

\noi The mass matrix in the $\pi^0$, $\eta$ basis is not diagonal, because of a
small admixture between these two fields. The appropriate diagonalization leads
to the mass eigenvalues:
\bea M_{\pi^{0}}^2 & = & 2\hat{m}B -\epsilon +{\cal O}(\epsilon^2),  \\
M_{\eta}^2 & = & \frac{2}{3}(\hat{m}+2m_s)B
 +\epsilon  +{\cal O}(\epsilon^2),
\eea

where
\be
\epsilon = \frac{B}{4}\frac{(m_u-m_d)^2}{(m_s-\hat{m})},
\ee

and
\be
\hat{m} \equiv \frac{1}{2}(m_u+m_d).
\ee

\noi There are a number of important relations which follow from these lowest
order results:

\begin{itemize}

\item{The Gell-Mann--Oakes--Renner relation~\cite{GOR68}}
\be\label{eq:GOR} f_{\pi}^2M_{\pi}^2=-\hat{m}<0|\bar{u}u+\bar{d}d|0>.
\ee

\noi
\item{The Current Algebra mass ratios~\cite{GOR68}$^{,}$~\cite{W77}}
\be B=\frac{M_{\pi^{\pm}}^2}{2\hat{m}}=
\frac{M_{K^{+}}}{(m_u+m_d)}=\frac{M_K^0}{(m_d+m_s)}
\approx \frac{3M_{\eta_{8}}^2}{(2\hat{m}+4m_s)}.
\ee

\noi
\item{The Gell-Mann--Okubo mass formula~\cite{G57}$^{,}$~\cite{O62}}
\be 3M_{\eta_{8}}^2=4M_{K}^2-M_{\pi}^2.
\ee

\end{itemize}

Barring the possibility that the parameter $B$ may be
unexpectedly small (see however~\cite{KS93}, and references thereof,) and
neglecting the small ${\cal O}(\epsilon)$ effects, we are led to two quark
mass ratios estimates~\cite{W77}:
\be
\frac{m_{d}-m_{u}}{m_{d}+m_{u}}=\frac{ M^2 _{K^0 }-M^2 _{K^+ }-(M^2 _{\pi^0
}-M^2_{\pi^+} )_{EM}} {M^2 _{\pi^0}}=0.29\,;
\ee

and
\be
\frac{m_{s}-\hat{m}}{2\hat{m}}=\frac{M^2 _{K^0 }-M^2 _{\pi^0}}{M^2
_{\pi_{0}}}=12.6\,,
\ee

\noi where we have subtracted the pion squared mass difference, to take into
account the electromagnetic contribution to the charged pseudoscalar mesons
self--energies. This electromagnetic contribution, in the chiral limit
($m_{u,d,s}=0$,) gives a common mass to the charged $K^+$ and $\pi^+$ mesons.

As already pointed out, in the case of the Standard Model, not all the external
gauge field sources $l_{\mu}$ and $r_{\mu}$ correspond to  {\em physical} gauge
fields. In the same way that the scalar and pseudoscalar sources are frozen to
the quark mass matrix, as indicated in (\ref{eq:spmasses}), the explicit chiral
symmetry breaking induced by the electroweak currents of the Standard Model
corresponds to the following choice:
\bea r_{\mu} & = & eQ_R[A_{\mu}-\tan\vartheta_{W}\,Z_{\mu}]; \\ l_{\mu} & = &
eQ_R[A_{\mu}-\tan\vartheta_{W}\,Z_{\mu}]+
\frac{e}{\sin\vartheta_{W}}Q_L^{(3)}\,Z_{\mu} \no \\ & + &
\frac{e}{\sqrt{2}\sin\vartheta_{W}}
[Q_L^{(+)}W_{\mu}^{(+)}+Q_L{(-)}W_{\mu}^{(-)}].
\eea

\noi Here the $Q$'s are the electroweak matrices:
\be \label{eq:charges} Q_{L}=Q_{R}=Q=\frac{1}{3}\hbox{\rm diag}(2,-1,-1),\qquad
Q_{L}^{(3)}=\hbox{\rm diag}(1,-1,-1);
\ee

\noi and
\be Q_{L}^{(+)}=(Q_{L}^{(-)})^\dagger = \left(
\begin{array}{ccc} 0 & V_{ud} & V_{us} \\ 0 & 0 & 0 \\ 0 & 0 & 0
\end{array} \right).
\ee

We can now compute, in a straightforward way, the chiral realization --to
lowest order in the chiral expansion--  of the electroweak currents of the
Standard Model. They follow by taking appropriate variations of the lowest
order effective action $\Gamma^{(2)}=\int d^{4}x{\cal L}^{(2)}(x)$, with
respect to the external field sources:
\be \label{eq:lc} J_{L}^{\mu}\equiv \bar{q}_{L}\gamma^{\mu}q_{L}\doteq
\frac{\delta\cL^{(2)}}{\delta l_{\mu}}  =
\frac{i}{2}f_{\pi}^2(D_{\mu}U^{\dagger})U\,,
\ee
\be \label{eq:rc} J_{R}^{\mu}\equiv \bar{q}_{R}\gamma^{\mu}q_{R}\doteq
\frac{\delta\cL^{(2)}}{\delta r_{\mu}}  =
\frac{i}{2}f_{\pi}^2(D_{\mu}U)U^{\dagger}\,.
\ee

\noi Expanding $U$ in powers of $\Phi$-matrix fields [see Eqs.(\ref{eq:phi})
and (\ref{eq:U}),] we have:
\bea J_{L}^{\mu} & = & f_{\pi}\frac{1}{\sqrt{2}}D^{\mu}\Phi -\frac{i}{2}[\Phi(
D^{\mu}\Phi)-(D^{\mu}\Phi)\Phi]+ {\cal O}(\Phi^3), \\ J_{R}^{\mu} & = &
-f_{\pi}\frac{1}{\sqrt{2}}D^{\mu}\Phi -\frac{i}{2}[\Phi(
D^{\mu}\Phi)-(D^{\mu}\Phi)\Phi]+ {\cal O}(\Phi^3).
\eea

\noi Let us now consider the decay $\pi^{+}\rightarrow \mu^{+}\nu_{\mu}$ we
mentioned earlier as an example. The hadronic matrix element can be readily
computed using the form of the axial current deduced from the expressions just
above, with the result
\be <0|(J_{A}^{\mu})_{12}|\pi^{+}>=i\sqrt{2}f_{\pi}p^{\mu},
\ee
\noi explicitly showing the physical meaning of the $f_{\pi}$ constant in the
chiral Lagrangian.

You are now in the position of being able to calculate any observable you wish
to lowest order in the chiral expansion; but I am sure you are more ambitious.
{\it How does one calculate higher order chiral corrections?} This is what we
shall now discuss in the next subsection.

%%%%%%%%%%%%%%%%%%%%%%%%%%%%%%%%%%%%%%%%%%%%%%%
\subsection{Chiral Perturbation Theory to ${\cal O}(p^4)$}
\label{subsec:LLC}

In QCD, the generating functional $Z[v,a,s,p]$ of the Green's functions of
colour singlet quark currents, is defined via the path--integral formula
\be \label{eq:qcdpi}
\exp\left\{iZ[v,a,s,p]\right\}=\int {\cal D}q{\cal D}\bar{q}{\cal D}G_{\mu}
\exp\left\{i\int d^4x{\cal L}_{QCD}\right\},
\ee
\noi with ${\cal L}_{QCD}$ defined as in eq.(\ref{eq:LQCD}). The physical
Green's functions of a specific flavour are then  obtained by functional
derivatives with respect to the appropriate external field sources $v,a,s,$ and
$p$. The chiral symmetry properties which we have discussed  imply that, at
sufficiently small energies, there exists an effective Lagrangian
${\cal L}_{\rm eff}$ of the  Nambu--Goldstone field modes alone, in the
presence of external field sources, such that
\be \label{eq:EFLPI}
\exp\left\{iZ[v,a,s,p]\right\}=\int {\cal D}U[\Phi(x)]
\exp\left\{i\int d^{4}x{\cal L}_{\rm eff}[U;v,a,s,p]\right\}.
\ee
\noi To lowest order in the chiral expansion, ${\cal O}(p^2)$ as we have seen,
the generating functional reduces to the classical action
\be Z^{(2)}[v,a,s,p]=\int d^{4}x\,{\cal L}_{\rm eff}^{(2)}(U;v,a,s,p),
\ee
\noi with ${\cal L}_{\rm eff}^{(2)}$ as defined in eq.(\ref{eq:LOp2}).

To next--to--leading order in the chiral expansion, i.e.${\cal O}(p^4)$, the
generating functional $Z[v,a,s,p]$ gets contributions from three different
sources:
\begin{itemize}
\item{The most general local effective chiral Lagrangian of ${\cal O}(p^4)$.}
\item{The one--loop functional generated from the lowest--order ${\cal L}_{\rm
eff}^{(2)}$ Lagrangian.}
\item{The well known Wess--Zumino--Witten
functional~\cite{WZ71}$^{,}$\cite{W83} induced by the non--abelian chiral
anomaly~\cite{B69}}.
\end{itemize}

%%%%%%%%%%%%%%%%%%%%%%%%%%%%%%%%%%%%%%%%%%%%%%
\subsubsection{\sl The Chiral Lagrangian of ${\cal O}(p^4)$}
\label{subsubsec:CL4}

The ingredients we have at our disposal to build this Lagrangian, their chiral
transformation properties, as well as  their chiral power counting are as
follows:
\beqan U\, ,  & \qquad V_{R}UV_{L}^{\dagger}\, , & \qquad {\cal O}(p^0)\, ; \\
D_{\mu}U\, , & \qquad V_{R}D_{\mu}UV_{L}^{\dagger}\, ,
& \qquad {\cal O}(p)\, ; \\
\chi\, ,   & \qquad V_{R}\chi V_{L}^{\dagger}\, ,  &
\qquad {\cal O}(p^2)\, ;
\\ F_{L}^{\mu\nu}\, ,   & \qquad  V_{L}F_{L}^{\mu\nu}V_{L}^{\dagger}\, , &
\qquad {\cal O}(p^2)\, ; \\  F_{R}^{\mu\nu}\, ,  & \qquad
V_{R}F_{R}^{\mu\nu}V_{R}^{\dagger}\, ,  &
\qquad {\cal O}(p^2)\, .
\eeqan
Here, $F_{L}^{\mu\nu}$ and $F_{R}^{\mu\nu}$ are the strength field
tensors associated to the external $l_{\mu}$ and $r_{\mu}$ sources:
\be\label{eq:fstes} F_{L}^{\mu\nu}  =
\partial^{\mu}l^{\nu}-\partial^{\nu}l^{\mu} -i[l^{\mu},l^{\nu}],  \qquad {\rm
and} \qquad F_{R}^{\mu\nu}  =  \partial^{\mu}r^{\nu}-\partial^{\nu}r^{\mu}
-i[r^{\mu},r^{\nu}].
\ee

\noi Furthermore, since we shall only use ${\cal L}_{\rm eff}^{(4)}$ at tree
level, we can eliminate possible terms using the ${\cal O}(p^2)$ equations of
motion obeyed by the $U$-functional:
\be\label{eq:eqsm} (\Box U)U^{\dagger}-U(\Box U^{\dagger})=
\chi U^{\dagger}-U\chi^{\dagger}-\frac{1}{3}{\rm tr} (\chi
U^{\dagger}-U\chi^{\dagger})\, .
\ee

\noi Remember also that traceless $3\times 3$ matrices $A_i$ obey the
constraint
\bea
\tr A_1A_2A_3A_4+\tr A_1A_3A_2A_4+\tr A_1A_4A_3A_2 & &
\no \\
+\tr A_1A_2A_4A_3+\tr A_1A_3A_4A_2+\tr A_1A_4A_2A_3+  & & \no \\
-\tr A_1A_2\tr A_3A_4-\tr A_1A_3\tr A_2A_4-\tr A_1A_4\tr A_2A_3
 & = & 0 \, .
\eea

\noi Other constraints which we must also impose are that only terms which are
invariant under parity and charge conjugation should be allowed. The most
general Lagrangian which satisfies all these conditions is the
following~\cite{GL85}
\beqn\label{eq:ELOP4} {\cal L}^{(4)}_{\rm eff} & = & L_1\, {\rm tr}\left(
D_{\mu}U^{\dagger} D^{\mu}U\right)^2+L_2\,{\rm tr}  D_{\mu} U^{\dagger} D_{\nu}
U{\rm tr} D^{\mu}U^{\dagger}D^{\nu}U+\no \\ & &\no \\
                       &   & L_3\,{\rm tr}  D_{\mu}U^{\dagger}
D^{\mu}UD_{\nu}U^{\dagger}D^{\nu}U+\no \\ & &\no \\
                       &   & L_4\, {\rm tr} D_{\mu}U^{\dagger}D^{\mu} U {\rm
tr}\left(\chi^{\dagger}U+U^{\dagger}\chi\right)+L_5\, {\rm tr}
D_{\mu}U^{\dagger}D^{\mu}U\left(\chi^{\dagger}U +U^{\dagger}\chi\right)+\no \\
& &
\no\\
                       &   & L_6\,\left[ {\rm tr}\left(\chi^{\dagger}{\cal
U}+U^{\dagger}\chi\right)\right]^2 +L_7\,\left[{\rm
tr}\left(\chi^{\dagger}U-U^{\dagger}
\chi\right)\right]^2 + \no\\  & &
\no \\
                       & & L_8\, {\rm tr}\left(U\chi^{\dagger}U
\chi^{\dagger} + U^{\dagger}\chi U^{\dagger}
\chi\right) + \no \\ & & \no \\
                       &   & iL_9\, {\rm tr}\left(F^{\mu\nu}_R D_{\mu}U
D_{\nu}U^{\dagger} +F^{\mu\nu}_L D_{\mu}U^{\dagger} D_{\nu}U\right)+L_{10}\,
{\rm tr} U^{\dagger} F^{\mu\nu}_R UF_{L\mu\nu}+ \no \\ & & \no \\
                       &   & H_1\, {\rm tr}\left(F^{\mu\nu}_R
F_{R\mu\nu}+F^{\mu\nu}_L F_{L\mu\nu}\right)+H_2\, {\rm tr}
\chi^{\dagger}\chi\, .
\eeqn

\noi In this Lagrangian the parameters $L_{i}$, $i=1,2,3,\dots,10$ are
dimensionless coupling constants, which like $f_{\pi}$ and $B$ in the lowest
order effective Lagrangian, are not fixed by chiral symmetry requirements
alone. The terms proportional to the coupling constants
$H_1$ and $H_2$ involve only the external fields. As a result these coupling
constants cannot be fixed from low--energy observables alone. Only if we had a
detailed knowledge of the dynamics of how the effective chiral Lagrangian
emerges from the underlying QCD Lagrangian could we give an unambigous
definition of $H_1$ and $H_2$. By contrast, as we shall later discuss, most
of the other couplings can be fixed from low energy observables. Of course, in
QCD, the $L_{i}$ constants, much the same as
$f_{\pi}$ and
$B$, are {\it in principle} calculable parameters in terms of the intrinsic
$\Lambda_{QCD}$ scale only. We shall come back to this important question later
in {\bf Sec.\,4}. For the time being we shall only be interested in the
phenomenological  determination of these constants. However, for that, we need
to discuss first the possible contributions to low--energy observables from
chiral loops.

%%%%%%%%%%%%%%%%%%%%%%%%%%%%%%%%%%%%%%%%%%%%%%

\subsubsection{\sl Chiral Loops}
\label{subsubsec:CLOOPS}

Simple power counting tells us that loops generated by the lowest order
Lagrangian are badly divergent. This is not a surprise: the non--linear sigma
model in 4--dimensions is not renormalizable i.e., an infinite number of local
counter--terms are required. Here, however, we are considering the chiral
Lagrangian as an effective field theory for low energies. Order by order in the
momentum expansion of the effective theory, it is possible to specify a
renormalizable framework. Of course, as we go to higher and higher powers of
momentum, more and more local counterterms will appear, with couplings which
are not fixed by chiral symmetry properties alone, and eventually the
predictive power of the effective field theory formulation will therefore
disappear. To define the loop integrals it is necessary to fix a regularization
which preserves the symmetries of the Lagrangian. The well known dimensional
regularization technique does the job beautifully. Since by construction, the
$\cO (p^4)$ Lagrangian
$\cL_{\rm eff}^{(4)}$ contains all possible terms which are allowed by chiral
invariance, all the one loop divergences  --which by power counting can only
give rise to local $\cO (p^4)$ terms--  can all be absorbed by suitable
renormalizations of the $L_{i}$ and $H_{1,2}$ constants.  This program has been
explicitly realized in the papers of Gasser and
Leutwyler~\cite{GL84}$^{,}$\cite{GL85}$^{,}$\cite{GL85bc}. In fact, these
authors have explicitly constructed the one--loop functional
$Z_{\rm loop}^{(4)}[v,a, s, p]$ at the required level of non--locality needed
for all practical calculations. They do the path integral around the classical
functional $U(\Phi_{\rm cl}[v,a,s,p])$, solution of the equations of motion in
(\ref{eq:eqsm}) with the boundary condition
$U(0)=1$. The method consists in expanding $\cL_{\rm eff}^{(2)}(U;v,a,s,p)$
around
$\Phi=\Phi_{\rm cl}$ and then doing the functional integral over the
fluctuations $\xi=\Phi-\Phi_{\rm cl}$. In order to obtain the one--loop
effective action it is sufficient to expand $\cL_{\rm eff}^{(2)}$ to
$\cO [(\xi)^2]$:
\be
\Gamma_{\rm eff}^{(2)}=\int d^4x \cL_{\rm eff}^{(2)}[\Phi_{\rm
cl}]-\frac{f_{\pi}^2}{2}(\xi,D\xi)+\cO (\xi^3),
\ee

\noi  where $D$ is a known, but rather complicated, differential operator
acting on the fluctuating
$\xi$--variables. The one--loop functional is then embodied in the master
formula:
\be Z_{\rm loop}^{(4)}[v,a, s, p]=\frac{i}{2}{\rm tr}\log D.
\ee

It is relatively easy to extract the singular part of
$Z_{\rm loop}^{(4)}[v,a, s, p]$. It corresponds to a local effective
Lagrangian, exactly like the one in eq.(\ref{eq:ELOP4}), with couplings
$L_{i}$ and $H_{j}$:
\be L_{i}^{\rm loop}=\Gamma_{i}\,\Lambda^{\rm loop}, \quad  i=1,2,3,...10;
\qquad H_{i}^{\rm loop}=\tilde{\Gamma}_{j}\,\Lambda^{\rm loop},  \quad j=1,2,
\ee

where
\be
\Lambda^{\rm loop}=\frac{\mu^{d-4}}{16\pi^2}\left\{\frac{1}{d-4}-
\frac{1}{2}[\log(4\pi)+\Gamma^{'}(1)+1]\right\}, \quad j=1,2;
\ee

and $\Gamma_{i}$, $\tilde{\Gamma}_{j}$ have the following rational values\, :
\beqn
\Gamma_1 = {3\over 32}, & \Gamma_2 = \dfrac{3}{16},\; & \Gamma_3 = 0\,\, ,
\qquad \Gamma_4 = {1\over 8} ,
\nonumber\\
\Gamma_5 = {3\over 8}\,\: , & \Gamma_6 = \dfrac{11}{144} , & \Gamma_7 = 0\,\, ,
\qquad \Gamma_8 = {5\over 48},
\no \\
\Gamma_9 = {1\over 4}\,\: , & \Gamma_{10} = -\dfrac{1}{4} , &
\widetilde\Gamma_1 = -{1\over 8}, \quad\,\, \widetilde\Gamma_2 = {5\over 24} .
\eeqn

\noi  The renormalized couplings $L_{i}^{\rm r}(\mu)$ and
$H_{j}^{\rm r}(\mu)$, are then defined by subtracting from the tree level
$L_{i}$ and
$H_{j}$ the one--loop singular contributions:
\be L_{i}=L_{i}^{\rm r}(\mu)+\Gamma_{i}\,\Lambda^{\rm loop}, \quad {\rm and}
\quad M_{j}=H_{j}^{\rm r}(\mu)+\tilde{\Gamma}_{j}\,\Lambda^{\rm loop}.
\ee

\noi  The renormalized coupling constants depend of course on the scale
$\mu$ introduced by the dimensional regularization. The running in
$\mu$ is governed by the coefficients $\Gamma_{i}$ (and
$\tilde{\Gamma_{j}}$), which play the r\^{o}le of one--loop
$\beta$--functions:
\be \label{eq:slaw} L_{i}^{\rm r}(\mu)= L_{i}^{\rm
r}(\mu^{'})+\frac{\Gamma_{i}}{16\pi^2}
\log\frac{\mu^{'}}{\mu}.
\ee

The $\mu$--scale dependence cancels however in the full $\cO (p^4)$ calculation
of a given physical observable. The non-polynomial contribution to a specific
physical process will in general have a logarithmic $\mu$--scale dependence
--the so called chiral logarithms-- which cancels with the $\mu$--dependence of
the tree level contribution modulated by the
$L_{i}(\mu)$--constants. Let us consider a typical example to illustrate this
feature: {\em the electromagnetic mean squared radius of the pion.} The pion
electromagnetic form factor, when expanded in a Taylor series in powers of the
momentum transfer, has the following structure:
\be  F_{\pi^{\pm}}(q^2)=1+\frac{1}{6}<r^2>_{\pi^{\pm}}\, q^2 + \cdots .
\ee

\noi  The result of the calculation of $<r^2>_{\pi^{\pm}}$, to lowest
non--trivial $\cO (p^4)$ in $\chi$PT is~\cite{GL85}
\be  <r^2>_{\pi^{\pm}}=12\frac{L_9^{\rm r}(\mu)}{f_{\pi}^2}-
\frac{1}{16\pi^{2}f_{\pi}^2}[\log\frac{M_{\pi}^2}{\mu^2}+\frac{1}{2}\log
\frac{M_{K}^2}{\mu^2}+\frac{3}{2}].
\ee

\noi  The first term in the r.h.s. shows the tree level contribution from the
renormalized $\cL_{\rm eff}^{(4)}$ Lagrangian; the rest of the terms are
generated by the --finite though $\mu$--dependent-- contribution from the
Feynman diagram loops (one--pion loop and one--kaon loop,) generated by the
lowest $\cL_{\rm eff}^{(2)}$ Lagrangian. The scaling law of the
$L_{9}^{\rm r}(\mu)$--constant, as shown in (\ref{eq:slaw}), is the same as the
one of the chiral logarithm, and therefore the whole contribution is scale
invariant, as it should be.

There are a number of interesting generic features which emerge from this
example, and which we next wish to point out:

\begin{itemize}

\item{The factor $\frac{q^2}{16\pi^{2}f_{\pi}^2}$ is a characteristic factor of
the loop--expansion. More precisely, this factor appears modulated by the
number of active flavour loops $n_{\rm f}$. Therefore, we expect chiral loops
to contribute $\cO [n_{\rm f}\frac{q^2}{16\pi^{2}f_{\pi}^2}\times
\log^{,}{\rm s}]$ to physical processes in general.}

\item{Experimentally~\cite{Aetal86}
$<r^2>_{\pi^{\pm}}=(0.439\pm 0.008){\rm fm}^2$, showing that the contribution
of the $L_{9}^{\rm r}$--constant, for any reasonable value of $\mu$, dominates
the electromagnetic mean squared radius of the pion. In other words, in this
example, the tree level $\cO (p^4)$ contribution largely dominates the ``chiral
logs'' induced from the lowest order Lagrangian. This is a fact which can be
easily understood within the framework of the
$1/N_c$--expansion\footnote{\ This is the approximation proposed by 't
Hooft~\cite{'tH74}, where the number of colours $N_c$ in QCD is large with the
product
$\alpha_{s}N_c$ kept fixed.}\hspace*{1em}: in the large--$N_c$ limit,
$L_9$ and
$f_{\pi}^2$ are
${\cO (N_c)}$; therefore the chiral loop contribution is
$1/N_c$--suppressed with respect to the tree level contribution.}

\item{The phenomenological value we get, from this observable, for the
$L_9$ coupling constant at the scale of the $\rho$--meson mass:
$L_{9}^{\rm r}(770{\rm MeV})=(6.9\pm 0.7)\times 10^{-3}$ is in the range of the
expected order of magnitude if, as we have assumed a priori, the chiral
Lagrangian is a good effective theory for energies below the S$\chi$SB scale
$\Lambda_{\chi}$. With a factor $\frac{1}{4}f_{\pi}^2$ pulled out from
$\cL_{\rm eff}^{(4)}$, (the same $\frac{1}{4}f_{\pi}^2$--factor which modulates
the lowest order $\cL_{\rm eff}^{(2)}$ Lagrangian,) the dimensionless couplings
$L_{i}$, can then be traded by
$\frac{1}{\Lambda_{i}^2}$ couplings, such that
$L_{i}^{\rm r}\equiv \frac{\frac{1}{4}f_{\pi}^2}{\Lambda_{i}^2}$. It seems
reasonable to expect the $\Lambda_{i}$--constants to be of the same order of
magnitude as $\Lambda_{\chi}$. Then, from (\ref{eq:Lscale}), we expect
$M_{\rho}(770{\rm MeV})\leq \Lambda_{i}
\leq M_{A_{1}}(1260{\rm MeV})$; i.e., $L_{i}^{\rm r}\sim 10^{-3}$ to
$5\times 10^{-3}$. We see that $L_9^{\rm r}$ falls in this expected range; and
in fact, as we shall soon see, so do all the other
$L_{i}$--couplings.}

\end{itemize}

There is a wealth of experimental information on low energy hadron physics,
which has permitted the phenomenological determination of the
$L_{i}$--constants. The power of the chiral approach is that once these
constants have been fixed from some experiment, or spectral function sum rules,
there are of course predictions --and therefore tests-- for other observables.
In Table 1, I have collected the most recent compilation of the
$L_{i}$'s. One can also read in the same Table 1, the experimental source which
has been used for the determination of the appropriate constant. It would be
nice to improve the accuracy of some of the low--energy experiments. Hopefully,
the DA$\Phi$NE--project at Frascati will eventually provide some of this
improvement. A lot of theoretical work has recently been made in view of this
project. This work has been published as a reference guide~\cite{DAFNE92}. One
can find there un update of many recent phenomenological applications of
$\chi$PT.

%%%%%%%%%%%%%%%%%%%%%%%%%%% TABLE %%%%%%%%%%%%%%%%%

\begin{table}
\label{tab:Lcouplings}
\caption{Phenomenological values of the renormalized couplings
$L_i^r(M_\rho)$.}
%taken from Ref. \protect\cite{ref:BEG92}.
\vspace{0.2cm}
\begin{center}
\begin{tabular}{|c|c|c|}
\hline\hline
%& &\\
$i$ & $L_i^r(M_\rho) \times 10^3$ & Source
%\\ & &
\\ \hline
   1 & $\hphantom{-}0.7\pm0.5$ & $K_{e4}$, $\pi\pi\to\pi\pi$
\\ 2 & $\hphantom{-}1.2\pm0.4$ & $K_{e4}$, $\pi\pi\to\pi\pi$
\\ 3 & $-3.6\pm1.3$ & $K_{e4}$, $\pi\pi\to\pi\pi$
\\ 4 & $-0.3\pm0.5$ &  Zweig rule
\\ 5 & $\hphantom{-}1.4\pm0.5$ & $F_K : F_\pi$
\\ 6 & $-0.2\pm0.3$ & Zweig rule
\\ 7 & $-0.4\pm0.2$ & Gell-Mann--Okubo, $L_5$, $L_8$, Sum Rules
\\ 8 & $\hphantom{-}0.9\pm0.3$ & $M_{K^0} - M_{K^+}$, $L_5$,
$(m_s - \hat{m}) : (m_d-m_u)$
\\ 9 & $\hphantom{-}6.9\pm0.7$ & $\langle r^2\rangle^\pi_{\rm em}$
\\ 10 & $-5.5\pm0.7$ & $\pi\to e\nu\gamma$
%\\ & &
\\ \hline\hline
\end{tabular}
\end{center}
\end{table}

%%%%%%%%%%%%%%%%%%%%%%%%%%%%%%%%%%%%%%%%%%%%%
\subsubsection{\sl The Non-Abelian Chiral Anomaly}
\label{subsubsec:NACA}

Although the QCD Lagrangian with external sources is formally invariant under
local chiral transformations, this is no longer true for the associated
generating functional. The anomalies of the fermionic determinant break chiral
symmetry at the quantum level. The anomalous change of the generating
functional under an infinitesimal chiral transformation
\be  V_{L,R} = 1 + i \alpha \mp i \beta + \ldots
\ee

\noi  is given by~\cite{B69}:
\be\label{eq:anomaly}
\delta Z[v,a,s,p]  \, = \, -{N_c\over 16\pi^2} \, \int d^4x \, {\rm tr}
\beta(x) \,\Omega(x) ,
\ee
where
\beqn
\Omega(x) & = & \varepsilon^{\mu\nu\sigma\rho} \,
 [ v_{\mu\nu} v_{\sigma\rho} + {4\over 3} \,\nabla_\mu a_\nu
\nabla_\sigma a_\rho + {2\over 3} i \,\{ v_{\mu\nu},a_\sigma a_\rho\}
 \no \\ & &  + {8\over 3} i
\, a_\sigma v_{\mu\nu} a_\rho + {4\over 3} \, a_\mu a_\nu a_\sigma  a_\rho ] ,
\qquad \varepsilon_{0123}=1;
\eeqn and
\be v_{\mu\nu}  =
\partial_\mu v_\nu - \partial_\nu v_\mu - i \, [v_\mu,v_\nu] ,
\qquad
\nabla_\mu a_\nu   =
\partial_\mu a_\nu - i \, [v_\mu,a_\nu] .
\ee

\noi  This anomalous variation of $Z$ is an $\cO (p^4)$ effect in the chiral
counting.

Chiral Symmetry is the basic requirement to construct the effective
$\chi$PT Lagrangian. Since chiral symmetry is explicitly violated by the
anomaly at the fundamental QCD level, one is forced to add an effective
functional with the property that its change under chiral gauge transformations
reproduces (\ref{eq:anomaly}). Such a functional was first constructed by Wess
and Zumino~\cite{WZ71}. An interesting topological interpretation was later
found by Witten~\cite{W83}. The functional in question, has the following
explicit form:
$$
\Gamma [U,\ell,r]_{WZW} =-\,\dfrac{i N_c}{240 \pi^2}
\int d\sigma^{ijklm}\,{\rm tr} \left\{ \Sigma^L_i
\Sigma^L_j \Sigma^L_k \Sigma^L_l \Sigma^L_m \right\}
$$
\be \label{eq:WZW}
 -\,\dfrac{i N_c}{48 \pi^2} \int d^4 x\,
\varepsilon_{\mu \nu \alpha \beta}\left( W (U,\ell,r)^{\mu \nu
\alpha \beta} - W ({\bf 1},\ell,r)^{\mu \nu \alpha \beta} \right),
\ee
with
\beqn W (U,\ell,r)_{\mu \nu \alpha \beta} & = & {\rm tr}
\left\{ U \ell_{\mu} \ell_{\nu} \ell_{\alpha}U^{\dg} r_{\beta} +
\frac{1}{4} U \ell_{\mu} U^{\dg} r_{\nu} U \ell_\alpha U^{\dg} r_{\beta} + i U
\partial_{\mu} \ell_{\nu} \ell_{\alpha} U^{\dg} r_{\beta}
\right.\no  \\ & & +~ i \partial_{\mu} r_{\nu} U \ell_{\alpha} U^{\dg}
r_{\beta} - i \Sigma^L_{\mu} \ell_{\nu} U^{\dg} r_{\alpha} U
\ell_{\beta} + \Sigma^L_{\mu} U^{\dg} \partial_{\nu} r_{\alpha} U
\ell_\beta
\no \\ & & -~ \Sigma^L_{\mu} \Sigma^L_{\nu} U^{\dg} r_{\alpha} U
\ell_{\beta} + \Sigma^L_{\mu} \ell_{\nu} \partial_{\alpha}
\ell_{\beta} + \Sigma^L_{\mu} \partial_{\nu} \ell_{\alpha}
\ell_{\beta}
\no\\ & & -~ i\left.  \Sigma^L_{\mu} \ell_{\nu} \ell_{\alpha}
\ell_{\beta} + \frac{1}{2} \Sigma^L_{\mu} \ell_{\nu}
\Sigma^L_{\alpha} \ell_{\beta} - i \Sigma^L_{\mu} \Sigma^L_{\nu}
\Sigma^L_{\alpha} \ell_{\beta}
\right\} \no \\ & & -~ \left( L \leftrightarrow R \right) ,
\eeqn
where
\be
\Sigma^L_\mu = U^{\dg} \partial_\mu U , \qquad\qquad
\Sigma^R_\mu = U \partial_\mu U^{\dg} ,
\ee

\noi and
$\left( L \leftrightarrow R \right)$ stands for the interchanges
$U \leftrightarrow U^\dg $, $\ell_\mu \leftrightarrow r_\mu $ and
$\Sigma^L_\mu \leftrightarrow \Sigma^R_\mu $. The integration in the first term
of Eq.~(\ref{eq:WZW}) is over a five--dimensional manifold whose boundary is
four--dimensional Minkowski space. The integrand is a surface term; therefore
both the first and the second terms of $\Gamma_{WZW}$ are $\cO (p^4)$ according
to the chiral counting rules.

Since the effect of anomalies is
completely calculable, their translation from the fundamental quark--gluon
level
to the effective chiral level is unaffected by hadronization problems. The
anomalous action (\ref{eq:WZW}) has no free parameters.
It is responsible for the $\pi^0\to 2\gamma$,
$\eta\to 2 \gamma$ decays, and the $\gamma 3\pi$,
$\gamma\pi^+\pi^-\eta$ interactions among others. The five--dimensional surface
term generates interactions among five or more Goldstone bosons.

The variation of the anomalous action $\Gamma [U,\ell,r]_{WZW}$ with respect to
appropriate external field sources, generates the chiral realization of the
anomalous electroweak currents of the Standard Model, much the same as we
calculated the electroweak currents in eqs.(\ref{eq:lc}), (\ref{eq:rc}) induced
{}from the lowest order chiral Lagrangian. With
\be  L_\mu \equiv iU^{\dg}D_{\mu}U,
\ee

\noi  and $F_{L}^{\mu\nu}$,  $F_{R}^{\mu\nu}$ the strength field tensors
(\ref{eq:fstes}) associated to $l_\mu$ and $r_\mu$, we have
\be
\frac{\delta \Gamma_{WZW}}{\delta (l_{\mu})_{ji}}=
\frac{N_c}{48\pi^2}\varepsilon^{\mu\nu\rho\sigma}
\left[iL_{\nu}L_{\rho}L_{\sigma}+
\left\{F_{\nu\rho}^{L}+\frac{1}{2}U^{\dg}F_{\nu\rho}^{R}U, \,
L_{\sigma}\right\}\right]_{ij}.
\ee

\noi  This anomalous current is in fact defined, up to a chirally
non--covariant polynomial in the external fields $l$, $r$. Only the covariant
anomalous current above is measurable. Of particular interest for semileptonic
decays of $K$--mesons, is the strangeness changing current which, in the
presence of external  electromagnetic interactions, couples to the charged
$W^{\pm}$ in the Standard Model:
$$ J_{\rm anom.}^{\mu}  =   -\frac{e}{\sqrt{2}\sin
{\theta_W}}\frac{N_c}{48\pi^2}\times$$
\be \label{eq:acc}
\varepsilon^{\mu\nu\rho\sigma}{\rm tr}Q_{\pm}
\left[D_{\nu}U^{\dg}D_{\rho}UD_{\sigma}U^{\dg}U+
ieF_{\nu\rho}\left\{U^{\dg}D_{\sigma}U,
\, Q+\frac{1}{2}U^{\dg}QU\right\}\right] +{\rm h.c.},
\ee

\noi   where, here,
\be D_{\mu}=\partial_{\mu}U+ieA_{\mu}[Q,U]
\ee

\noi  is the covariant derivative with respect to electromagnetism only, ($Q$
is the charge matrix in (\ref{eq:charges}),) and
$F_{\mu\nu}=\partial_{\mu}A_{\nu}-\partial_{\nu}A_{\mu}$ the electromagnetic
field strength tensor. The first term in the r.h.s. contributes to
$K_{l4}$--decays such as
\be  K^{+}(p)\rightarrow \pi^{+}(p_{+})\pi^{-}(p_{-})e^{+}\nu_{e}\,,
\qquad q=p-p_{+}-p_{-}\,,
\ee

\noi  via the vector current, ($H$ denotes the conventional phenomenological
parametrization of the vector form factor,)
\be  V^{\mu}=-H(q^2)\frac{1}{M_{K}^3}\varepsilon^{\mu\nu\rho\sigma}
q^{\nu}(p_{+}+p_{-})^{\rho}(p_{+}-p_{-})^{\sigma},
\ee

\noi with~\cite{B90}$^{,}$\cite{RGDH91}
\be  H(0)=-\frac{N_c}{16\pi^2f_{\pi}^2}\,\frac{4}{3}\frac{M_K^3}
{\sqrt{2}f_{\pi}}=-2.7.
\ee

\noi  Experimentally~\cite{Retal77}
\be  H_{\rm threshold}=-2.68\pm 0.68.
\ee

\noi  The second term in (\ref{eq:acc}) contributes to radiative semileptonic
$K$--decays. An update of the rich phenomenology of these processes can be
found in the DA$\Phi$NE physics handbook~\cite{BEG95}.

%%%%%%%%%%%%%%%%%%%%%%%%%%%%%%%%%%%%%%%%%%%%%%%%%%%%%%%%%%%%

\section{Weak Interactions and Chiral Perturbation Theory}
\label{sec:LEWI}

In this lecture we shall study the weak interactions of $K$--particles within
the framework of $\chi$PT. We shall be  particularly concerned with processes
induced by virtual $W^{\pm}$--exchange between quark currents, in the presence
of the strong and electromagnetic interaction. A central issue in the study of
the non--leptonic weak decays of $K$--mesons, is the question of the origin of
the $\Delta I=1/2$ selection rule. Let me start by explaining what the problem
is.

There is experimental evidence, both from $K$--decays and hyperon decays, that
the rates of non--leptonic strangeness changing transitions $\Delta S=1$ with
isospin change $\Delta I=1/2$ are particularly enhanced. This phenomenological
fact is referred to as ``the $\Delta I=1/2$ rule''. The explanation of this
selection rule has been a continuous challenge to theorists for the last three
decades! There are good qualitative indications that in the Standard Model, the
observed $\Delta I=1/2$ rule has a dynamical origin; but no clear quantitative
description of the enhancement has yet been exhibited.

To illustrate the problem let us try to make a simple guess of the strength of
the transitions expected for $K\ra \pi\pi$ decays in the Standard Model. For
such processes the $W$--mass can be considered as infinitely heavy, and the
exchange of the
$W^\pm$--field between two charged quark currents, ignoring for the moment
gluonic interactions between the two quark currents, is described by a simple
effective four--quark Hamiltonian:
\be {\cH}_{\rm eff}^{\Delta S=1}=\frac{G_F}{\sqrt{2}}V_{\rm ud}V_{\rm
us}^{\ast}4(\bar{s}_L\gamma^{\mu}u_L)(\bar{u}_L\gamma_{\mu}d_L)+ {\rm h.c.}
\ee

\noi with e.g.,
\be (\bar{s}_L\gamma^{\mu}u_L)\equiv \sum_{\alpha}
\bar{s}^{(\alpha)}(x)\gamma^{\mu}\left(\frac{1-\gamma_5}{2}\right)
u^{(\alpha)}(x),
\ee

\noi and where $\alpha$ denotes colour indices. Here $G_F$ is the Fermi
coupling constant $(1.166\times 10^{-5}{\rm GeV}^{-2})$ and $V_{\rm ud}$,
$V_{\rm us}^{\ast}$ are matrix elements of the Cabibbo--Kobayashi--Maskawa
flavour mixing matrix, (see the lectures of Roberto Peccei~\cite{Peccei}.)  At
this level of approximation, the effective Hamiltonian appears as the
factorized product of two quark currents. In the previous section, and to
lowest order in the chiral expansion,  we have explicitly worked out the
realization of these currents in terms of pseudoscalar fields, [see
Eqs.(\ref{eq:lc}) and (\ref{eq:rc}),] with the result:
\beqn (\bar{s}_L\gamma_{\mu}u_L) & \Ra & -\frac{1}{\sqrt{2}}f_{\pi}
\partial_{\mu}K^{+} \no \\
 & & -\frac{i}{2\sqrt{2}}\left[\left(\pi^0
\partial_{\mu}K^+\right)+\sqrt{3}\left(\eta
\partial_{\mu}K^+\right)+\sqrt{2}\left(\pi^+
\partial_{\mu}K^0\right)\right] \no \\
 & & +\frac{i}{2\sqrt{2}}\left[\left(K^+
\partial_{\mu}\pi^0\right)+\sqrt{3}\left(K^+
\partial_{\mu}\eta\right)+\sqrt{2}\left(K^0
\partial_{\mu}\pi^+\right)\right]+\cdots ;  \\
 & & \no \\
 & & \no \\ (\bar{u}\gamma_{\mu}d_L) & \Ra & -\frac{1}{\sqrt{2}}f_{\pi}
\partial_{\mu}\pi{-} \no \\
 & &  -\frac{i}{2}\left[\sqrt{2}\left(\pi^-\partial_{\mu}\pi^0\right)+
\left(K^0\partial_{\mu}K^-\right)\right] \no \\
 & & +\frac{i}{2}\left[\sqrt{2}\left(\pi^0\partial_{\mu}\pi^-\right)+
\left(K^-\partial_{\mu}K^0\right)\right]+\cdots .
\eeqn

\noi We have all the ingredients now to calculate the $K\ra \pi\pi$ isospin
amplitudes $A_I$; $I=0,2$ which we introduced in the {\bf Sec.1.}, [see
Eqs.(\ref{eq:KI}) and (\ref{eq:KBI}),] with the result:
\beqn
\label{eq:A0F} A_0 & = & -\frac{G_F}{\sqrt{2}}V_{\rm ud}V_{\rm us}^{\ast}
\frac{2}{3}\sqrt{2}f_{\pi}\left(M_K^2-M_{\pi}^2\right),  \\
\label{eq:A1F} A_2 & = & -\frac{G_F}{\sqrt{2}}V_{\rm ud}V_{\rm us}^{\ast}
\frac{1}{3}2f_{\pi}\left(M_{K}^2-M_{\pi}^2\right).
\eeqn

\noi As already mentioned in {\bf Sec.1} as well, experimentally, from the
ratio of the decay rates $\Gamma(K_{S}\ra \pi^{+}\pi^{-})$ and
$\Gamma(K_{S}\ra \pi^{0}\pi^{0})$, and with neglect of radiative corrections,
one finds
\be {\rm Re}\left(\frac{A_{0}}{A_{2}}\right)\Big|_{\rm exp} \simeq 22.2\, ,
\ee

\noi i.e., a factor of {\it sixteen} larger than our factorization estimate!
More precisely, taking also into account the experimental rate for
$K^{+}\ra \pi^{+}\pi^{0}$, we are led to the conclusion that our first guess
{\it underestimates} the $\Delta I=1/2$ amplitude by a factor of {\it eight},
and {\it overestimates} the $\Delta I=3/2$ amplitude by a factor of {\it two}.

One may be tempted to conclude that the factorization assumption we have used
to make our estimate must indeed be a very na\"{\i}ve picture. It turns out
however, that this is precisely the result predicted by the leading behaviour
of the $1/N_c$--expansion in QCD; i.e., the limit where the number of colours
$N_c$ is taken to be large with $\alpha_{s}N_c$ fixed~\cite{'tH74}. We shall
see later why the large--$N_c$ estimate fails in this case to give the right
order of magnitude.

%%%%%%%%%%%%%%%%%%%%%%%%%%%%%%%%%%%%%%%%%%%%%
\subsection{Short--Distance Reduction to an Effective Four--Quark Hamiltonian}
\label{subsec:EFQH}

In QCD, the composite operator
\be
\label{eq:Q2op} Q_2\equiv 4(\bar{s}_L\gamma^{\mu}u_L)(\bar{u}_L
\gamma_{\mu}d_L),
\ee

\noi  is not multiplicatively renormalizable. As first pointed out by Gaillard
and Lee~\cite{GL74a} and by Altarelli and Maiani~\cite{AM74}, the effect of
gluon exchanges generates a new
$\Delta S=1$ operator ($\alpha$ and $\beta$ are quark colour
indices; summation over repeated Greek letters is
understood)
\be
\label{eq:four_quark_operators_1} Q_1  \,= \, 4 \,
\bar s^\alpha_L \gamma^\mu u^\beta_L \,\, \bar u^\beta_L
\gamma_\mu d^\alpha_L\, ,
\ee

\noi which mixes with the standard $Q_2$ operator
in~(\ref{eq:Q2op}) under renormalization. It was later noticed by Vainshtein,
Zakharov and Shifman~\cite{VZS75}, that the effect of the so-called
``penguin'' diagrams where {\it one light quark line is attached by
gluon exchange to a weak $s-d$ self--energy like diagram},
brings further $\Delta S=1$ operators [$q_R
\,=\, {1\over 2}\, (1+\gamma_5)
q(x)$]:
\beqn
\label{eq:four_quark_operators_3} Q_3 & = &
4 \, \left(\bar s_L \gamma^\mu d_L\right)
\, \sum_{q=u,d,s} \left(\bar q_L \gamma_\mu q_L\right)\, , \\
\label{eq:four_quark_operators_4} Q_4 & = & 4
\, \bar s^\alpha_L
\gamma^\mu d^\beta_L
\,
\sum_{q=u,d,s} \bar q^\beta_L \gamma_\mu q^\alpha_L\, , \\
\label{eq:four_quark_operators_5}Q_5 & =
& 4
\, \left(\bar s_L \gamma^\mu d_L\right)
\sum_{q=u,d,s} \, \left(\bar q_R \gamma_\mu q_R\right)\, , \\
\label{eq:four_quark_operators_6} Q_6 & = & 4
\,
\bar s^\alpha_L \gamma^\mu d^\beta_L
\, \sum_{q=u,d,s} \bar q^\beta_R \gamma_\mu q^\alpha_R\,
,
\eeqn

\noi which also mix with $Q_2$ and among themselves under
renormalization.
The operators
\be \label{eq:four_quark_relations_1}
Q_-  \,= \, Q_2 - Q_1
\ee

\noi and $Q_i, \; i=3,4,5,6,$ induce pure $\Delta I={1\over 2}$ transitions,
but only four of these operators are independent, since
\be
\label{eq:four_quark_relations_2}
Q_- + Q_3 - Q_4 = 0\ .
\ee

\noi Under $SU(3)_L \times SU(3)_R$ transformations, $Q_-$ and the
operators $Q_i, \; i=3,4,5,6$, transform like $(8_L,\, 1_R)$ operators,
while the combination
\be \label{eq:four_quark_relations_3}
Q^{(27)}\,= \, 2 Q_2 + 3 Q_1 - Q_3
\ee

\noi transforms like a $(27_L,\, 1_R)$ operator which induces both
$\Delta I={1\over 2}$ and $\Delta I={3\over 2}$ transitions via its
components
\be
\label{eq:four_quark_27}
Q^{(27)} = {4\over 3} \left( Q_{1/2}^{(27)} + 5 Q_{3/2}^{(27)}
\right),
\ee

\noi where
\beqn
\label{eq:four_quark_27,1/2}
Q_{1/2}^{(27)} & = & \left(\bar s_L \gamma^\mu d_L\right)
\left(\bar u_L \gamma_\mu u_L\right) + \left(\bar s_L
\gamma^\mu u_L\right)
\left(\bar u_L \gamma_\mu d_L\right) \no \\
& & + 2 \left(\bar s_L
\gamma^\mu d_L\right)
\left(\bar d_L \gamma_\mu d_L\right) -3 \left(\bar s_L
\gamma^\mu d_L\right) \left(\bar s_L \gamma_\mu
s_L\right),
\eeqn

\noi and
\beqn
\label{eq:four_quark_27,3/2}
Q_{3/2}^{(27)}& = &\left(\bar s_L \gamma^\mu d_L\right)
\left(\bar u_L \gamma_\mu u_L\right) + \left(\bar s_L
\gamma^\mu u_L\right)
\left(\bar u_L \gamma_\mu d_L\right) \no \\
&  & -\left(\bar s_L
\gamma^\mu d_L\right)
\left(\bar d_L \gamma_\mu d_L\right)\, .
\eeqn

\noi The operator $Q^{(27)}$ is multiplicatively renormalizable and does not
mix with the o\-thers.

The procedure to go from the Lagrangian of the Standard Model to an effective
electroweak Hamiltonian, where only the degrees of freedom of the light quark
fields $u,\,d,\,s$ appear~\cite{WWG79}, consists in using the
asymptotic freedom property~\cite{GWP73} of QCD to successively integrate out
the fields with heavy masses down to scales $\mu^2 < m^2_c$, i.e. below the
charm quark mass. The appropriate technique is the operator product
expansion~\cite{WZ69} and the use of renormalization group
equations~\cite{SPGLCS} to compute the various Wilson coefficient functions of
the four--quark operators $Q_i, \; i=1,\dots ,\, 6$. The inclusion of virtual
electromagnetic interactions in the process of integrating out the fields with
heavy masses brings in~\cite{BW84,LU89} four new four--quark operators to the
effective electroweak Hamiltonian. With $e_q$ denoting the corresponding quark
charges in units of the electric charge, the new operators
are:
\beqn
\label{eq:four_quark_operators_7} Q_7 & = &
6 \, \left(\bar s_L \gamma^\mu d_L\right)
\sum_{q=u,d,s}
\, e_q \, \left(\bar q_R \gamma_\mu q_R\right)\, ,
\\ \label{eq:four_quark_operators_8} Q_8 & = & 6
\, \bar s^\alpha_L \gamma^\mu d^\beta_L
\, \sum_{q=u,d,s} e_q \, \bar q^\beta_R \gamma_\mu q^\alpha_R\, ,
\\ \label{eq:four_quark_operators_9} Q_9 & = & 6 \,
\left(\bar s_L \gamma^\mu d_L\right)
\sum_{q=u,d,s} \, e_q \, (\bar q_L \gamma_\mu q_L)\, ,
\\ \label{eq:four_quark_operators_10} Q_{10} & = & 6
\, \bar s^\alpha_L \gamma^\mu d^\beta_L
\,
\sum_{q=u,d,s} e_q \, \bar q^\beta_L \gamma_\mu q^\alpha_L\ .
\eeqn

\noi Under the action of the chiral group $SU(3)_L \times SU(3)_R$ the
operators $Q_7$ and $Q_8$ transform like combinations of
$(8_L,\, 1_R)$ and $(8_L,\, 8_R)$ operators, while $Q_9$ and
$Q_{10}$ transform like combinations of $(8_L,\, 1_R)$ and $(27_L,\, 1_R)$.

Owing to the big value of the top quark mass, higher--order  electroweak
contributions turn out to be relevant~\cite{IL81} when analyzing some
CP--violation effects.  These additional corrections do not generate new
operators, but contribute to the Wilson coefficient functions of the operators
$Q_i, \; i=1,\dots ,\, 10$.

In processes with leptons in the final state, three more operators need still
to be considered~\cite{GW80}:
\beqn\label{eq:q11}  Q_{11} &= & 4 \, (\bar s_L \gamma^\mu
d_L)
\sum_{l=e,\mu} \, (\bar l_L \gamma_\mu l_L) \, , \\
\label{eq:q12} Q_{12} & = & 4 \, (\bar s_L \gamma^\mu
d_L)
\, \sum_{l=e,\mu}  (\bar l_R \gamma_\mu l_R) \, , \\
\label{eq:q13} Q_{13} & = & 4 (\bar s_L\gamma^\mu d_L)\,
(\bar\nu_L\gamma_\mu\nu_L)\, .
\eeqn

\noi Since these operators are bilinear in the quark--fields, they have to be
considered separately whenever they may be relevant. The operators $Q_{11}$ and
$Q_{12}$ play an important r\^{o}le in the transition $K_{L}\ra
\pi^{0}e^{+}e^{-}$ which we shall study in {\bf Sec.5.}

In the Standard Model, the $\Delta S=1$ non--leptonic interactions can then be
described by an effective Hamiltonian
\be
\label{eq:ds_hamiltonian}
{\cal H}^{\Delta S = 1}_{\mbox{\rms eff}}\, = \, {G_F \over
\sqrt{2}}            V_{\rm ud}^{\hphantom{*}} V_{\rm us}^* \,
\sum^{10}_{i=1} C_i(\mu) \, Q_i \, +
\, \mbox{h.c.} \, ,
\ee

\noi where $Q_i$ are the ten local four--quark operators introduced above, and
$C_i(\mu)$ the modulating Wilson coefficients which are functions of the masses
of the fields which have been integrated out, i.e.
$t,\, Z,\, W,\, b$, and $c$, as well as of the overall renormalization scale
$\mu$. The Wilson coefficients of the ten four--quark operators take into
account the effect of the strong interactions down to scales of ${\cO}(\mu)$,
in the presence of virtual electroweak interactions, which in practice are kept
to ${\cO}(\alpha)$.

The Wilson coefficients $C_{i}(\mu)$ obey the renormalization group equation
$\left(\frac{g_{s}^{2}}{4\pi}=\alpha_{s}\right)$
\be
\label{eq:rge}
\left[\mu\frac{\partial}{\partial\mu} +
  \beta(\alpha_s) \,\alpha_s\frac{\partial}{\partial
\alpha_s}\right]
  \,\vec{C}\left(\frac{M_{W}^{2}}{\mu^{2}}, \alpha_s,\alpha\right)
  \, = \,
  \gamma^{T}(\alpha_{s},\alpha)
\,
  \vec{C}\left(\frac{M_{W}^{2}}{\mu^{2}},
\alpha_s,\alpha\right),
\ee

\noi where $\beta(\alpha_s)$ denotes the QCD beta function
\beqn
\label{eq:beta_function}
\beta(\alpha_s) & = & \beta_{1}\frac{\alpha_{s}}{\pi}+
  \beta_{2}\left(\frac{\alpha_{s}}{\pi}\right)^{2}+
\cdots\, ,
\\
\beta_1 & = &
\frac{1}{6}(-11N_c+2n_f),
\\
\beta_2 & = &
\frac{-17}{3}\left(\frac{N_c}{2}\right)^2+\frac{1}{2}
\frac{N_c^2-1}{2N_c}\frac{n_f}{2}+
\frac{5}{12}N_cn_f.
\eeqn

\noi The effect of electromagnetic corrections in the
$\beta$ function can be neglected at the level of precision that the
Wilson coefficients are needed at present. The anomalous--dimension matrix
$\gamma_{ij}$ is defined by the mixing of the operators $Q_i$ under
renormalization:
\be
\label{eq:renor_eq_1}
\mu^2{d \over d \mu^2} \left< Q_i \right> =
-{1\over                             2}\gamma_{ij}(\alpha_s,\alpha)
\left< Q_j \right>\ .
\ee

\noi In perturbation theory, the matrix elements of $\gamma$ which are needed
for a calculation of the Wilson coefficients at the leading and
next--to--leading logarithmic approximation, are the coefficients of the first
four terms in the expansion
\be
\label{eq:anomalous_dimension}
\gamma (\alpha_s,\alpha) =  {\alpha_s N_c \over \pi}  {1\over 2}
\gamma^{(0)}_s + {\alpha \over \pi}  {1\over 2} \gamma^{(0)}_e +
\left({\alpha_s N_c
\over \pi}\right)^2 {1\over 4} \gamma^{(1)}_s + {\alpha_s N_c
\over
\pi} \, {\alpha \over \pi} \, {1\over 4} \gamma^{(1)}_{se} + \dots \,
,
\ee

\noi The most recent calculations of the full $10\times 10$ anomalous dimension
matrix corresponding to this expansion, can be found in
Refs.~\cite{BJLW93,CFMR93}.

The general solution of the renormalization group equation above is given
by
\be
\label{eq:w_coef}
 \vec{C}\left(\frac{M_{W}^{2}}{\mu^{2}},\alpha_s(\mu),\alpha\right) \,
=\,
 \left\{ {\cT}_{\alpha_s}
\exp\int_{\alpha_s(M_W)}^{\alpha_s(\mu)}
  dz\frac{\gamma^{T}(z,\alpha)}{z
\beta(z)}\right\}
  \vec{C}(1,
\alpha_s(M_W),\alpha),
\ee

\noi where ${\cT}_{\alpha_s}$ denotes ordering in the QCD coupling constant
increasing from right to left. The vector
$\vec{C}(1,\alpha_s(M_W),\alpha)$, which defines the initial
boundary conditions, has to be extracted from the calculation of
the full Feynman diagrams at the $M_W$ mass scale. In the
limit $\alpha_{s}=\alpha=0$: $C_{2}(M_W)=1$ and all the
other $C_{i}$ vanish. To $O(\alpha)$, the $C_{i}$ with
$i=1,2,3,7$ and $9$ get contributions, while $C_{8}=C_{10}=0$.
The two--loop final expressions for the Wilson coefficient
functions are rather elaborate. I shall not reproduce them here.
They can be found in the references quoted earlier~\cite{BJLW93,CFMR93}.

We can now understand why our factorization estimate,earlier, failed. The
effect of gluon exchanges between quark current bilinears brings in, by mixing,
new four--quark operators. The Wilson coefficients of the full set of
four--quark operators has to be scaled from the large $t$--mass and
$W$--mass scales down to a renormalization mass scale $\mu\simeq 1$GeV, below
which the QCD perturbative evaluation is no longer trustworthy. This long
running of scales brings in large logarithms, the effect of their size being
governed by the anomalous dimension matrix of the four--quark operators. It is
only when terms of at least ${\cO}(N_c)$ are taken into account in the weak
amplitudes, that the effect of the long evolution from short--distances to
long--distances is incorporated. To see this more explicitly, it suffices to
compute the lowest order Wilson coefficients in the sector of the $Q_1$ and
$Q_2$ four--quark operators. In the basis of the $Q_{-}=Q_1-Q_2$ and
$Q_{+}=Q_1+Q_2$ operators, you will find
\be C^{\pm}(\mu^2)= \left(\frac{\alpha_s(M_W^2)}{\alpha_s(\mu^2)}\right)^
{\left[\frac {\pm 3/2(1\mp 1/N_c)}{-\beta_1}\right]} .
\ee

\noi The net effect is that transitions induced by the $Q_{-}$--operator are
{\it enhanced} by a factor $\left(\log\frac{M_W^2}{\Lambda_{\bar{\rm MS}}^2}/
\log\frac{\mu^2}{\Lambda_{\bar{\rm MS}}^2}\right)^{4/7}$; while the transitions
induced by the $Q_{+}$--operator are {\it depressed} by a factor
 $\left(\log\frac{\mu^2}{\Lambda_{\bar{\rm MS}}^2}/
\log\frac{M_W^2}{\Lambda_{\bar{\rm MS}}^2}\right)^{2/7}$. Although this does
not explain the observed $\Delta I=1/2$ enhancement, it goes well in the right
direction, and points towards a solution of the problem: {\it the possibility
that the enhancement due to the anomalous dimensions continues in the evolution
of the hadronic matrix elements at low energies.}

Another interesting issue, which appears at the two loop level of the
perturbative calculations, is the question of renormalization scheme
dependence. The Wilson coefficient functions have been calculated in two
schemes: the 't Hooft--Veltman renormalization scheme~\cite{HV72}, and
dimensional regularization with an anticommuting $\gamma_{5}$. The scheme
dependence can only be removed by matching the calculation of the $C_{i}$'s
with a similar calculation of the matrix  elements of the $Q_{i}(\mu)$
operators in the same scheme. Unfortunately, the technology to calculate low
energy matrix elements of light four--quark operators is not yet developed to
the degree of sophistication of perturbative QCD. Only approximate methods have
been developed so far: lattice models of QCD (see the lectures of Steven
Sharpe~\cite{Sharpe}); various versions of QCD sum
rules~\cite{PR85}$^{,}$\cite{PDPPR91}$^{,}$\cite{PR91}$^{,\rm and}$\cite{JP94}
; the $1/N_c$ approach developed by Bardeen, Buras and
G\'{e}rard~\cite{BBG87,BBG87a,BBG88} ; and more recently, the  QCD low--energy
effective action approach~\cite{PR91}.  We postpone the discussion of some of
these methods to the following sections, where we shall review estimates of
various $\chi$PT coupling constants relevant to
$K$--decays.

%%%%%%%%%%%%%%%%%%%%%%%%%%%%%%%%%%%%%%%%%%%%%%%%%%%

\subsection{The Effective Four--Quark Hamiltonian in $\chi$PT}
\label{subsec:EFQHCPT}

The effective four--quark Hamiltonian that we have derived in the previous
subsection contains only light--quark fields degrees of freedom. The light
quark fields still have strong interactions mediated by the gluon fields of the
QCD--Lagrangian. Now we want to find the effective chiral realization of this
Hamiltonian in terms of (pseudo) Nambu--Goldstone degrees of freedom only;
i.e., we want a description in terms of an effective chiral Lagrangian  --with
the same chiral transformation properties as the four--quark Hamiltonian--
which will be incorporated as a weak perturbation to the strong chiral
effective Lagrangian we discussed in {\bf Sec.\,2.} We also want to construct
the weak effective Lagrangian in terms of a chiral expansion in powers of
derivatives and quark masses, much the same as we did for the sector of the
strong interactions.

As we saw in the last subsection, there are essentially\footnote{\ There
is also the combination of the operators $Q_7$ and $Q_8$ which transforms like
$(8_{L},8_{R})$ which has to be considered
separately~\cite{BW84}.}\hspace*{1em}two types of four--quark operators,
according to their transformation properties under chiral--$SU(3)$: those which
transform as $(8_{L},1_{R})$ and those which transform as $(27_{L}, 1_{R})$. To
lowest order in powers of derivatives, all the possible operators one can
construct with these transformation properties are the following: With
${\cL}_{\mu}(x)$ the
$3\times 3$ flavour matrix field
\be
\label{eq:Lweak} {\cL}_{\mu}(x)\equiv
-i\frac{f_{\pi}^2}{2}U^{\dg}(x)D_{\mu}U(x),
\ee

\noi which by itself transforms as ${\cL}_{\mu}\to
V_{L}{\cL}_{\mu}V_{L}^{\dg}$, we can construct the two operators
\beqn {\cL}_{8}(x) & = & \sum_i\left({\cL}_{\mu}\right)_{2i}
\left({\cL}_{\mu}\right)_{i3}; \\ {\cL}_{27}(x)  & = &
\frac{2}{3}\left({\cL}_{\mu}\right)_{21}
\left({\cL}_{\mu}\right)_{13}+\left({\cL}_{\mu}\right)_{23}
\left({\cL}_{\mu}\right)_{11}.
\eeqn

\noi The second operator induces both $\Delta I=1/2$ and $\Delta I=3/2$
transitions via its components~\cite{PR91}:
\be {\cL}_{27}(x) = \frac{1}{9}{\cL}_{27}^{(1/2)}(x)+\frac{5}{9}
{\cL}_{27}^{(3/2)}(x),
\ee
\noi

where
\beqn {\cL}_{27}^{(1/2)}(x) & = & \left({\cL}_{\mu}\right)_{2I}
\left({\cL}_{\mu}\right)_{13}+\left({\cL}_{\mu}\right)_{13}
\left[4\left({\cL}_{\mu}\right)_{11}+5\left({\cL}_{\mu}\right)_{22}
\right], \\ {\cL}_{27}^{(3/2)}(x) & = & \left({\cL}_{\mu}\right)_{2I}
\left({\cL}_{\mu}\right)_{13}+\left({\cL}_{\mu}\right)_{23}
\left[\left({\cL}_{\mu}\right)_{11}-\left({\cL}_{\mu}\right)_{22}\right].
\eeqn

\noi The most general effective Lagrangian we are looking for, to lowest order
in the chiral expansion, is then\footnote{\ A coupling $\sim  (\chi
U^{\dg}+U\chi^{\dg})_{23}$ is also possible a priori. However, for on--shell
processes, this term can be rotated away so as to maintain the normalization
condition $<U>=1$. The physical effect is then of higher order. This term has
still physical relevance, even to lowest order, for off--shell non--leptonic
Green's functions, and therefore brings in an extra coupling constant which is
not fixed by symmetry requirements alone.}\
\be
\label{eq:WL2} {\cL}_{\rm eff}^{\Delta S=1}(x)=-\frac{G_F}{\sqrt{2}} V_{\rm
ud}V_{\rm us}^{\ast}\, 4\left[{\bf g}_{8}{\cL}_{8}+{\bf
g}_{27}{\cL}_{27}^{(3/2)}+\frac{1}{5} {\bf g}_{27}{\cL}_{27}^{(1/2)}\right] +
{\rm h.c.}\, .
\ee

\noi The two constants ${\bf g}_{8}$ and ${\bf g}_{27}$ are dimensionless
constants, real constants in as far as CP--violation effects are neglected,
which like
$f_{\pi}$, $B$, and the $L_{i}$ coupling constants of the strong Lagrangian,
cannot be determined from symmetry arguments alone. With the factors of
$f_{\pi}$ included in the definition of ${\cL}_{\mu}$ in (\ref{eq:Lweak}), the
constants
${\bf g}_{8,27}$ are of ${\cO}(1)$ in the large--$N_c$ expansion. We can get
their phenomenological values from a comparison between the expressions for the
$K\ra \pi\pi$ isospin amplitudes $A_{I}$ calculated with the Lagrangian above:
\beqn A_0 & = & -\frac{G_F}{\sqrt{2}}V_{\rm ud}V_{\rm us}^{\ast} ({\bf
g}_{8}+\frac{1}{5}{\bf g}_{27})\sqrt{2}f_{\pi}
\left(M_K^2-M_{\pi}^2\right), \\ A_2 & = & -\frac{G_F}{\sqrt{2}}V_{\rm
ud}V_{\rm us}^{\ast} {\bf g}_{27}2f_{\pi}\left(M_{K}^2-M_{\pi}^2\right),
\eeqn

\noi and experiment; with the result~\cite{PR91}:
\be
\label{eq:g8g27} |{\bf g}_{8}+\frac{1}{5}{\bf g}_{27}|_{\rm exp.}\simeq 5.1,
\qquad  |{\bf g}_{27}|_{\rm exp.}\simeq 0.16\, .
\ee

\noi From a comparison between the $A_{I}$-amplitudes calculated above, and the
results of our factorization estimate in Eqs.(\ref{eq:A0F}) and
(\ref{eq:A1F}), we can also read the values of the coupling constants
${\bf g}_{8,27}$ corresponding to this approximation:
\be {\bf g}_{8}|_{\rm fact.}= \frac{3}{5}, \qquad   {\bf g}_{27}|_{\rm fact.
}=\frac{1}{3}.
\ee

\noi rather {\it far away}, as already discussed, from the experimental values.

Once the couplings ${\bf g}_{8}$ and ${\bf g}_{27}$ have been fixed
phenomenologically, there follow a wealth of non--trivial predictions for
$K\ra \pi\pi\pi$ decays and {\it some} radiative $K$--decays as well.
Concerning the latter, there appear some interesting features, which I think
are worth pointing out, because they reveal the power and the simplicity of the
chiral approach:

\begin{itemize}

\item{\it $K$-decay amplitudes with any number of real or virtual photons and
at most one pion in the final state vanish to lowest ${\cO}(p^2)$ in the chiral
expansion~\cite{EPR87,EPR88}. }

The reason for it is due to the fact that electromagnetic gauge invariance
requires physical amplitudes to have a number of chiral powers higher than just
the two powers allowed by the lowest order effective Lagrangian. Processes like
\be K_S\ra \gamma\gamma, \qquad K_L\ra \pi^0\gamma\gamma;
\ee
\be K^+\ra \pi^+\gamma\gamma, \qquad K_{S}\ra \pi^0\gamma\gamma;
\ee
\be K^+\ra \pi^{+}e^{+}e^{-}, \qquad K_{L}\ra \pi^{0} e^{+} e^{-},
\ee

\noi are all of this type. They are at least ${\cO}(p^4)$ in the chiral
expansion.

\item{\it $K$-decay amplitudes with two pions and any number of real or virtual
photons in the final state factorize, at ${\cO}(p^2)$, into the corresponding
$K\ra \pi\pi$ on--shell amplitude times a universal brem{\ss}trahlung--like
amplitude.}

The proof consists in a simple adaptation of a well known theorem
due to F. Low~\cite{L58}, to lowest order in $\chi$PT. [A sketch of the
proof for one photon is given in Ref.~\cite{deR89}. Applications to
$K\ra
\pi\pi\gamma$ can be found in Ref.~\cite{ENP94}.]

\end{itemize}

%%%%%%%%%%%%%%%%%%%%%%%%%%%%%%%%%%%%%%%%%%%%%%%

\subsubsection{\sl Weak Amplitudes to ${\cO}(p^4)$ in $\chi$PT}
\label{subsubsec:WAOP4}

The full analysis of the one loop divergences generated by the lowest order
weak Lagrangian, in the presence of the strong effective chiral Lagrangian; as
well as the classification of the possible local terms of ${\cO}(p^4)$ was
first made in~\cite{KMW90}; and, using different techniques in~\cite{E90,EF91}
as well. The number of terms is too large to do a phenomenological
determination of the couplings, as it has been done in the strong interaction
sector. Even restricting the attention to the effective realization of the
four--quark operators which transform like $(8_L,1_R)$, still leaves {\it
twenty two} possible terms that contribute to non--leptonic $K$--decays with
possible external photons or virtual $Z$--bosons. To determine
phenomenologically these couplings is not a question of calculation complexity;
it is simply that there is not enough available experimental information to do
the job! The only possible way to do $\chi$PT usefully in the sector of the
non--leptonic weak interactions is to combine the chiral expansion with other
approximation methods, like e.g., the $1/N_c$--expansion; or to resort to
models of the low--energy effective action of QCD that one can first test in
the strong interaction sector.

The art of the game in making {\it clean} $\chi$PT predictions for
non--leptonic
weak processes, consists in finding subsets of observables which to
${\cO}(p^4)$ are fully given by a chiral loop only, like e.g., $K_S\ra
\gamma\gamma$~\cite{DEG}, and $K_L\ra \pi^0\gamma\gamma$~\cite{EPRa87}; or
which involve a small number of unknown ${\cO}(p^4)$ local couplings, like
$K\ra \pi l^{+}l^{-}$ decays~\cite{EPR87,EPR88}. For a recent review on the the
state of the art see Ref.~\cite{Detal94}. I shall discuss some of these
processes in {\bf Sec.\,5.}

Another sector where it has been possible to test the validity of the chiral
expansion in non--leptonic weak interactions is in $K\ra 2\pi$ and $K\ra 3\pi$
decays. The description of these decays to ${\cO}(p^4)$ in $\chi$PT involves
{\it seven} linear combinations of coupling constants. Altogether, imposing
isospin and Bose symmetries, these decays can be parametrized in terms of {\it
twelve} observables. Five of these parameters, the quadratic slopes in the
Dalitz plot for the various $K\ra 3\pi$ modes, vanish to lowest order in the
chiral expansion. The resulting {\it five} constraints can be formulated in
terms of neat predictions~\cite{KDHMW92} for the slopes. The five predictions
are compatible with experiment within errors. Another important result which
also emerges from the $K\ra 2\pi,3\pi$ amplitude analysis~\cite{KMW91}, and
which is relevant for the understanding of the underlying dynamics of the
$\Delta I=1/2$ rule, is the fact that, in the presence of the ${\cO}(p^4)$
corrections, the fitted value for
$|{\bf g}_{8}|$ is $\simeq 30\%$ smaller than the lowest order determination in
Eq.(\ref{eq:g8g27}). This is, again, another little factor which, like the
short--distance enhancement that we discussed earlier in {\it subsec.\,3.1},
helps towards explaining the $\Delta I=1/2$ rule, but the bulk of the
enhancement remains still unravelled.  Let me note that, at the same order of
approximation, the coupling
${\bf g}_{27}$ has no large corrections.

%%%%%%%%%%%%%%%%%%%%%%%%%%%%%%%%%%%%%%%%%%%%%%%
\subsubsection{\sl Penguins in the Large--$N_c$ Limit}
\label{subsubsec:PLNL}

It has often been speculated that the bulk of the origin of the
$\Delta I=1/2$ enhancement comes from the {\it large} matrix elements of the
Penguin--generated $Q_{6}$ operator in  (\ref{eq:four_quark_operators_6}). By
Fierz reordering, this operator can also be written as
\be
\label{eq:Q6Fierz} Q_{6}= -8\sum_{q=u,d,s}\left(\bar{s}_{L}q_{R}\right)
\left(\bar{q}_{R}d_{L}\right).
\ee

\noi I find it very instructive to discuss the chiral realization of this
operator~\cite{CFG86}, because it reveals a number of interesting features; and
also because it provides an excellent example of the way that, when
systematically combined, the chiral expansion and the large--$N_c$ expansion
may eventually help us to make substantial progress in the understanding of
low--energy QCD.

Using Eqs.(\ref{eq:LQCD}) and (\ref{eq:chies}), the term which couples quark
bilinears to external scalar and pseudoscalar sources can be written as follows
\be -\bar{q}(s-i\gamma_5 p)q=-\frac{1}{2B}(\bar{q}_{R}\chi q_{L}
+\bar{q}_{L}\chi^{\dg}q_{R}).
\ee

\noi In the large--$N_c$ limit, the operator $Q_{6}$ in (\ref{eq:Q6Fierz})
factorizes into the product of two $\bar{q}q$--densities. The chiral
realization of these densities to ${\cO}(p^2)$ in the chiral expansion and
keeping only those terms which may eventually contribute to the lowest order
${\bf g}_8{\cL}_8$--piece in the Lagrangian in (\ref{eq:WL2}), proceeds as
follows:
\beqn (\bar{s}_{L}q_{iR})\doteq 2B\frac{\delta\cL_{\rm eff}}{\delta
\chi^{\dg}_{3i}}
& = & 2B\left\{\frac{1}{4f_{\pi}^2}U_{i3}+
L_{5}(UD_{\mu}U^{\dg}D^{\mu}U)_{i3}+\cdots \right\} , \\
(\bar{q}_{Ri}d_{L})\doteq 2B\frac{\delta\cL_{\rm eff}}{\delta \chi_{i2}} & = &
2B\left\{\frac{1}{4f_{\pi}^2}U^{\dg}_{2i}+
L_{5}(D_{\mu}U^{\dg}D^{\mu}UU{^\dg})_{2i}+\cdots \right\}.
\eeqn

\noi Since $UU^{\dg}=U^{\dg}U=1$, there is no contribution to $Q_{6}$ of
${\cO}(p^0)$, and to leading order, both in the chiral expansion and in the
large--$N_c$ expansion, we find
\be Q_6\Rightarrow -8\times 4B^{2}\, \frac{1}{4f_{\pi}^2}\, 2\, L_5\,
(D^{\mu}U^{\dg} D_{\mu}U)_{23}.
\ee

\noi We can cast this result in terms of the contribution to the ${\bf
g}_{8}$--coupling, induced by the term in the effective four--quark weak
Hamiltonian proportional to the $Q_6$--operator:
\be
\label{eq:g8penguin}
\lbrack{\bf g}_{8}\rbrack_{6^{\rm th}-{\rm Penguin}} = -16L_{5}C_{6}(\mu^2)\,
\left[\frac{<\bar{\psi}\psi>}{f_{\pi}^3}\right]^{2}.
\ee

\noi As I said before, there are a number of interesting features emerging from
this result:

\begin{itemize}

\item Remember that in the large--$N_c$ limit $f_{\pi}^2$ ,$L_5$ and
$<\bar{\psi}\psi>$ are all ${\cO}(N_c)$. Because of the fact that $C_{6}\sim
\alpha_{s}$, $\lbrack{\bf g}_{8}\rbrack_{6^{\rm th}-{\rm Penguin}}$ is
next--to--leading in the large--$N_c$ limit. It is precisely this {\it
next--to--leading} contribution that we have succeeded in calculating
explicitly.

\item The Wilson coefficient $C_{6}$ has an imaginary part induced by the
CP--violation phase in the Cabibbo--Kobayashi--Maskawa matrix~\cite{Peccei}. It
is precisely this contribution that, in the Standard Model, induces the
$\epsilon'$--amplitude we discussed in {\bf Sec.\,1.} The fact that we can say
{\it something quantitative} about the size of the modulating coupling
$\lbrack{\bf g}_{8}\rbrack_{6^{\rm th}-{\rm Penguin}}$ is of course welcome for
phenomenology. We shall come back to this in {\bf Sec.\,5.}

\item In QCD the matrix element $<\bar{\psi}\psi>$ is $\mu$--scale dependent.
The scale dependence is given by the anomalous dimension of the
$\bar{\psi}\psi$--operator. On the other hand, the coupling constant ${\bf
g}_{8}$ is a scale independent quantity. This example exhibits explicitly the
cancellation~\cite{deR89} between the $\mu$--scale dependence of the
short--distance Wilson coefficient $C_{6}(\mu^2)$ and the $\mu$--scale
dependence of the long--distance four--quark operator bosonization, which
appears via the constant
$(<\bar{\psi}\psi>)^2$. In the large--$N_c$ limit the operator $Q_6$ does not
mix with the others because, in that limit, the anomalous dimension matrix in
Eq.(\ref{eq:anomalous_dimension}):
$\left[\gamma_{s}^{(0)}\right]_{ij}\to -\frac{3}{2}N_{c}\delta_{66}$. In that
limit, $C_6$ depends on $\mu$ via $C_6\sim
\alpha(\mu^2)^{9/11}$, which exactly cancels the $\mu$--dependance of
$(<\bar{\psi}\psi>)^2$ evaluated in the same large--$N_c$ limit.

\item Of the parameters which appear in the r.h.s. of Eq.(\ref{eq:g8penguin}),
the one which has the largest uncertainty is $<\bar{\psi}\psi>$. The most
recent
determination from an update of QCD sum rules methods gives the
value~\cite{BPR94}:
\be
\label{eq:npsipsi}
<\bar{\psi}\psi>(1{\rm GeV}^2)= -(0.013\pm0.003){\rm GeV}^3.
\ee

\noi It is also questionable which value of $f_{\pi}$ one should use in the
r.h.s. of Eq.(\ref{eq:g8penguin}). At the level of approximation that the
calculation has been made, it seems appropriate to take the value of $f_{\pi}$
in the chiral limit; i.e.,
$f_{\pi}^{(0)}\simeq 86{\rm MeV}$. Then, using this value for $f_{\pi}$, the
value $L_5\simeq 1.4\times 10^{-3}$ which is in Table 1, and the result of the
$C_6$--calculation in~\cite{BJLW93}, I find
\be
\lbrack{\bf g}_{8}\rbrack_{6^{\rm th}-{\rm Penguin}}\simeq 0.9.
\ee

{}From this result, one is led to conclude that, unless the next--to--leading
order
$1/N_c$ corrections to this calculation are huge, the bulk of the
$\Delta I=1/2$ enhancement does not come from the Penguin $Q_6$--operator.
Barring this possibility, the long--distance contribution from the
$Q_{-}$--operator remains then the most likely candidate to provide the bulk of
the enhancement.

\end{itemize}

%%%%%%%%%%%%%%%%%%%%%%%%%%%%%%%%%%%%%%%%%%%%%%%
\subsection{$\Delta S=2$ Non-Leptonic Weak Interactions}
\label{subsec:D2NLWI}

The Standard Model predicts strangeness changing transitions with
$\Delta S=2$ via two virtual $W$--exchanges between quark lines, the so--called
box diagrams. The reduction via the operator product expansion results in an
effective Hamiltonian~\cite{GW83} which is proportional to the local
four--quark operator
\be
\label{eq:deltas2} Q_{\Delta S=2}\equiv
(\bar{s}_L\gamma^{\mu}d_L)(\bar{s}\gamma_{\mu}d_L),
\ee

\noi modulated by products of the flavour mixing matrix elements
\be
\lambda_{q}=V_{\rm qd}^{\ast}V_{\rm qs}, \qquad {\rm q}={\rm u,c,t}\, ;
\ee

\noi times functions $F_{1,2,3}$ of the heavy masses $m_{t}^2, M_{W}^2,
m_{b}^2,
m_{c}^2$ of the fields which have been integrated out:
\be {\cH}_{\rm eff}^{\Delta S=2}=\frac{G_{F}^{2}M_{W}^2}{4\pi^2}
\left[\lambda_{c}^2F_1+\lambda_{t}^2F_2+2\lambda_{c}\lambda_{t}F_3\right]
C_{\Delta S=2}(\mu^2)Q_{\Delta S=2}\, .
\ee

\noi The operator $Q_{\Delta S=2}$ is multiplicatively renormalizable and has
an anomalous dimension $\gamma(\alpha_{s})$ defined by the equation
\be
\mu^2\frac{d}{d\mu^2}<Q_{\Delta S=2}>=-\frac{1}{2}\gamma(\alpha_{s}) <Q_{\Delta
S=2}>
\ee

\noi At the one--loop level
\be
\gamma(\alpha_s)=\frac{\alpha}{\pi}\gamma_1 +
{\cO}\left(\frac{\alpha}{\pi}\right)^2, \quad
\gamma_1= \frac{3}{2}(1-1/N_c);
\ee

\noi and
\be C_{\Delta
S=2}(\mu^2)=\alpha_{s}(\mu^2)^{\left[-\frac{\gamma_1}{-\beta_1}=-2/9\right]}.
\ee

The matrix element
\be
\label{eq:Bpar} <\bar{K}^0|Q_{\Delta s=2}(0)|K^0>\equiv
\frac{4}{3}f_{K}^2M_{K}^2B_{K}(\mu^2),
\ee

\noi defines the so--called $B_{K}$--parameter, which governs $K^0-\bar{K}^0$
mixing at short distances, and is one of the crucial unknown parameters in the
phenomenological studies of CP--violation in the Standard Model. The definition
above is such that in the so--called {\it vacuum saturation
approximation}~\cite{GL74}
\be (B_K)_{\rm VS}=1.
\ee

In the large--$N_c$ limit, the  four quark operator $Q_{\Delta S=2}$ factorizes
into a product of the two left--handed current operators:
$(\bar{s}_L\gamma^{\mu}d_L)$ times $(\bar{s}_L\gamma_{\mu}d_L)$. Each of these
currents, to lowest order in $\chi$PT, has a bosonic realization as indicated
in Eq.(\ref{eq:lc}); i.e.,
\be (\bar{s}_L\gamma^{\mu}d_L)\Ra -\frac{1}{\sqrt{2}}f_{\pi}\partial_{\mu}K^{0}
+\cdots.
\ee

\noi This approximation results then in a value: $B_{K}=\frac{3}{4}
\frac{f_{\pi}^2}{f_K^2}$. There are however chiral corrections of
${\cO}(p^4)$ to this determination which are still leading in the large--$N_c$
limit. The leading $1/N_c$--corrections come from the $L_5$ term in ${\cL}_{\rm
eff}^{(4)}$; as well as from the $K^0$ wave--function renormalization, with the
result
\be (\bar{s}_L\gamma^{\mu}d_L)\Ra -\frac{1}{\sqrt{2}}f_{\pi}
\left(1+\frac{8M_{K^0}^2}{f_{\pi}^2}L_5\right)
\left(1-\frac{4M_{K^0}^2}{f_{\pi}^2}L_5\right)\partial_{\mu}K^{0} +\cdots.
\ee

\noi where the first factor in the r.h.s. is the one coming from
${\cL}_{\rm eff}^{(4)}$. (Notice that the contribution from chiral logarithms
is $1/N_c$ suppressed.) The net result is a renormalization of
$f_{\pi}$. This renormalized $f_{\pi}$--coupling is a first approximation to
the physical
$f_{K}$--coupling constant which is measured in $K^{+}\ra \mu^{+}\nu_{\mu}$
decays: $f_{K}\simeq 113{\rm MeV}$. To the extent that the two are identified,
one then obtains the result which is often quoted in the literature as the {\it
large--$N_c$ prediction}:
\be B_{K}|_{N_c\ra \infty}=\frac{3}{4}.
\ee

%%%%%%%%%%%%%%%%%%%%%%%%%%%%%%%%%%%%%%%%%%%%%%%%%%%%%%%%%%%%%
\section{Models of the QCD Low--Energy Effective Action}
\label{sec:models}

The purpose of
this lecture is to give an overview on models of the low--energy hadronic
interactions, which have been developed during the last few years, and which
try to focus on general properties that the QCD low--energy effective action is
expected to have. The aim here is to find {\it generic} features of these
models  which, eventually, one may be able to promote to the rank of a
low--energy effective action  derived from QCD, hopefully within a well defined
set of approximations.

Let me first formulate the problem which one would like to solve. For this
purpose, it is convenient to promote the global chiral--$SU(3)$ symmetry to a
local gauge symmetry, in the same way that has already been discussed in {\bf
Sec.\,2.} It is also convenient to use a path integral representation for the
generating functional $\Gamma(v,a,s,p)$ of the Green's functions of quark
currents:
\be
\label{eq:genfunc} e^{i\Gamma(v,a,s,p)}=\frac{1}{Z}\int \cD G_{\mu}\det \!\not
\!\! D \: \exp(-i\int \! d^4
x\:\frac{1}{4}\,\vec{G}_{\mu\nu}\vec{G}^{\mu\nu})\, ,
\ee

\noi with $\not \!\! D$ the Dirac operator
\be
\label{eq:dirac}
\not \!\! D=\gamma^{\mu}(\partial_{\mu}+ig_{s}G_{\mu})
-i\gamma^{\mu}(v_{\mu}+\gamma_{5} a_{\mu})+i(s-i\gamma_{5}p)\, ;
\ee

\noi where $G_{\mu}$ is the gluon field, $\vec{G}^{\mu\nu}$ the gluon field
strength tensor, and $v_{\mu},a_{\mu},s,p$ external field sources. The
normalization factor
$Z$ is such that
$\Gamma (0,0)=1.$

The chiral symmetry of the underlying QCD theory implies that
$\Gamma(v,a,s,p)$ admits a low--energy representation
\be
\label{eq:lefr} e^{i\Gamma(v,a,s,p)}=\frac{1}{Z}\int \cD U \:
\exp [i\int \! d^4 x\: \cL _{\rm eff}(U;v,a,s,p)],
\ee

\noi in terms of an effective Lagrangian $\cL _{\rm eff}(U;v,a,s,p)$ with
$U(x)$ a $3\times 3$ unitary matrix, with
$\det U=1$, which collects the octet of pseudoscalar fields $(\pi ,K,\eta )$.

There is only one term in $\cL _{\rm eff}$ which is known from first
principles. It is the term associated with the existence of anomalies in the
fermionic determinant~\cite{B69}. The corresponding effective action is the
Wess and Zumino~\cite{WZ71} functional that  we have discussed in {\bf
Sect.\,2.} All possible other terms  in $\cL _{\rm eff}$, are not fixed by
symmetry requirements alone. It would be nice to find, at least
approximatively, a dynamical scheme rooted in QCD, with as few free parameters
as possible (ideally $\Lambda_{QCD}$ only!), which allows for a derivation of
the coupling constants in the effective chiral Lagrangian. As we have seen in
{\bf Sec.\,3}, the need for such an approximate dynamical scheme is urgently
needed at present to make progress in the phenomenology of non--leptonic
flavour
dynamics.

The various types of models which have been discussed in the literature can be
classified, roughly speaking, in one of the following entries:

\begin{itemize}

\item QCD in the Large--$N_c$ Limit.

\item Low--Lying Resonances Dominance Models.

\item The Constituent Chiral Quark Model.

\item Effective Action Approach Models.

\item The Extended Nambu and Jona-Lasinio Model ( ENJL--model.)

\end{itemize}

\subsection{QCD in the Large--$N_c$ limit}

It would be a major breaktrough, if one could derive the low--energy effective
Lagrangian of the interactions between Nambu--Goldstone modes in the
large--$N_{c}$ limit of QCD. So far, it has only been possible to obtain  {\it
constraints} among various coupling constants in this limit; but not their {\it
values} in terms, say, of $\Lambda_{\rm QCD}$. A typical example is the
relation:
$2L_{1}=L_{2}$, which, as first noticed by  Gasser and Leutwyler~\cite{GL85},
follow in the large--$N_{c}$ limit of QCD. Unfortunately, nobody can claim as
yet to be able to {\it compute}, say $L_{2}$, in that limit. Often in the
literature, there appear statements about `` large--$N_{c}$ predictions'' but,
in fact, they have been all derived with some extra {\it ad hoc} assumptions.

An interesting approach to do approximate calculations within the framework of
the $1/N_c$--expansion is the one proposed by Bardeen, Buras and
G\'{e}rard~\cite{BBG87,BBG87a,BBG88}, which they have applied extensively to
the calculation of non--leptonic weak matrix elements. The basic idea is to
start with the factorized form of the four--quark operators in the effective
weak Hamiltonian, and to do one--loop chiral perturbation theory, keeping track
of the quadratic divergences which appear. If one was able to work with the
{\it full} hadronic low--energy effective Lagrangian, it would be possible to
obtain a smooth matching between the scale dependence of the Wilson
coefficients, calculated at short--distances, and the hadronic matrix elements
calculated with the {\it full} hadronic low--energy effective Lagrangian. The
hope with the approach proposed by Bardeen, Buras and G\'{e}rard is that the
{\it numerical} matching of the {\it quadratic} long--distance scale with the
{\it logarithmic} short--distance scale, may turn out to be already a good
first approximation to the problem one would like to solve. The technology of
their approach is explained with detail in their papers.

\subsection{Low--Lying Resonances Dominance Models}

There has been quite a lot of progress during the last few years in
understanding the r\^{o}le of resonances in  $\chi $PT. At the phenomenological
level~\cite{EGPR89,DRV89}, it turns out that the observed values of the
$L_{i}$--constants are practically saturated by the contribution from the
lowest resonance exchanges between the pseudoscalar particles; and particularly
by vector--exchange, whenever vector mesons can contribute. The specific form
of an effective chiral invariant Lagrangian describing the couplings of vector
and axial--vector particles to the (pseudo) Nambu--Goldstone modes is not
uniquely fixed by chiral symmetry requirements alone. When the vector fields
describing heavy vector particles are integrated out, different field theory
descriptions may lead to different predictions for the
$L_{i}$--couplings. It has been shown however that if a few QCD short--distance
constraints are imposed, the ambiguities of different formulations are then
removed~\cite{EGLPR89}. The most compact effective Lagrangian formulation,
compatible with the short--distance constraints, has two free parameters:
$f_{\pi}$ and $M_{V}$. When the vector and axial--vector fields are integrated
out, it leads to specific predictions for {\it five} of the $L_i$ constants:
\be L_{1}^{(V)}=L_{2}^{(V)}/2=-L_{3}^{(V)}/6=L_{9}^{(V)}/8=-L_{10}^{(V+A)}/6
=\frac{f_{\pi}^2}{16M_{V}^2}\simeq 0.6\times 10^{-3}\,,
\ee

\noi in good agreement, within errors, with experiment. [See Table 1.]

It is fair to conclude that the old phenomenological concept of {\it vector
meson dominance} (VMD)~\cite{NS62} can now be formulated in a way compatible
with the chiral symmetry properties and the short--distance behaviour of QCD.

In view of this success, there have been various suggestions of {\it
extensions} of VMD--models in the literature. In particular, one would like to
extend the idea of VMD to the non--leptonic sector of the weak interactions, in
order to have a useful low--energy chiral effective Lagrangian formulation.
(Remember the problem of the proliferation of couplings we have discussed in
{\bf Sec.\,3.}) At the strict phenomenological level, one is forced to
introduce weak couplings between vector (axial--vector) fields and/or the
(pseudo) Nambu--Goldstone fields. The fact that these weak coupling constants
are experimentally unknown, forces one to resort to models if the idea of VMD
is to be pursued.

Models like the {\it Geometric Model}~\cite{EPR90,E90} and the {\it Quark
Resonance Model}~\cite{PP93} are attempts to reduce the number of free coupling
parameters which are allowed in principle by chiral symmetry requirements
alone. However, the precise relation of these models to the specific
assumptions which one is making within the underlying QCD--theory to justify
them, remains unclear.
%%%%%%%%%%%%%%%%%%%%%%%%%%%%%%%%

\subsection{The Constituent Chiral Quark Model}
\label{subsec:CCQM}

This model was introduced by Georgi and
Manohar~\cite{GM84}, in an attempt to reconcile the successful features of the
Constituent Quark Model~\cite{DGG75}, with the chiral symmetry requirements of
QCD. The basic assumption of the model is the idea that between the scale of
chiral symmetry breaking $\Lambda_{\chi}$ and the confinement scale $\sim
\Lambda_{\rm QCD}$ the underlying QCD--theory, may admit a useful effective
Lagrangian realization in terms of {\it constituent quark fields $Q$}; {\it
pseudoscalar particles}; and, perhaps, {\it ``gluons''}. The Lagrangian in
question has the form
\beqn
\label{eq:GML}
\cL _{\rm eff}^{\rm GM} & = &
i\bar{Q}\gamma_{\mu}(\partial_{\mu}+ig_{s}G_{\mu}+\Gamma_{\mu} )Q +
\no
\\ & &
\frac{i}{2}g_{A}\bar{Q}\gamma_{5}\gamma^{\mu}\xi_{\mu}Q -M_{Q}\bar{Q} Q +
\no \\
& & \frac{1}{4} f_{\pi}^2\, trD_{\mu}UD^{\mu}U^{\dagger}
-\frac{1}{4}\,\vec{G}_{\mu\nu}\vec{G}^{\mu\nu}.
\eeqn

\noi Some explanations about the notation here are in order. Remember that
under chiral rotations $(V_{L},V_{R})$, $U$ transforms like: $U\ra
V_{R}UV_{L}$. The unitary matrix
$U$ is the product of the so--called left and right coset representatives:
$U=\xi_{R}\xi_{L}^{\dg}$ and, without lost of generality, one can always choose
the gauge where
$\xi_{L}^{\dg}=\xi_{R}\equiv \xi$. The coset representative
$\xi$, ($U=\xi\xi$,) transforms like:
\be
\xi\ra V_{R}\xi h^{\dagger}=h\xi V_{L}^{\dagger} \qquad h\in SU(3)_{V}\, ,
\ee

\noi where $h$ denotes the rotation induced by the chiral transformation
$(V_{L},V_{R})$  in the diagonal
$SU(3)_{V}$. In Eq.(\ref{eq:GML}) the constituent quark fields $Q$  transform
like
\be
\label{eq:cqt} Q\ra hQ, \qquad h\in SU(3)_{V}.
\ee

\noi In the presence of external sources\footnote{\ The original formulation of
the model of Georgi and Manohar \cite{GM84} was in fact made without external
fields.}\, ,
\be
\label{eq:gamu}
\Gamma_{\mu}=\frac{1}{2}\{\xi^{\dagger} [\partial_{\mu}-i(v_{\mu}+a_{\mu})]\xi+
\xi[\partial_{\mu}-i(v_{\mu}-a_{\mu})]\xi^{\dagger}\}
\ee

\noi and
\be
\label{eq:ximu}
\xi_{\mu}=i\xi^{\dagger}D_{\mu}U\xi^{\dagger}.
\ee

\noi The free parameters of the theory are $f_{\pi}$, $M_{Q}$, and
$g_{A}$. The QCD coupling constant is assumed to have entered a regime (below
$\Lambda_{\chi}$,) where its running is frozen and is taken to be constant.

The merit of this model is that it automatically digests the phenomenological
successes of the constituent quark model, in a way compatible with chiral
symmetry. We shall in fact see, that effective Lagrangians of the
Georgi--Manohar type, do indeed appear in practically all QCD low--energy
models where quarks are not confined. The weak point of the model is its
``vagueness'' about the gluonic sector. In the absence of a dynamical
justification for the ``freezing'' of the QCD running coupling constant, it is
very unclear what the ``left out'' gluonic interactions mean; and in fact, in
most applications they are simply ignored.

%%%%%%%%%%%%%%%%%%%%%%%%%%%%%%%%%%%%%

\subsection{Effective Action Approach Models}
\label{subsec:eaam}

The basic idea	in this class of models is
to make some kind of drastic approximation to compute the  non--anomalous part
of the QCD--fermionic determinant in the presence of external $v_{\mu}$ and
$a_{\mu}$ fields, but with the external $s$ and $p$ fields frozen to the quark
matrix
$$s+ip=\cM = {\rm diag}(m_{u},m_{d},m_{s}).$$ For this purpose, it is
convenient to perform a chiral rotation of the quark fields
$q_{L,R}(x)\equiv
\frac{1}{2}(1\pm
\gamma_5 )q(x)$ in the initial QCD--Lagrangian:
\be
\label{eq:chrot}
q_{L}(x)\Ra Q_{L}(x)=\xi(x)\, q_{L}(x), \qquad
q_{R}\Ra Q_{R}(x)=\xi^{\dagger}(x)\, q_{R}(x),
\ee

\noi with $\xi$ chosen so that $\xi\xi = U$. Under chiral rotations
$(V_{L},V_{R})$, the quark fields of the rotated basis transform like the
constituent chiral quarks of the Georgi--Manohar Lagrangian:
\be
\label{eq:chcqt} Q_{L,R}\ra h(x)Q_{L,R},
\ee

\noi with $h[\phi (x)]$ the rotation  in $SU(3)_{V}$ induced by the chiral
transformation $(V_{L},V_{R})$; $h$ is the same object which appears in
Eq.(\ref{eq:cqt}). The QCD--Lagrangian in the rotated field basis, and in
Euclidean space has then the following form:
\be
\label{eq:qcdeu}
\cL _{QCD}^{(E)}= -\frac{1}{4}\, G_{\mu\nu}^{(a)}G_{\mu\nu}^{(a)}  +\bar{Q}
D_{E}Q,
\ee

\noi with $D_{E}$ the Euclidean Dirac operator ($\tilde{\gamma}_{\mu}\equiv
-i\gamma_{\mu}$ are Hermitian Dirac matrices with positive metric)
\be
\label{eq:diraceu} D_{E}=\tilde{\gamma}_{\mu}(\partial_{\mu} + ig_{s}G_{\mu}
+\Gamma_{\mu} -\frac{i}{2}\gamma_5
\xi_{\mu})-\frac{1}{2}(\Sigma-\gamma_{5}\Delta),
\ee

\noi where $\Gamma_{\mu}$ and $\xi_{\mu}$ are the same as in
Eqs.(\ref{eq:gamu}) and (\ref{eq:ximu}); and
\be
\label{eq:sigdelta}
\Sigma = \xi^{\dagger}\cM \xi^{\dagger}+\xi \cM \xi, \quad
\Delta =  \xi^{\dagger}\cM \xi^{\dagger}-\xi \cM \xi.
\ee

\noi The $\Sigma$ and $\Delta$ terms break explicitly the chiral symmetry.

The Euclidean effective action $\Gamma_{E}(U,v,a,\cM )$ associated to $\cL
_{QCD}^{(E)}$ is then given by
\beqn
\label{eq:effac}
\exp \Gamma_{E}(U,v,a,\cM ) & = &
\frac{1}{Z}\int \cD G_{\mu} \exp[-\int\! d^4 z\:
\frac{1}{4}\,G_{\mu\nu}^{(a)}G_{\mu\nu}^{(a)}]
\exp \Gamma_{E}(G_{\mu};U,v,a,\cM ) \no \\
 & \equiv & \frac{1}{Z}\int [\cD G_{\mu}]
\exp \Gamma_{E}(G_{\mu};U,v,a,\cM ),
\eeqn

\noi with
\be
\label{eq:effacg}
\exp \Gamma_{E}(G_{\mu};U,v,a,\cM )=
\int \cD \bar{Q}\cD Q\, exp(\int\! d^4 z\:\bar{Q}D_{E}Q)=
\det D_{E}.
\ee

\noi In fact, for the  non--anomalous part of the effective action, it is
sufficient to consider the modulus of the determinant. Then,
\be
\label{eq:meffacg}
\Gamma_{E}(G_{\mu};U,v,a,\cM )=
\frac{1}{2}\log\, \det\, D_{E}^{\dagger}D_{E}.
\ee

It is in trying to evaluate the modulus of the fermionic determinant that one
encounters the {\it first problem}. There is need of a regularization; and if
possible, a regularization which shows the explicit cut--off dependence. The
simplest regularization in that respect is {\it the proper--time} cut--off
method, where
\be
\label{eq:ptreg}
\Gamma_{E}=-\frac{1}{2}\int_{1/\Lambda^2 }^{\infty}
\frac{d\tau}{\tau}{\rm Tr} \exp (-\tau  D_{E}^{\dagger}D_{E}),
\ee

\noi where Tr stands for trace over Dirac $\gamma$--matrices,
$SU(3)$--colour matrices, Gell-Mann's flavour--$SU(3)$ matrices, and Euclidean
space.

Since $ D_{E}^{\dagger}D_{E}$ is a second--order elliptic operator, the
proper--time integrand in (\ref{eq:ptreg}) has a well defined power series
expansion in powers of $\tau$; this is the so called {\it heat-kernel}
expansion:
\be
\label{eq:seedew} (x\mid e^{-\tau   D_{E}^{\dagger}D_{E}}\mid y)=
\frac{1}{16\pi^2 \tau^2 }e^\frac{-(x-y)^2}{4\tau}
\sum_{0}^{\infty}H_{n}(x,y)\tau^{n}.
\ee

\noi The functions $H_{n}(x,y)$ are known in the literature as Seeley-DeWitt
coefficients. Here we encounter the {\it second problem}: for $n\geq 2$, the
proper--time integrals are infrared divergent. We need an infrared regulator to
integrate the large--$\tau$ behaviour.  Now, it just happens that, with neglect
of the gluonic coupling  in the Dirac operator in (\ref{eq:diraceu}), and for
{\it some} of the
$\cO (p^4)$ terms in the effective action, the proper time integral in
(\ref{eq:ptreg}) is convergent both in the ultraviolet and infrared domains.
Therefore, the values of the corresponding
$L_{i}$--coupling constants in this brutal approximation turn out to be finite
quantities. The results in question are~\cite{N84}:
\be
\label{eq:lvls} 8L_{1}=4L_{2}=L_{9}=\frac{N_c}{48\pi^2}; \qquad
L_{3}=L_{10}=-\frac{N_c}{96\pi^2}.
\ee

\noi Numerically, these results, when compared to the phenomenological
determinations listed in Table 1, are surprisingly good.  Any attempt however,
to improve on these results, and/or to compute other couplings of the
low--energy effective Lagrangian, necessarily brings in the question of the
infrared behaviour of the underlying theory.

A simple suggestion which has been proposed~\cite{ERT90}, in connection with
the infrared behaviour, is to parametrize the effect of spontaneous symmetry
breaking, by adding to the QCD--Lagrangian a phenomenological {\it order
parameter} like--term:

\be\label{eq:MQterm}
\label{eq:mqterm}\Delta \cL_{QCD}\equiv -M_{Q}(\bar{q}_{R}Uq_{L}+
\bar{q}_{L}U^{\dagger}q_{R}),
\ee

\noi which introduces at the same time the $U$--field in a way non--invariant
under
$U\ra -U$, and the mass parameter $M_{Q}$ which provides the infrared regulator
needed in the evaluation of the low energy effective action. It is easy to see
that in the presence of this term, and with$M_{Q}>0$, there is quark
condensation~\cite{ERT90}:
$<\bar{\psi}\psi> \neq 0$ and negative. Furthermore, in the presence of this
term, it is also possible to evaluate the effect of large--$N_c$ gluonic
interactions  which appear as inverse powers of $M_{Q}$ times the appropriate
vacuum expectation values of gauge invariant gluonic operators. For example, in
the chiral limit, and including the leading gluonic contributions, one
gets~\cite{ERT90}:
\be
\label{fpiert} f_{\pi}^2 =\frac{N_c}{16\pi^2 }4M_{Q}^2 \left[
\log\frac{\Lambda^2}{M_{Q}^2} +
\frac{\pi^2}{6N_c}\frac{<\frac{\alpha_s}{\pi}GG>}{M_{Q}^4}+
\frac{1}{360N_c}\frac{<g^3 GGG>}{M_{Q}^6}+\cdots\right]\, .
\ee

\noi There also appear gluonic corrections to some of the previous results for
the $L_{i}$--couplings:
\be
\label{lvlsg} L_{3}=L_{10}=-\frac{N_c}{96\pi^2}\left[1+\frac{\pi^2}{5N_c}
\frac{<\frac{\alpha_s}{\pi}GG>}{M_{Q}^4}+\cO (\frac{1}{M_{Q}^6})\right].
\ee

\noi The positive sign of this correction helps towards the agreement with the
experimental values. Notice however that the gluon condensate which appears
here, is an average over configurations from
$1/\Lambda^2$ to roughly $1/M_{Q}^2$ distances; and therefore  cannot be
related easily to the usual gluon condensate which appears in the phenomenology
of QCD--sum rules. The predictions for the other
$L_{i}$--coupling constants to
$\cO (\frac{1}{M_{Q}^4})$. remain the same as those in (\ref{eq:lvls}).

Applications of this approach to the non--leptonic weak interactions have also
been made. The problem here is to find the effective action of a given
four--quark operator. For example, we have seen in {\bf Sec.\,3}, that the
$\Delta S=2$ transitions, after the heavy degrees of freedom have been
integrated out, are governed by the  four--quark operator
$$Q_{\Delta S=2}\equiv (\bar{s}_{L}\gamma^\mu d_{L})  (\bar{s}_{L}\gamma_\mu
d_{L}),$$ modulated by products of flavour--mixing matrix elements, times
Wilson coefficient functions resulting from the short--distance integration.
The Euclidean effective action of $Q_{\Delta S=2}$ in an external gluonic
background, is given by the expression (see Ref.~\cite{PR91} for technical
details):
\beqn
\label{eq:eff4op} <Q_{\Delta S=2}>\Big|_{G_{\mu};U,v,a,\cM ,M_{Q}} & = &  -{\rm
Tr}\left(D_{E}^{-1}\frac{\delta_{\mu}D_{E}}{\delta l_{\mu}(x)_{32}}\right) {\rm
Tr}\left(D_{E}^{-1}\frac{\delta_{\mu}D_{E}}{\delta l_{\mu}(x)_{32}}\right) \no
\\
 & & + {\rm Tr}\left(D_{E}^{-1}\frac{\delta_{\mu}D_{E}}{\delta
l_{\mu}(x)_{32}}D_{E}^{-1}\frac{\delta_{\mu}D_{E}}{\delta
l_{\mu}(x)_{32}}\right).
\eeqn

\noi The first term in the r.h.s. is the one induced by the configuration where
the two currents in $Q_{\Delta S=2}$ are factorized, the term leading in the
$1/N_c$--expansion; the second term contains the non--factorizable
contributions, which are next--to--leading in the $1/N_c$--expansion. These
contributions have been evaluated in Ref.~\cite{PR91} including, in addition to
the well known $\cO (N_{c}^2)$ factorizable contributions, the subleading $\cO
(N_c)$ and $\cO (\alpha_{s}N_c)$ terms. The corresponding result for the
$B_{K}$--parameter defined in {it subsec.\,3.3} reads as follows:
\be
\label{eq:bfpr} B_{K}=\frac{3}{4} \left[1+\frac{1}{N_c}\left(1-\frac{N_c}{2}
\frac{<\frac{\alpha_{s}}{\pi}GG>} {16\pi^2 f_{\pi}^4}
 + \cO (\alpha_{s}N_c)^2\right)\right].
\ee

\noi The corresponding results for the other coupling constants of the
$\cO (p^2)$ non--leptonic effective Lagrangian discussed in {\bf Sect.\,3},
evaluated at the same approximation~\cite{PR91} as the $B_{K}$--parameter
above, are the following:
\be
\label{eq:g27pr} {\bf g}_{27}^{(3/2)} =  C_{+}(\mu^2) \frac{2}{3}
\left[1+\frac{1}{N_c}\left(1-\frac{N_c}{2} \frac{
<\frac{\alpha_{s}}{\pi}GG>}{16\pi^2 f_{\pi}^4} + \cO (\alpha_{s}N_c)^2\right)
\right].
\ee

\noi and
\beqn
\label{eq:g8pr} {\bf g}_{8} & = & C_{-}(\mu^2)\left
[1-\frac{1}{N_c}\left(1-\frac{N_c}{2}
\frac{<\frac{\alpha_{s}}{\pi}GG>} {16\pi^2 f_{\pi}^4} + \cO
(\alpha_{s}N_c)^2\right)\right]
\no \\ & &  +
C_{+}(\mu^2)\frac{1}{5}\left[1+\frac{1}{N_c}\left(1-\frac{N_c}{2}
\frac{<\frac{\alpha_{s}}{\pi}GG>} {16\pi^2 f_{\pi}^4} + \cO
(\alpha_{s}N_c)^2\right)\right]
\no \\ & & +  C_{4}(\mu^2)-C_{6}(\mu^2)16L_{5}\,
\left[\frac{<\bar{\psi}\psi>}{f_{\pi}^3}\right]^2 + \cO (\alpha_{s}N_c).
\eeqn

There are a number of interesting features which emerge from these results,
worth commenting upon.

\begin{itemize}

\item The results of the so called {\it vacuum saturation approximation}
(VSA), often used in the literature, are obtained from those above when the
terms $\cO (\alpha{s}N_c)$ are dropped. This model calculation shows however
that, numerically, the neglected VSA--terms are as important as those
retained.

\item The results in eqs.(\ref{eq:bfpr}) and (\ref{eq:g27pr}) satisfy indeed
the chiral limit symmetry relation between $\Delta S=2$ and
$\Delta I=3/2$, $\Delta S=1$ transitions first observed by Donoghue, Golowich
and Holstein~\cite{DGH82}.

\item Equations (\ref{eq:bfpr}), (\ref{eq:g27pr}), and (\ref{eq:g8pr}) show an
interesting correlation between the $1/N_c$--corrections to the different
couplings. {\it The same correction} which decreases the $C_{+}$--modulated
terms, and hence the
$B_{K}$--parameter as well, increases the $C_{-}$--modulated term; and hence
the $\Delta I=1/2$--transitions. It can be shown, in full generality, that this
is in fact a general property of the full
$\cO (N_c)$--corrections in the chiral limit~\cite{PR95}.

\item The term proportional to $C_6(\mu^2)$ in Eq.(\ref{eq:g8pr}) is the result
of the next--to--leading $1/N_c$ calculation we have already discussed in the
{\sl subsubsec.\,3.2.2.}

\item The cancellation of the $\mu^2$-dependence of the other Wilson
coefficients with the bosonization of the corresponding four--quark operators,
is more involved. To exhibit this cancellation explicitly, requires the
knowledge of  the full dynamics of next--to--leading order in the
$1/N_c$--expansion.    It is clear that, in as far as one doesn't know the
origin of the phenomenological term in eq.(\ref{eq:mqterm}), this cannot be
shown. It is possible however to show that the logarithmic
$\alpha_{s}$--corrections to the bosonized operators are weighted by the same
anomalous dimension factors as those of the original four--quark operators;
i.e.,
$$<Q_{i}>\sim  -\frac{1}{2}\gamma_{ij}^{(1)}\frac{\alpha_{s}}{\pi}
\log(\frac{\mu^2}{M_{Q}^2})<Q_{j}>.$$

\end{itemize}

Further calculations within the framework of the {\it Effective Action Approach
Model} discussed here can be found in
references~\cite{B92}$^{,}$\cite{BrP93}$^{,\rm and}$\cite{B94}. A possible
generalization of the constituent mass ansatz term in (\ref{eq:MQterm}) to a
non--local form has also been suggested~\cite{HTV90}. Phenomenological
applications using a non--local constituent mass term can be found in
Ref.~\cite{H90}.

%%%%%%%%%%%%%%%%%%%%%%%%%%%%%%%%%%%%%

\subsection{The Extended Nambu and Jona-Lasinio Model (ENJL--model)}
\label{subsec:enjlm}

Since the early work of Nambu and Jona-Lasinio~\cite{NJL61}, there have been
many suggestions in the literature proposing models of the type first
discussed by these authors, as relevant models for low--energy hadron dynamics.
[For a recent review where earlier references can be found see~\cite{M93}.] The
scenario suggested in Refs.~\cite{BBR93,BRZ94}, which I shall follow here,
assumes that at intermediate energies below or of the order of the spontaneous
chiral symmetry breaking scale $\Lambda_{\chi}$, the leading operators of
higher dimension  which, after integration of the high frequency modes of the
quark and gluon fields down to the scale
$\Lambda_{\chi}$, become relevant in the QCD--Lagrangian, are those which can
be cast in the  form of four--fermion operators, i.e.,
\be
\label{eq:qcdnjl}
\cL _{QCD}\Ra \cL _{QCD}^{\chi}+\cL _{S,P}+ \cL _{V,A} + \cdots ,
\ee

\noi where
\be
\label{eq:enjlsp}
\cL _{S,P}=
\frac{1}{N_{c}}\frac{8\pi^2 }{\Lambda_{\chi}^2 }{\bf G}_{S}
\sum_{i,j}(\bar{q}_{R}^i q_{Lj})(\bar{q}_{L}^j q_{Ri} ),
\ee

\noi and
\be
\label{eq:enjlva}
\cL _{V,A}=-\frac{1}{N_{c}}\frac{8\pi^2 }{\Lambda_{\chi}^2 }{\bf G}_{V}
\sum_{i,j}[(\bar{q}_{L}^i \gamma^{\mu}q_{Lj})(\bar{q}_{L}^j
\gamma_{\mu} q_{Li}) + L \leftrightarrow R] .
\ee

\noi Here $i$,$j$ denote $u$, $d$, and $s$ flavour indices and summation over
colour degrees of freedom within each bracket is understood;
$q_{L,R}\equiv \frac{1}{2}(1\pm \gamma_{5})q$. The couplings
${\bf G}_{S,V}$ are dimensionless functions of the ultraviolet integration
cut--off $\Lambda$. They are expected to grow as
$\Lambda$ approaches the critical value $\Lambda_{\chi}$, where spontaneous
chiral symmetry breaking occurs. (This is the reason why the operators $\cL
_{S,P}$ and $\cL _{V,A}$ become relevant.) In QCD, and with the factor
$N_{c}^{-1}$ pulled out, both couplings ${\bf G}_{S}$ and ${\bf G}_{V}$  are
$\cO (1)$ in the  large--$N_{c}$ limit. These constants are in principle
calculable functions of the ratio $\Lambda / \Lambda_{\rm QCD}$. In practice
however, the calculation requires non--perturbative knowledge of QCD in the
region where $\Lambda \simeq \Lambda_{\chi}$, and we shall take ${\bf G}_{S}$
and ${\bf G}_{V}$, as well as
$\Lambda_{\chi}$, as independent unknown parameters. The $\chi$ index
in $\cL _{QCD}^{\chi}$ means that only the low--frequency modes
$\Lambda \leq
\Lambda_{\chi}$ of the quark and gluon fields are to be considered from now
onwards.

Notice that in QCD, couplings of the type $\cL _{S,P}$ and $\cL _{V,A}$ appear
naturally from gluon exchange between two QCD colour currents. Using Fierz
rearrangement, one has in the large--$N_{c}$ limit:
$$
\begin{array}{ccc} g_{s}^2 \sum _{a} (\bar{q}\gamma^{\mu}
\frac{\lambda^{a}}{2}q)
(\bar{q}\gamma_{\mu}\frac{\lambda^{a}}{2}q) & \Rightarrow &
\frac{1}{N_{c}}\frac{8\pi^2 } {\Lambda_{\chi}^2 } 4\frac{\alpha_{s}N_{c}}{\pi}
\sum_{i,j} (\bar{q}_{R}^i q_{Lj})(\bar{q}_{L}^j q_{Ri} ) \\ & &
-\frac{1}{N_{c}}\frac{8\pi^2 }{\Lambda_{\chi}^2 }
\frac{\alpha_{s}N_{c}}{\pi}
\sum_{i,j}  [(\bar{q}_{L}^i \gamma^{\mu}q_{Lj}) (\bar{q}_{L}^j \gamma_{\mu}
q_{Li}) + L \leftrightarrow R];
\end{array}
$$
i.e; ${\bf G}_{V}={\bf G}_{S}/4=\frac{\alpha_{s}N_{c}}{\pi}$ in this case.
The two operators $\cL _{S,P}$ and $\cL _{V,A}$ have however different
anomalous dimensions, and it is therefore not surprising that
${\bf G}_{S}\neq 4{\bf G}_{V}$ for the corresponding physical values.

If furthermore, one assumes that the relevant gluonic effects for low--energy
physics are those already absorbed in the new couplings
${\bf G}_{S}$ and ${\bf G}_{V}$, then
$$
\cL _{QCD}^{\chi}\Rightarrow i\bar{q}\not \!\! Dq
$$
in Eq.(\ref{eq:qcdnjl})
with $\not \!\! D$ the Dirac operator given in Eq.~(\ref{eq:dirac}), where now
the gluon field $G_{\mu}$ plays the r{\^o}le of an external colour field
source. There is no gluonic kinetic term any longer.

As is well known from the early work of Nambu and Jona-Lasinio
\cite{NJL61}, the operator $\cL _{S,P}$, for values of
${\bf G}_{S}>1$, is at the origin of the spontaneous chiral symmetry breaking.
This can best be seen following the standard procedure of introducing auxiliary
field variables to convert the four--fermion coupling operators into bilinear
quark operators. For this purpose, one introduces a $3\times 3$ auxiliary field
matrix $M(x)$ in flavour space; the so called collective field variables, which
under chiral--$SU(3)$ transform as
$$M\ra V_{R}MV_{L}^{\dagger};$$ and uses the functional integral identity:
$$\exp \,[ i\int \! d^4 x\, \frac{1}{N_{c}}
\frac{8\pi^2 }{\Lambda_{\chi}}{\bf G}_{S}
\sum_{i,j} (\bar{q}_{R}^i q_{Lj})(\bar{q}_{L}^j q_{Ri} )] = $$
\be
\label{eq:fiisp}
\int \!\cD M\, \exp\,[ i\int \! d^4 x\{-(\bar{q}_{L}M^{\dagger}q_{R} + h.c.)
-N_{c}\frac{\Lambda_{\chi}^2 }{8\pi^2 }\frac{1}{{\bf G}_{S}}trM
M^{\dagger}\}].
\ee

\noi By polar decomposition
$$M=\xi H\xi,$$ with $\xi\xi=U$ unitary and $H$ hermitian.

Next, we look for translational-invariant solutions, which minimize the
effective action;
$$\frac{\partial \Gamma_{\rm eff}}{\partial M}
\Big|_{H=<H>=M_{Q},\xi=1; v=a=s=p=0.} =0.$$ The minimum is reached when all the
eigenvalues of $<H>$ are equal, i.e., $<H>=M_{Q}1$; and the minimum condition
leads to
\be
\label{eq:mincon} {\rm Tr}(x\mid \frac{1}{\not \!\! D}\mid x)=
-2M_{Q}N_{c}\frac{\Lambda_{\chi}^2 }{8\pi^2 }\frac{1}{{\bf G}_{S}}
\int \! d^4 x\,.
\ee

\noi The trace in the l.h.s. of this equation is formally proportional to
$<\bar{\psi}\psi>$. The calculation however requires a regularization, with
$\Lambda_{\chi}$ the ultraviolet cut--off. We choose the proper time
regularization. [See e.g., Ref.~\cite{BBR93} for technical details.] Then
\be
\label{eq:fercon} <\bar{\psi}\psi>= -\frac{N_{c}}{16\pi^2 }4M_{Q}^3
\Gamma (-1,\frac{M_{Q}^2 }{\Lambda_{\chi}})\,;
\ee

\noi and the minimum condition in Eq.~(\ref{eq:mincon}) leads to the so-called
{\it gap equation}:
\be
\label{eq:gapeq}
\frac{M_{Q}}{{\bf G}_{S}}=M_{Q}\left\{ \exp(-
\frac{M_{Q}^2 }{\Lambda_{\chi}^2 })
-\frac{M_{Q}^2 }{\Lambda_{\chi}^2 }\Gamma (0,\frac{M_{Q}^2 } {\Lambda_{\chi}^2
}) \right\}\,.
\ee

\noi The functions
$$\Gamma(n-2,x\equiv \frac{M_{Q}^2 }{\Lambda_{\chi}^2 })=
\int_{x}^{\infty} \frac{dz}{z}e^{-z}z^{n-2}; \quad n=1,2,3,\dots,$$
are incomplete gamma functions. Equations~(\ref{eq:fercon}) and
(\ref{eq:gapeq}) show the existence of two phases with regards to chiral
symmetry. The unbroken phase corresponds to the trivial solution $M_{Q}=0$,
which implies $<\bar{\psi}\psi>=0$. The broken phase corresponds to the
possibility that the coupling ${\bf G}_{S}$ increases as we decrease the
ultraviolet cut--off $\Lambda$ down to $\Lambda_{\chi}$, allowing for solutions
to Eq.~(\ref{eq:gapeq}) with $M_{Q}>0$ and therefore $<\bar{\psi}\psi>\neq 0$
and negative. In this phase the Hermitian auxiliary field $H(x)$ develops a
non--vanishing vacuum expectation value, which is at the origin of a
constituent chiral quark mass term [see the r.h.s. of  Eq.(\ref{eq:fiisp})]:
$$-M_{Q}(\bar{q}_{L}U^{\dagger}q_{R}+\bar{q}_{R}Uq_{L})= -M_{Q}\bar{Q}Q\,,$$

\noi like the one which appears in the Georgi--Manohar model;
and like the one proposed in the effective action approach of
Ref.~\cite{ERT90}.

In the presence  of the operator $\cL _{V,A}$, we need two more auxiliary
$3\times 3$ complex field matrices
$L_{\mu}(x)$ and $R_{\mu}(x)$ to rearrange the Lagrangian in (\ref{eq:qcdnjl})
into an equivalent Lagrangian which is only quadratic in the quark fields.
Under chiral $(V_{L},V_{R})$ transformations these collective field variables
are chosen to transform as follows:
$$L_{\mu}\ra V_{L}L_{\mu}V_{L}^{\dagger}\,, \qquad R_{\mu}\ra
V_{R}R_{\mu}V_{R}^{\dagger}\,.$$
Then, the following functional identity
follows:
$$\exp (-i\int \!d^4 x \,
\frac{1}{N_{c}}\frac{8\pi^2 }{\Lambda_{\chi}^2 } {\bf G}_{V}
\sum_{i,j}\, [(\bar{q}_{L}^i \gamma^{\mu}q_{Lj})(\bar{q}_{L}^j
\gamma_{\mu} q_{Li}) + L \leftrightarrow R]\}) =$$
\be
\label{eq:fiiva}
\int \!\cD L_{\mu}\,\cD R_{\mu}\, \exp [i\int \!d^4 x\,
 \{\bar{q}_{L}\gamma^{\mu}L_{\mu}q_{L}+N_{c}
\frac{\Lambda_{\chi}^2 }{8\pi^2 }\frac{1}{{\bf G}_{V}}
\frac{1}{4}trL^{\mu}L_{\mu} + L\leftrightarrow R \}]\,.
\ee

\noi It is convenient to trade the auxiliary field matrices $L_{\mu}(x)$ and
$R_{\mu}(x)$ by new vector field matrices
$$W_{\mu}^{(\pm)}=\xi L_{\mu}\xi^{\dagger}
\pm \xi^{\dagger}R_{\mu}\xi,$$ which transform homogeneously under chiral
transformations $(V_{L},V_{R})$; i.e,
$$W_{\mu}^{(\pm)}\ra hW_{\mu}^{(\pm)}h^{\dagger},$$
with $h$ the $SU(3)_{V}$ rotation induced by $(V_{L},V_{R})$. The fermionic
determinant can then be obtained using standard techniques, like for example
the {\it heat kernel} expansion we described earlier. When computing the
resulting effective action, there appears a mixing term between the fields
$W_{\mu}^{(-)}$ and $\xi_{\mu}$. One needs a new redefinition of the auxiliary
field
$W_{\mu}^{(-)}$:
$$W_{\mu}^{(-)}\ra \hat{W}_{\mu}^{(-)} + (1-g_{A})\xi_{\mu}\,,$$
in order to diagonalize the quadratic form in the variables
$W_{\mu}^{(-)}$ and $\xi_{\mu}$. It is this mixing which is at the origin of an
effective axial coupling of the constituent quarks with the Nambu--Goldstone
modes:
$$\frac{1}{2}ig_{A}\bar{Q}\gamma^{\mu}\gamma_{5}\xi_{\mu}Q\,,$$
a term like the axial coupling which appears in the Georgi--Manohar model. but
with a specific form for the axial coupling constant $g_{A}$:
\be
\label{eq:ga} g_{A}=\frac{1}{1+{\bf G}_{V}\frac{4M_{Q}^2 }{\Lambda_{\chi}^2 }
(\Gamma(0,\frac{M_{Q}}{\Lambda_{\chi}^2 })}\,.
\ee

\noi In terms of Feynman diagrams this result can be understood as an {\it
infinite} sum of constituent quark bubbles, with a coupling  at the end to the
pion field. These are the diagrams generated by the
${\bf G}_{V}$ four-fermion coupling to leading order in the
$1/N_{c}$--expansion.  The quark propagators in these diagrams are constituent
quark propagators, solution of the Schwinger-Dyson which is at the origin of
the {\it gap equation} in (\ref{eq:gapeq}). In the limit where ${\bf G}_{V}=0$,
$g_{A}=1$; but in general~\cite{PR93}
$g_{A}\neq 1$ to leading order in the $1/N_c$--expansion.

Kinetic terms for the auxiliary field variables are also generated by the
functional integral over the quark fields
$Q$ and $\bar{Q}$. The resulting Lagrangian, after wave--function rescaling of
the auxiliary fields, has the form of a constituent chiral quark model, with
scalar $S(x)$, vector $V(x)$, and axial--vector $A(x)$ field couplings:
\beqn
\label{eq:cqmsva}
\cL _{\rm eff}^{ENJL} & = & i\bar{Q} \gamma^{\mu}(\partial_{\mu}+
\Gamma_{\mu}-\frac{i}{\sqrt{2}f_{V}}V_{\mu})Q -M_{Q}\bar{Q}Q  \no \\ & + &
\frac{i}{2}g_{A}\bar{Q}\gamma_{5}\gamma^{\mu}
(\xi_{\mu}-\frac{\sqrt{2}}{f_{A}}A_{\mu})Q -\frac{1}{\lambda_{S}}\bar{Q}S(x)Q
\no \\ & + & \frac{1}{2}tr[\partial_{\mu}S\partial^{\mu}S-M_{S}^2 SS]
\no \\ & - & \frac{1}{4}tr[(\partial_{\mu}V_{\nu}-\partial_{\nu}V_{\mu})
(\partial^{\mu}V^{\nu}-\partial^{\nu}V^{\mu})-2M_{V}V_{\mu}V^{\mu}]
  \no \\ & - &
\frac{1}{4}tr[(\partial_{\mu}A_{\nu}-\partial_{\nu}A_{\mu})
(\partial^{\mu}A^{\nu}-\partial^{\nu}A^{\mu})-2M_{A}^2 A_{\mu}A^{\mu}]\no \\ &
+ & \frac{1}{4}f_{\pi}^2{\rm tr}D_{\mu}UD^{\mu}U^{\dg} + \cO (p^4) {\rm
terms}\,,
\eeqn

\noi where $\Gamma_{\mu}$ and $\xi_{\mu}$ are the same as those
defined in Eqs.(\ref{eq:gamu}) and (\ref{eq:ximu}), and the coupling constants
and masses are now expressed in terms of only three input parameters. As input
parameters, we can either fix:
${\bf G}_{S}$, ${\bf G}_{V}$, and
$\Lambda_{\chi}$; or the more physical parameters:
$$M_{Q}, \qquad \Lambda{\chi}, \qquad  g_{A}.$$ The coupling constants are
then:
\be
\label{eq:fpi} f_{\pi}^2 =\frac{N_{c}}{16\pi^2 }4M_{Q}^2 g_{A}
\Gamma(0,M_{Q}^2/\Lambda_{\chi}^2 ),
\ee

\noi
\be
\label{eq:fv} f_{V}^2 =\frac{N_{c}}{16\pi^2 }
\frac{2}{3}\Gamma(0,M_{Q}^2/\Lambda_{\chi}^2 ),
\ee
\be
\label{eq:fa} f_{A}^2 =\frac{N_{c}}{16\pi^2 }\frac{2}{3}g_{A}^2
[\Gamma(0,M_{Q}^2/\Lambda_{\chi}^2 ) -\Gamma(1,M_{Q}^2/\Lambda_{\chi}^2 )],
\ee
\be
\label{eq:ls}
\lambda_{S}^2 = \frac{N_{c}}{16\pi^2 }\frac{2}{3}
[3\Gamma(0,M_{Q}^2/\Lambda_{\chi}^2 ) -2\Gamma(1,M_{Q}^2/\Lambda_{\chi}^2 )];
\ee

\noi and the masses:
\be
\label{eq:mv} M_{V}^2 =6M_{Q}^2 \frac{g_{A}}{1-g_{A}},
\ee
\be
\label{eq:ma} M_{A}^2 =6M_{Q}^2 \frac{1}{1-g_{A}}
\frac{1}{1-\frac{\Gamma(1,M_{Q}^2 /\Lambda_{\chi}^2 )} {\Gamma(0,M_{Q}^2
/\Lambda_{\chi}^2 )}},
\ee

\noi and
\be
\label{eq:ms} M_{S}^2 =4M_{Q}^2
\frac{1}{1-\frac{2}{3}\frac{\Gamma(1,M_{Q}^2 /\Lambda_{\chi}^2 )}
{\Gamma(0,M_{Q}^2 /\Lambda_{\chi}^2 )}}.
\ee

\noi In the absence of the vector  and axial--vector four--fermion like
coupling i.e., when ${\bf G}_{V}=0$: $g_{A}=1$,
$M_{V}\ra \infty$ and $M_{A}\ra \infty$. Then the vector and axial--vector
interactions decouple, and the model becomes  equivalent to the {\it
Constitutuent Chiral Quark Model} of Georgi and Manohar, with $g_{A}=1$ and a
non--trivial coupling to a scalar field.

The functional integration over the quark fields and the auxiliary
$S(x)$, $V(x)$, and $A(x)$ fields results in an effective action among the
Nambu--Goldstone boson particles, with all the couplings fixed by the three
parameters $M_{Q}$, $\Lambda_{\chi}$, and $g_{A}$. The explicit results one
gets for the $L_{i}$ constants which appear in the  large--$N_{c}$ limit at
$\cO (p^4 )$ in the  chiral expansion are shown in Table 2. The reason why the
constant $L_{7}$ does not appear in this Table is that, phenomenologically,
this constant gets a large contribution from the integration of the {\it heavy}
singlet $\eta'$ particle. However, in the chiral limit, the mass of the $\eta'$
is induced by the axial--$U(1)$ anomaly, which only appears to
next--to--leading order in the $1/N_c$--expansion. By definition, the
ENJL--model, as formulated here, ignores this effect. In order to take these
next--to--leading effects in $1/N_c$ systematically, together with the chiral
expansion, one has to resort to a $U(3)\times U(3)$ formulation of the
effective theory~\cite{W79}. The constants $L_{4}$ and $L_{6}$ are of
next--to--leading order in the
$1/N_c$--expansion; this is the reason why they do not appear in Table 2
either. We also show in Table 2 the numerical results of the fit 1 discussed in
Ref.~\cite{BBR93}. These results correspond to the set of input parameter
values :
\be
\label{eq:inpars} M_Q=265\ MeV,\quad \Lambda_{\chi}=1165\ MeV,  \quad g_A=0.61.
\ee

%%%%%%%%%%%%%%%%%%%%%%%%%%%%%%%%%%%%%%%%%%%%%%%
%%%%%%%TABLE%%%%%%%%

\begin{samepage}
{\bf Table 2 :} The $L_{i}$-coupling constants in the
 ENJL--model of Ref.\cite{BBR93}, with $g_A$ defined in Eq.(\ref{eq:ga}), and
$\Gamma_n\equiv\Gamma (n, M^2_Q/\Lambda^2_{\chi})$. The second column gives the
results corresponding to the input parameter values in (\ref{eq:inpars}). The
third column gives the experimental values of Table 1.
\vglue 5mm

\begin{tabular}{|l|r|r|} \hline \hline & & \\ {\em The $L_{i}$ couplings of
$\cO (p^4)$ in the  ENJL--model} &  {\em Fit 1} & {\em Experiment} \\
\hline  &  &  \\
$L_{1}=\frac{N_c}{16\pi^2}\frac{1}{48} [(1-g_{A}^2 )^2 \Gamma_0 +4g_{A}^2
(1-g_{A}^2 )\Gamma_1 + 2g_{A}^4 \Gamma_2 ]$ & $0.85$ & $0.7\pm0.5$ \\ & & \\
$L_{2}=2L_{1}$ & $1.7$ & $1.2\pm0.4$ \\ & & \\
$L_{3}=-\frac{N_c}{16\pi^2}\frac{1}{8}
\{[(1-g_{A}^2 )^2 \Gamma_0 +4g_{A}^2 (1-g_{A}^2 )\Gamma_1+ $
 & & \\
$\qquad \qquad \quad  -\frac{2}{3}g_{A}^4 [2\Gamma_1 -4\Gamma_2 +
3\frac{1}{\Gamma_0 }(\Gamma_0 -\Gamma_1 )^2 ]\}$ & $-4.2$ &
$-3.6\pm 1.3$ \\  &  &  \\ \hline & & \\
$L_{5}=\frac{N_c}{16\pi^2}\frac{1}{4}g_{A}^3  [\Gamma_0 -\Gamma_1 ]$ & $1.6$ &
$1.4\pm 0.5$ \\ & & \\
$L_{8}=\frac{N_c}{16\pi^2}\frac{1}{16}g_{A}^2 [\Gamma_0 -\frac{2}{3}\Gamma_1 ]$
& $0.8$ & $0.9\pm 0.3$ \\ & & \\
 \hline  &  &  \\
$L_{9}=\frac{N_c}{16\pi^2}\frac{1}{6} [(1-g_{A}^2)\Gamma_0 +2g_{A}^2
\Gamma_1 ]$ & $7.1$ & $6.9\pm 0.7$ \\ & & \\
$L_{10}=-\frac{N_c}{16\pi^2 }\frac{1}{6} [(1-g_{A}^2)\Gamma_0 +g_{A}^2
\Gamma_1 ]$ & $-5.9$ & $-5.5\pm 0.7$
\\ & & \\
\hline \hline
\end{tabular}
\end{samepage}

%%%%%%%%%%%%%%%%%%%%%%%%%%%%%%%%%%%%%%%%%%%%%%%%
\vspace{5 mm}
\noi The overall picture which emerges from this simple model is quite
remarkable. The main improvement with respect to the results obtained in the
{\it effective action approach model} discussed in {\it
subsec.\,4.4} is on the constants $L_{5}$ and $L_{8}$, where the combined
effect of the vector and scalar degrees of freedom leads to rather simple
results modulated by powers of the $g_{A}$--constant, which agree very well
with the phenomenological determinations. One of the characteristic features of
the  ENJL--model, is that it interpolates successfully between pure VMD--type
predictions and those of the constituent chiral quark model. A nice
illustration is the result for $L_{9}$ in Table 2 , where the first term is the
one coming from vector--exchange, while the second one comes from the chiral
quark loop integral.

There is no difficulty to reproduce the anomalous Wess--Zumino--Witten
functional within the ENJL--model~\cite{BP94a}

QCD two--point functions, beyond the low--energy expansion, have  also been
evaluated in the  ENJL--model~\cite{BRZ94}. This involves calculations to
leading order in the
$1/N_c$--expansion (i.e., an infinite number of chains of fermion bubbles; but
no loops of chains,) and to all orders in powers of momenta $Q^2
/\Lambda_{\chi}^2 $. As a result, vector and axial--vector correlation
functions have a VMD--like form, but with slowly varying couplings and masses.
For the transverse invariant functions for example, the results are
\be
\label{pivv1}
\Pi_{V}^{(1)}(Q^2 )=
\frac{2f_{V}^2 (Q^2 )M_{V}^2 (Q^2 )} {M_{V}^2 (Q^2 )-Q^2 },
\ee

\noi and
\be
\label{piaa1}
\Pi_{A}^{(1)}(Q^2 )= \frac{2f_{\pi}^2 (Q^2 )}{Q^2 }+
\frac{2f_{A}^2 (Q^2 )M_{A}^2 (Q^2 )} {M_{A}^2 (Q^2 )-Q^2 },
\ee

\noi where
$$f_{V}^2 (Q^2 )= 4\frac{N_c}{16\pi^2 }\int_0 ^1 dx x(1-x)
\Gamma(0,x_{Q}\equiv [M_{Q^2 }+x(1-x)Q^2 ]/\Lambda_{\chi}^2 ]).$$
The product
$$2f_{V}^2 (Q^2 )M_{V}^2 (Q^2 )=N_c
\frac{\Lambda_{\chi}^2 }{8\pi^2 }\frac{1}{{\bf G}_{V}}$$
is scale invariant.

With
$$g_{A}(Q^2 )= \frac{1} {1+{\bf G}_{V}\frac{4M_{Q^2 }}{\Lambda_{\chi}^2 }
\int_0 ^1 dx \Gamma(0,x_{Q})},$$ the other couplings are fixed by
$$f_{A}^2 (Q^2 )= g_{A}^2 (Q^2 ) f_{V}^2 (Q^2 ),$$ and the relations

\be
\label{1stwsr} f_{V}^2 (Q^2 )M_{V}^2 (Q^2 )=f_{A}^2 (Q^2 )M_{A}^2 (Q^2 )+
f_{\pi}^2 (Q^2 ),
\ee

\noi and
\be
\label{2ndwsr} f_{V}^2 (Q^2 )M_{V}^4 (Q^2 )=f_{A}^2 (Q^2 )M_{A}^4 (Q^2 ).
\ee

\noi The last two equations are the $Q^2$--dependent version of the
$1^{\rm st}$-- and $2^{\rm nd}$--Weinberg sum rules~\cite{W67}.

In the case of the scalar two--point function there appears a pole in the
$Q^2$--summed expression at $M_{S}=2M_{Q}$. The case of the other two--point
functions is somewhat more involved because they mix through the four--fermion
interaction terms. The corresponding  results can be found in Ref.\cite{BRZ94}.
Calculations of the low--energy behaviour of some three--point functions in the
ENJL--model have also been made~\cite{BP94}. Corrections due to possible
four--quark operators of higher dimension involving derivative terms, have also
been studied recently~\cite{PaP94}.

In principle, the ENJL--model can be applied to obtain a systematic calculation
of the low--energy constants of the weak non--leptonic Lagrangian, like the
$B_{K}$--parameter, ${\bf g}_{8}$ and ${\bf g}_{27}$. The difficulties are
mainly technical; but progress is underway.

%%%%%%%%%%%%%%%%%%%%%%%%%%%%%%%%%%%%%%%%%%%%%%%%%%%%%%%%%%%%%

\section{Rare Kaon Decays and CP--Violation}
\label{sec:rkdcpv}

This last lecture is dedicated to a discussion of the theoretical status,
within the Standard Model, of the CP--violation parameters $\epsilon$ and
$\epsilon'$ which we have introduced in {\bf Sec.\,1.} and also to the
study of the decay $K_{L}\ra \pi^{0} e^{+}e^{-}$. Other than the
$K\ra \pi\pi$ decays which we have discussed, this rare decay mode seems to
be the most promising candidate to observe direct CP--violation in
$K$--physics in the near future.

%%%%%%%%%%%%%%%%%%%%%%%%%%%%%%%%%%%%%%%%%%%%%%%%
\subsection{The Parameters $\epsilon$ and $\epsilon'$ Revisited}
\label{subsec:eps_epsp}

In {\bf Sec.\,1.} we have seen that CP--violation in
$K\ra \pi\pi$ decays is governed by the parameters $\epsilon$ and
$\epsilon'$, [cf. Eqs.(\ref{eq:eps}) and (\ref{eq:epsilonprime2}).] We
have also seen that, to a good approximation, these parameters can be
written as follows:
\be
\label{eq:epsa}
\epsilon\simeq \frac{1}{\sqrt{2}}e^{i\pi /4}
\left(\frac{{\rm Im}M_{12}}{\Delta m} + \frac{{\rm Im}A_0}{{\rm
Re}A_0}\right)\,,
\ee

\noi and
\be
\label{eq:epspa}
\epsilon'\simeq \frac{1}{\sqrt{2}}e^{i(\delta_2 -\delta_0 + \pi /2)}
\frac{{\rm Re}A_2}{{\rm Re}A_0}\left(\frac{{\rm Im}A_2}{{\rm Re}A_2} -
\frac{{\rm Im}A_0}{{\rm Re}A_0}\right)\,.
\ee

\noi The question we want to discuss here, is the present
status in the determination of the various components which appear in
these expressions.

%%%%%%%%%%%%%%%%%%%%%%%%%
\subsubsection{\sl The mass difference $\Delta m$.}
\label{subsub:mlms}

The $K_{L}-K_{S}$ mass difference $\Delta m$ is well determined
experimentally~\cite{PDB94}:
\be
\label{eq:kl_ks_md}
\Delta m\equiv m_{K_{L}}-m_{K_{S}} =
(3.522 \pm 0.016)\times 10^{-12}MeV.
\ee

\noi
As we have shown in {\bf Sec.\,1.}, $\Delta m$ is
related to the real part of the off--diagonal $K^0 -\bar{K}^0$ mass
matrix, [cf. Eq.(\ref{eq:mdiff})]:
$$\Delta m \simeq 2{\rm Re}M_{12}.$$
In the Standard Model, there are both short--distance and long--distance
contributions to ${\rm Re}M_{12}$. The short--distance contributions, from
the effective $\Delta S =2$ Hamiltonian described in {\it subsec.\,3.3},
are modulated by the same unknown four--quark matrix element as the
contribution to
${\rm Im}M_{12}$; i.e., the so called $B_{K}$--parameter defined in
Eq.(\ref{eq:Bpar}). If one was able to calculate the long--distance
component to ${\rm Re}M_{12}$, it would be possible to fix the value of the
$B_{K}$--factor from the measured value of $\Delta m$. Unfortunately, as I
shall next explain, this is not the case.

In principle, the long--distance contribution to
${\rm Re}M_{12}$, and hence to $\Delta m$, can be evaluated in
$\chi$PT. It appears at the one--loop level, from diagrams with two
vertex insertions of the effective
$\Delta S=1$ chiral Lagrangian of $\cO (p^2)$ described in {\it
subsec.\,3.2}, in the presence of the strong interactions. These chiral loops
are however divergent, and therefore they necessarily bring in new $\cO
(p^4)$,
$\Delta S=2$ local couplings which are not fixed by chiral symmetry
requirements alone. Here again, one has to resort to models of the
low--energy effective action to make further progress. Estimates of the
long--distance contributions to
$\Delta m$, show that they are important~\cite{BGK91}. This is not
surprising: the fact that nominally they are of higher order, $\cO (p^4)$ in
the chiral expansion, is here largely compensated by the relative
$\Delta I=\frac{1}{2}$ enhancement of the two lowest $\cO (p^2)$
vertices. In the absence of a reliable way to calculate the needed
couplings of the local $\cO (p^4)\: \Delta S=2$ effective Lagrangian,
there appears no way to get rid of the unknown
$B_{K}$--factor in the ratio
$\frac{{\rm Im}M_{12}}{{\rm Re}M_{12}}$; and this is the reason why, for
phenomenological purposes, one is forced to use the experimental value of
$\Delta m$ in Eq.(\ref{eq:epsa}).

%%%%%%%%%%%%%%%%%%%%%%%%%%%
\subsubsection{\sl The phase--shift difference $\delta_2 -\delta_0$.}
\label{subsub:d2_d0}

Contrary to the situation concerning $\Delta m$, the
$\pi-\pi$ phase shift difference
$\delta_2 -\delta_0$ is poorly known experimentally. (It is not
easy to extract $\pi-\pi$ elastic scattering amplitudes from
physical observables.) Here, $\chi$PT does better! One can extract
this phase shift difference from the calculation of the elastic
$\pi-\pi$ amplitudes to $\cO (p^4)$, without introducing new unknown
parameters, with the result~\cite{GM91}:
\be
\label{eq:psd}
\delta_{I=2}^{J=0}(M_{K}^2 ) -
\delta_{I=0}^{J=0}(M_{K}^2 ) =-45^{\circ} \pm 6^{\circ}.
\ee

%%%%%%%%%%%%%%%%%%%%%%%%
\subsubsection{\sl ${\rm Re}A_{0}$ and ${\rm Re}A_{2}$.}
\label{subsub:a0_a2}

There is very little we can do as yet, theoretically, concerning
${\rm Re}A_{0}$ and ${\rm Re}A_{2}$. As we have seen in {\bf Sec.\,3.},
their calculation to lowest order in
$\chi$PT, involves the couplings ${\bf g}_{8}$ and
${\bf g}_{27}$, which are not fixed by symmetry requirements
alone. The models discussed in {\bf Sec.\,4.} are not yet
developed to the required degree of precision for theory to be useful
here. With neglect of electromagnetic corrections, the ratio
$\frac{{\rm Re}A_{2}}{{\rm Re}A_{0}}$, can be extracted from the physical
branching ratio
$\frac{\Gamma (K_{S}\ra \pi^{+} \pi^{-})}{\Gamma (K_{S}\ra \pi^0
\pi^{0})}$, as already discussed in {\bf Sec.\,1.},
[cf. Eqs. (\ref{eq:NPSBR}) to (\ref{eq:A2vA0exp1}),] with the result
\be
\label{eq:rea2_rea0}
\frac{{\rm Re}A_{0}}{{\rm Re}A_{2}}= +22.2\,.
\ee

\noi
Since ${\rm Im}A_{I}\ll {\rm Re}A_{I}$, we can also extract ${\rm Re}A_{0}$
{}from the experimental $\Gamma (K\ra \pi\pi)$ rates with the result
\be
\label{eq:rea0}
|{\rm Re}A_{0}|=3.3\times 10^{-7}{\rm Ge}V.
\ee

To make predictions about
$\epsilon$ and $\epsilon'$ we still need to know ${\rm Im}M_{12}$ and
${\rm Im}A_{I}$ for $I=0,2$.

%%%%%%%%%%%%%%%%%%%%%%%%
\subsubsection{\sl The $\epsilon$--parameter.}
\label{subsub:epsend}

It is presently known that, experimentally\footnote{\ See the lectures of Sunil
Somalwar~\cite{Somal} for a description of the relevant
experiments.}\hspace*{1em}:
$\epsilon'\ll \epsilon$. On the other
hand, the Bell-Steinberger inequality we have discussed in {\it
subsec.\,1.2}, when restricted to the
$2\pi$ intermediate states, can be written as follows:
\be
(\frac{\Gamma_{S}+\Gamma_{L}}{2}+i\Delta m)\,2{\rm Re}\,\tilde{\epsilon}=
\epsilon \Gamma_{S} + \epsilon'[\Gamma_{S}(+-)-2\Gamma_{S}(00)].
\ee

\noi
The second term in the r.h.s. is suppressed in two ways: i) because
$\epsilon'\ll \epsilon$; ii) because of the $\Delta I=1/2$--rule.
Furthermore, using the empirical facts that $\Gamma_{L}\ll \Gamma_{S}$
and $\Delta m\simeq \Gamma_{S}/2$, we arrive at the constraint
\be
\label{eq:eps_epsp_const.}
(1+i){\rm Re}\,\tilde{\epsilon}\simeq \epsilon\,.
\ee

\noi
With $\tilde{\epsilon}$ given by Eq.(\ref{eq:epstilde1}) and $\epsilon$ by
Eq.(\ref{eq:epsa}), this constraint implies the neglect of
$\frac{{\rm Im}A_{0}}{{\rm Re}A_{0}}$ versus $\frac{{\rm Im}M_{12}}{\Delta
m}$. We conclude that, to a good approximation, $\epsilon$ is
completely governed by the size of $M_{12}$, evaluated in the usual
phase convention of flavour mixing in the Standard Model. With
$\frac{{\rm Im}A_{0}}{{\rm Re}A_{0}}$ versus $\frac{{\rm Im}M_{12}}{\Delta m}$
neglected, it is reasonable to neglect as well the long
distance effects contributing to ${\rm Im}M_{12}$. In fact, it has been argued
that the two effects largely cancel each other~\cite{BGK91}. To this
approximation,
$\epsilon$ is then entirely given by the local structure of the $\Delta
S=2$ box diagrams which generate the four--quark operator in
(\ref{eq:deltas2}).

It has become conventional in the phenomenology of flavour dynamics, to use
an approximate form of the flavour mixing matrix; the
Cabibbo--Kobayashi--Maskawa matrix\footnote{\ See the lectures of
Roberto Peccei~\cite{Peccei} for further details on this
parametrization. For a recent update of the Unitarity Triangle see
Ref.~\cite{PP94}}\hspace*{1em}:
\be
V=\left(
    \begin{array}{ccc}
    1- \frac{\lambda^2}{2} & \lambda & A\lambda^3\sigma e^{-i\delta} \\
    -\lambda & 1- \frac{\lambda^2}{2} & A\lambda^2 \\
    A\lambda^3(1-\sigma e^{i\delta}) & -A\lambda^2 & 1
           \end{array}
          \right) + \cO (\lambda^4)\,,
\ee

\noi
where $\lambda$ denotes the Cabibbo angle,
\be
\lambda \simeq |V_{\rm us}|= 0.2205\pm 0.0018\,;
\ee

\noi
and $A$ and $\sigma$ are parameters of order one. The phase $\delta$ is at
the origin of the CP--violation in the Standard Model. The merit of this
parametrization is that it clearly shows the orders of magnitude of the
various mixing matrix elements. Often, the notation
\be
\sigma e^{-i\delta}=\rho -i\eta\,,
\ee

\noi
is also used. In this parametrization, the Jarlskog invariant which
governs all the observables which violate CP--invariance~\cite{J85}, is
given by the product:
\be
|J_{\rm CP}|=2\eta\lambda^6 A^2.
\ee

\noi The
$A$--parameter is fixed from
$B$--decays into states containing charm:
\be
A\simeq \lambda^{-2}|V_{\rm cb}|= 0.82\pm 0.12\,;
\ee

\noi and $\sigma$ is fixed, once $|V_{\rm ub}|$ is measured:
\be
\lambda \sigma
\simeq \frac{\mid V_{\rm ub}\mid }{\mid V_{\rm cb}\mid }= 0.08\pm 0.02\,.
\ee

In terms of this
parametrization of the mixing matrix, and to the approximation that we have
finally adopted, the $\epsilon$--parameter can then be written as follows:
\beqn
\label{eq:epsfinal}
\epsilon &
= & \frac{1}{\sqrt{2}}e^{i\pi/4}\: \frac{1}{\Delta m}\:
\frac{4}{3}f_{K}^2 M_{K}^2\,\frac{1}{2M_{K}}\,
\frac{G_{F}^2}{2\pi^2}M_{W}^2
\times \hat{B}_{K}\,\times
\nonumber
\\ & &
\eta\lambda^6 A^2\,[\eta_{3}S(r_{c},r_{t})-\eta_{1}S(r_{c})
+A^2 \lambda^4 (1-\rho )\eta_{2}S(r_{t})]\,.
\eeqn

\noi
In this expression, ${\hat B}_{K}$ denotes the scale invariant
$B_{K}$--parameter:
\be
{\hat B}_{K}\equiv \alpha_{s}(\mu^2)^{-2/9}B_{K}(\mu^2)\,;
\ee

\noi
and $S(r_{ c},r_{t})$ and $S(r_{t})$ are the box--diagram
functions of the ratios
$r_{q}\equiv \frac{m_{q}^2}{M_{W}^2}$, with $m_{q}$ the corresponding
quark mass running at the pole value:
\be
S(r_{t})=\frac{r_{t}}{4}\left[1+\frac{9}{1-r_{t}}-\frac{6}{(1-r_{t})^2}
-\frac{6r_{t}^2\log r_{t}}{(1-r_{t})^3}\right]\,;
\ee

\noi and
\be
\label{eq:sfunc}
S(r_{c},r_{t})=r_{c}[\log (\frac{r_{t}}{r_{c}})-
\frac{3r_{t}}{4(1-r_{t})}(1+\frac{r_{t}}{1-r_{t}}\log r_{t})].
\ee

\noi The $\eta_{i}$ parameters
correct for short--distance gluon exchanges in the $\Delta S=2$
box--diagrams.~\cite{BH92} With
$\hat{B}$ factored out, they are renormalization scale
invariant, with the values\footnote{\ \ The
$\eta_{1}$-parameter is rather sensitive to $m_{c}$ and
$\Lambda_{QCD}$, which explains the different values found in\break the
literature. Here we use~\cite{A93} $\Lambda_{\overline{MS}}^{(5)}=
(240\pm 90) MeV$,
which corresponds to
$\Lambda_{\overline{MS}}^{(3)}=(350\pm 150) MeV$; for the charm quark
pole mass we take~\cite{N89} $m_{c}=(1.47\pm 0.05) MeV$. }\hspace*{1em} :
\be
\label{eq:etasqcd}
\eta_{1}=1.10, \qquad \eta_{2}=0.57, \qquad \eta_{3}=0.36.
\ee

\noi Equation(\ref{eq:epsfinal}), with the value of $\epsilon$ fixed from
experiment [$\mid \epsilon \mid =(2.26\pm 0.02)\times 10^{-3}$], and for a
{\it given}
 value of ${\hat B}_{K}$, determines then a hyperbola in the
$(\rho,\eta)$ plane. A useful approximation to Eq.(\ref{eq:epsfinal}), for
quick numerical estimates is
\be
|\epsilon|=(3.4\times 10^{-3})\eta\,A^2\,\hat{B}_{K}
\left[1+1.3A^2\,(1-\rho)\left(\frac{m_{\rm t}}{M_{W}}\right)^{1.6}\right]\,.
\ee

%%%%%%%%%%%%%%%%%%%%%%%%
\subsubsection{\sl Calculations of the ${\hat B}_{K}$--parameter.}

There exist several calculations of this parameter using very different
techniques. The problem is that, although the errors from some of the
calculations are rather small, the central values are still too
dispersed. Two simple calculations are the ones we have already discussed in
{\it subsec.\,3.3}: the vacuum saturation approximation~\cite{GL74},
which gives
\be
(B_{K})_{\rm VS}=1\,;
\ee

\noi and the large--$N_c$ prediction, which corresponds to
\be
B_{K}|_{N_c\to\infty}=3/4\,.
\ee

\noi None of these two calculations can distinguish between ${\hat B}_{K}$ and
$B_{K}(\mu^2)$.

We also pointed out in {\it subsec.\,4.4} that, as first observed by
Donoghue, Golowich and Holstein~\cite{DGH82}, there is a
symmetry relation in the chiral limit between
$\Delta S=2$ and $\Delta I=3/2$, $\Delta S=1$ transitions. Using the known
$K^{+}\ra \pi^{+}\pi^{0}$ decay rate, it is then possible to fix $B_{K}$ in
that limit, with the result
\be
{\hat B}_{K}\Big|_{\chi{\rm PT}\; {\rm to}\; \cO (p^2)} = 0.37\,.
\ee

\noi The calculation of $B_{K}$ within the effective action approach
model we have discussed in {\it subsec.\,4.4} shows how to reconcile this
result with the large--$N_c$ result above. The next--to--leading
corrections in the $1/N_c$--expansion appear to be large and negative. The
same pattern appears in the $1/N_c$--approach developped by Bardeen, Buras
and G\'{e}rard. From their matching of short--distances to long--distances,
these authors give the estimate~\cite{BBG88,G89}:
\be
{\hat B}_{K}=0.70\pm 0.10\,.
\ee

\noi This result includes the effect of the one loop $\cO (p^4)$ chiral
corrections as well, which are large and positive. Higher $\cO (p^4)$ chiral
corrections, including the effect of the local $\cO (p^4)$ terms,
evaluated within the effective action approach model, have also been recently
estimated~\cite{B94}, with the result
\be
{\hat B}_{K}=0.42\pm 0.06\,.
\ee

There are a variety of QCD Sum Rules that have been used to estimate
$B_{K}$ as well. The description of the techniques involved in these
calculations goes beyond the scope of these lectures. I shall however give
some results. The most recent estimate of $B_{K}$ using two--point function
sum rules of the
$\Delta	S=2$ four--quark local operator~\cite{PDPPR91} gives
\be
{\hat B}_{K}\Big|_{\rm 2p.f.\;sum\; rules}=0.39\pm 0.10\,.
\ee

\noi
The same approach reproduces rather well the observed $K^{+}\ra
\pi^{+}\pi^{0}$ decay rate. (It overestimates by less than
$15\%$ the coupling constant ${\bf g}_{27}^{(3/2)}$.)

The calculations based on QCD Sum Rules for three--point functions give a
large variety of outputs. I shall only quote the result of
Ref.~\cite{BDG88}, wherefrom one can trace earlier results:
\be
{\hat B}_{K}\Big|_{\rm 3p.f.\;sum\; rules}=0.5\pm 0.1\pm 0.2
\ee

Lattice--QCD simulations of the $B_{K}$--parameter tend to find results in
the higher range of the estimates we have presented here. They are
discussed by Steve Sharpe in his lectures at this TASI school.

In my opinion, the only safe claim that theorists can make at present is
that the value of ${\hat B}_{K}$ is likely to lie within the range:
\be
\label{eq:Brange}
0.35\le {\hat B}_{K}\le 0.80\,.
\ee

%%%%%%%%%%%%%%%%%%%%%%%%
\subsubsection{\sl The ratio $\epsilon'/\epsilon$.}
\label{subsub:eps/epsp_fnal}

As discussed in the lectures of Sunil Somalwar~\cite{Somal} , there is
experimental information on this ratio of parameters from the
measurement of the ratio of branching ratios [cf.
Eqs.(\ref{eq:etapm}),(\ref{eq:etapm1}) and
(\ref{eq:etazz1})]:
\be
\mid \frac{\eta_{+-}}{\eta_{00}}\mid \simeq
1+6{\rm Re}(\frac{\epsilon'}{\epsilon})\,,
\ee

\noi
with the results:
\be
\label{eq:eps/pesp_exps}
{\rm Re}(\frac{\epsilon'}{\epsilon})=\left\{
\ba{cc} (23\pm 6.5)\times 10^{-4} & {\rm CERN-NA}31\,, \\
(7.4\pm 6.0)\times 10^{-4} & {\rm FERMI\:lab-E}731\,. \ea \right.
\ee

\noi
The phases in the expressions of $\epsilon$ and $\epsilon'$ being
practically the same, the theoretical prediction for the ratio
$\frac{\epsilon'}{\epsilon}$ turns out to be a real number. If
furthermore, we use the experimental input in
Eq.(\ref{eq:rea2_rea0}) as well as the experimental determination of
$|\epsilon|\simeq 2.3\times 10^{-3}$, we get
\be
\label{eq:epsp_eps1}
\frac{\epsilon'}{\epsilon}=-\frac{1}{\sqrt{2}}\:
\frac{1}{\left(\frac{{\rm Re}A_{0}}{{\rm Re}A_{2}}\simeq 22.2\right)}
\: \frac{1}{\left(|\epsilon|\simeq 2.3\times 10^{-3}\right)}\:
\frac{{\rm Im}A_{0}}{{\rm Re}A_{0}}
\: (1-22.2\frac{{\rm Im}A_{2}}{{\rm Im}A_{0}}).
\ee

\noi
The dominant contribution to ${\rm Im}A_{0}$ originates in the diagrams
which give rise to the Penguin-operator $Q_{6}$. They bring the
factors $V_{{\rm td}}V_{{\rm ts}}^{*}$, $V_{{\rm cd}}V_{{\rm cs}}^*$, and
hence the CP--violation phase which contributes to ${\rm Im}A_{0}$. We have
seen in {\it subsec.\,3.2}, that to lowest order in $\chi$PT, the amplitude
$A_{0}$ takes the form,
$$A_{0}= -\frac{G_{F}}{\sqrt{2}}\, V_{{\rm ud}}V_{{\rm us}}^{*}\,
({\bf g}_{8}+{\bf g}_{27}^{(1/2)})\,\sqrt{2}f_{\pi}(M_{K}^2-m_{\pi}^2).$$

\noi
The operator $Q_{6}$ transforms like an ($\underline{8}_{L},
\underline{1}_{R})$
operator and contributes only to ${\bf g}_{8}$. In the
large--$N_{c}$ expansion, its contribution to
${\bf g}_{8}$ is {\it next--to--leading}. As we discussed in 3.2.2, it is
quite remarkable that, at this order of approximation, this
contribution can be calculated exactly in terms of known parameters!.
Remember the result [cf. Eq.(\ref{eq:g8penguin})]:
\be
\label{eq:chiral6}
{\rm Im}{\bf g}_{8}=
-{\rm
Im}C_{6}(\mu^2)\,16L_{5}\,\left[\frac{<\bar{\psi}\psi>}{f_{\pi}^3}\right]^2
\{ 1+\cO (1/N_c)  + \cO (p^2)\}.
\ee

\noi
With this result in hand, we can rewrite Eq.(\ref{eq:epsp_eps1}) in
the following way:
\beqn
\label{eq:epsp_eps2}
\frac{\epsilon'}{\epsilon} & = & \frac{1}{\sqrt{2}}\,\frac{1}{22.2}\,
\frac{\eta\lambda^4 A^2}{(2.3\times 10^{-3})}\,
\frac{(0.09\pm 0.01)}{\left({\rm Re}[{\bf g}_{8}+{\bf g}_{27}^{(1/2)}]\simeq
5.1\right)}\, 16L_{5}\,\left[\frac{<\bar{\psi}\psi>}{f_{\pi}^3}\right]^2 \no
\\
 & \times &
\left(1-22.2 \frac{{\rm Im}A_{2}}{{\rm Im}A_{0}}\right)
\left\{1+\cO (1/N_c)  +
\cO (p^2)\right\}.
\eeqn

\noi
Several comments concerning this result are in order:

\begin{itemize}

\item
The factor $(0.09\pm 0.01)$ in the r.h.s. is the value of the calculated
short--distance Wilson coefficient ${\rm Im}C_{6}$, once the overall
flavour factor $\eta\lambda^4\ A^2$ has been pulled out. The error
here comes from the uncertainties in the top--quark mass and in
$\Lambda_{\overline{MS}}$.

\item
As we already mentioned in {\it subsec.\,3.2}, the largest uncertainty
{}from the calculated matrix element of the ``penguin'' $Q_{6}$--operator is
due to the poorly known value of the quark condensate $<\bar{\psi}\psi>$. [Cf.
Eq.(\ref{eq:npsipsi}).] It is customary to trade this factor by the value of
the strange--quark mass, using the Gell-Mann--Oakes--Renner relation in
Eq.(\ref{eq:GOR}) of lowest order $\chi$PT. I prefer not to do that, because
then one should also discuss corrections to the Gell-Mann--Oakes--Renner
relation, which complicates things even further.

\item
Because of the large coefficient --$22.2$-- modulating ${\rm Im}A_2
/{\rm Im}A_0$ in the r.h.s. of the final expression
for $\epsilon'/\epsilon$ above, a reliable estimate of the ratio
${\rm Im}A_2 /{\rm Im}A_0$ is clearly needed. There are two sources of
contributions to this term:

i) Isospin breaking effects,

ii) Electroweak ``penguin'' effects, where the gluon exchange of the ordinary
``penguin''--like diagram is replaced by a photon or a
$Z$--vector boson. Because of the isovector component in the coupling of the
photon and the $Z$ to the quark--currents, the electroweak ``penguin''
diagrams lead to effective operators which can induce $\Delta
I=3/2$--transitions.

Both effects i) and ii) have been shown to go in the same direction, and they
decrease the leading effect of the $Q_6$--operator that we have calculated
above. The effect ii) has become particularly important because, as the
top--quark mass $m_t$ increases, for $m_t> M_W$, it grows roughly
as $m_t^2$. [See Refs.~\cite{BH92,BBH90} for a detailed discussion.]
Isospin breaking effects due to $\pi^{0}-\eta-\eta'$ mixing have been
estimated~\cite{DGHT86} to contibute
\be
\left(\frac{{\rm Im}A_2 }{{\rm Im}A_0}\right)_{\eta + \eta'}\simeq 1.4\times
10^{-2}\,.
\ee

\item
Some attempts to estimate the other $\cO (\frac{1}{N_c})$ and the other $\cO
(p^2)$ corrections in the r.h.s. of Eq.(\ref{eq:epsp_eps2}) have also been
made. The results however are controversial. New efforts in that direction
are urgently needed, if we want to compare usefully with the next
generation of $\epsilon'/\epsilon$ experiments.

\end{itemize}

Due to the various uncertainties we have discussed, it is
difficult to claim, at present, a theoretical prediction for
$\epsilon'/\epsilon$ better than the limits:
\be
\label{eq:epsp_eps3}
10^{-4}\leq\frac{\epsilon^{'}}{\epsilon}\leq 2\times 10^{-3}.
\ee

%%%%%%%%%%%%%%%%%%%%%%%%%%%%%%%%%%%%%%%%%%%%%%%%%%%%
%%%%%%%%%%%%%%%%%%%%%%%%%%%%%%%%%%%%%%%%%%%%%%%%%%%%%%%%%%%%%%%%%%%%%%%%%%%%%%
\subsection{The Decay $K_{L}\ra \pi^{0}e^{+}e^{-}$ and CP--Violation}

Transitions like $K\ra \pi \gamma$, with $\gamma$ a real photon, are
forbidden by gauge invariance. They are allowed, however, for virtual
photons $\gamma^*$, which can produce real lepton pairs. Because of the
absence of strangenes changing neutral currents in the Standard Model,
transitions like $K\ra \pi l^+ l^-$, with $l=e, \mu$, are then governed by
the interplay of weak non--leptonic and electromagnetic interactions. To
lowest order in the electromagnetic coupling constant they are expected to
proceed, in principle dominantly, via one--photon exchange. This is
certainly the case for the $K^{\pm}\ra \pi^{\pm} l^+ l^-$ and
$K_{S}\ra \pi^0 l^+ l^-$ decays. The transition
$K_{2}^{0}\ra \pi^0\gamma^*\ra \pi^0 l^+ l^-$, via one virtual photon, is
however forbidden by CP--invariance. It is then not obvious any longer
whether the physical decay $K_{L}\ra \pi^0 l^+ l^-$ will be still dominated
by the CP--suppressed
$\gamma^*$--virtual transition; or whether a transition via two virtual
photons, which is of a higher order in the electromagnetic coupling but
CP--allowed, may dominate. With the possibility of reaching branching
ratios for the mode $K_{L}\ra
\pi^{0}e^{+}e^{-}$  as small as $10^{-12}$ in the near future dedicated
experiments, the theoretical study of this mode has become of major
importance.

%%%%%%%%%%%%%%%%%%%%%%
\subsubsection{\sl The one--photon exchange amplitude}

The interesting thing about this transition amplitude is that it gets a
direct contribution coming from the terms in the $\Delta S=1$ short--distance
Hamiltonian proportional to the operators $Q_{11}$ and $Q_{12}$
 in {\it subsec.\,3.1}. These are the operators induced by ``penguin''--like
diagrams, where the gluon exchange of the ordinary
``penguin''--like diagram is replaced by a photon or a
$Z$--vector boson. Much the same as in the case of the
$Q_{6}$--operator which we have already discussed, these diagrams bring in
the factors $V_{\rm td}V_{\rm ts}^*$,$V_{\rm cd}V_{\rm cs}^*$, and hence the
CP--violation phase. The most recent calculation of the corresponding Wilson
coefficients, can be found in Ref.~\cite{BLMM94}. Another interesting issue
concerning this transition, is that the hadronic matrix element here is the one
of a quark bilinear operator; and by isospin symmetry is related to the well
known hadronic matrix element of charged
$K_{l3}$--decays. The main sources of errors in the calculation of this
direct transition amplitude are then the mass of the top quark, and the
CP--violation $\eta$--parameter [see Eq.(\ref{eq:epsfinal}),] which modulates
the overall amplitude. Letting the $\hat{B}_{K}$--parameter be within the
range in Eq.(\ref{eq:Brange}); and for a top mass $150{\rm GeV}\leq m_{\rm
t}\leq 200{\rm GeV}$, the expected branching ratio from this direct source of
CP--violation is
\be \label{eq:dirCP}
1.3\times 10^{-12}\leq  {\rm Br}(K_{L}\ra
\pi^{0}e^{+}e^{-})\Big|_{1\gamma \,{\rm dir.}}
\leq 5.8\times 10^{-12}\,.
\ee

\noi
Let me remind you that the present experimental upper limit
is~\cite{OBetal}
$(90\%\:{\rm C.L.})$
\be
{\rm Br}(K_{L}\ra \pi^{0}e^{+}e^{-})\Big|_{\rm exp.}< 5.5\times
10^{-9}\,,
\ee

\noi
still quite far away.

The other source of CP--violation, usually called ``indirect'', is the one
induced by the $K_{1}^{0}$--component of the $K_{L}$ state, which brings in
the CP--violation admixture parameter $\tilde{\epsilon}$, i.e., the parameter
$\epsilon$ to a very good approximation. The problem here is then reduced to
the evaluation of the CP--conserving transition $K_{S}\ra
\pi^{0}e^{+}e^{-}$.

The analysis of $K\ra \pi l^{+}l^{-}$ decays in general, within the
framework of $\chi$PT has been made in Refs.~\cite{EPR87,EPR88}. To $\cO
(p^4)$ in the chiral expansion, the corresponding decay amplitudes get
contributions both from one chiral loop graphs, and from tree level
contributions from local operators of $\cO (p^4)$. In fact, only two local
operators of the $\cO (p^4)$ non--leptonic effective Lagrangian contribute to
the CP--conserving amplitudes:
\be
\cL_{\rm eff}^{\Delta S=1}(x)\doteq -ieG_{8}\,\frac{2}{f_{\pi}^2}\,F^{\mu\nu}
\left\{{\bf w}_{1}{\rm tr}(Q\lambda_{6-i7}\cL_{\mu}\cL_{\nu})+ {\bf
w}_{2}{\rm tr}(Q\cL_{\mu}\lambda_{6-i7}\cL_{\nu})\right\}\,,
\ee

\noi
where $\cL_{\mu}(x)$ is the $3\times 3$ flavour matrix field we
introduced in Eq.(\ref{eq:Lweak}); $Q$ the electric charge matrix:
diag$(2/3,-1/3,-1/3)$; $F^{\mu\nu}$ the electromagnetic field strength
tensor;
$\lambda_{6-i7}$ the $SU(3)$ Gell-Mann matrix: $\lambda_{6}-i\lambda_{7}$;
and
${\bf w}_{1,2}$ are two dimensionless coupling constants not fixed by chiral
symmetry requirements alone. The overall constant $G_{8}$ denotes the
combination of couplings
\be
G_{8}=-\frac{G_{F}}{\sqrt{2}}\,V_{\rm ud}V^*_{\rm us}\,{\bf g}_{8}\,,
\qquad |G_{8}|\simeq 9\times 10^{-6}\,{\rm GeV}^{-2}\,.
\ee

The mode  $K^{+}\ra \pi^{+}e^{+}e^{-}$turns out to be particularly
interesting because both its branching ratio and the pion energy spectrum
have now been measured in the BNL--E777 experiment~\cite{Aetal92}. In full
generality, one can predict the $K^{+}\ra \pi^{+} l^{+}l^{-}$ decay rates as
a function of the unknown scale invariant combination of coupling constants
\be
\label{eq:kppiplplmw}
{\bf w}_{+}= -\frac{1}{3}(4\pi^2 ) [{\bf w}_{1}^{\rm
r}-{\bf w}_{2}^{\rm r}+3({\bf w}_{2}^{\rm r}-4L_{9}^{\rm r})]
-\frac{1}{6}\log\frac{M_{K}^2 m_{\pi}^{2}}{\nu^4}.
\ee

\noi
where ${\bf w}_{1,2}^{\rm r}$ and $L_{9}^{\rm r} $ are renormalized
couplings at the scale
$\nu$; $L_{9}^{\rm r}$ is the same strong coupling constant which governs the
mean squared radius of the pion [see the discussion in section 2.2.2].
Viceversa, the measured branching ratio~\cite{Aetal92}
\be
{\rm Br}(K^{+}\ra \pi^{+} e^{+}e^{-})=(2.99\pm 0.22)\times 10^{-7}\,
\ee

\noi
determines two possible solutions for the combination of constants ${\bf
w_{+}}$. The degeneracy between the two possible solutions can be lifted
{}from the measurement of the invariant mass distribution of the final lepton
pair, (the $q^2$--dependence; or the $\pi^{+}$ energy--spectrum in the
$K^{+}$ rest--system). A fit to the high--$q^2$ spectrum from the latest
BNL-E777 data~\cite{Aetal92} favours the positive solution, with the result:
\be
\label{eq:kppiplplm_expw}  {\bf w}_{+}=0.89 \ba{cc}  + & 0.24 \\
 - & 0.14 \ea;
\ee

\noi
while the positive solution, extracted from the measured decay rate  in
the same experiment corresponds to the value:
\be
\label{eq:kppiplplm_brw}
{\bf w}_{+}^{BR}=1.2 \ba{cc}  + & 0.4 \\
 - & 0.5 \ea.
\ee

\noi
The two numbers are consistent with each other within less than two
standard deviations; a fact which provides an independent check of the level
of accuracy of $\cO (p^4)$--$\chi$PT to describe this process.

Unfortunately, the determination of the combination of constants ${\bf
w}_{+}$ is not enough to predict the $K_{S}\ra \pi^{0}e^{+}e^{-}$ decay
rate. To the same order in the chiral expansion, the corresponding transition
amplitude brings in the combination of constants:
\be
{\bf w}_{s}= -\frac{1}{3}(4\pi^2)[{\bf w}_{1}^{\rm r}-{\bf
w}_{2}^{\rm r}]-\frac{1}{3}\log\frac{M_{K}^{2}}{\nu^2}\,.
\ee

\noi
Clearly, one has to resort to models to go any further in making a
prediction.

In Ref.~\cite{EPR87} it was pointed out that the relation ${\bf
w}_{2}-4L_{9}=0$ holds for the divergent parts of the corresponding
regularized coupling constants. It is also valid for the full couplings if,
as it happens in many models, the mesonic effective strangeness--changing
current which couples to the virtual photon is required to transform as a
pure
$SU(3)$ octet. In general however, this current is allowed to have terms
which transform as $10$ and $\bar{10}$ $SU(3)$--representations as well.
Therefore, as emphasized in~\cite{EPR87}, the constraint: ${\bf
w}_{2}^{\rm r}=4L_{9}$, is not required by general chiral invariance arguments
alone. It is nevertheless tempting to see what the prediction for the
$K_{S}$--mode is in the class of models which satisfy this relation. Using
the center value for ${\bf w}_{+}=1.2$ from the measurement of the
$K^{+}$--mode branching ratio quoted above, one gets
\be
{\rm Br}(K_{S}\ra \pi^{0}e^{+}e^{-})\Big|_{{\bf w}_{2}^{\rm r}=4L_{9}^{\rm
r}}\simeq 5.4\times 10^{-10}\,,
\ee

\noi
which in turn, corresponds to an ``indirect'' CP--violation branching
ratio
\be
{\rm Br}(K_{L}\ra \pi^{0}e^{+}e^{-})\Big|_{1\gamma\,{\rm ind.}\,({\bf
w}_{2}^{\rm r}=4L_{9}^{\rm r})}\simeq 1.6\times 10^{-12}\,,
\ee

\noi
on the lower range of the ``direct'' CP--violation prediction.

It is important to analyze the sensitivity of this result to models which do
not satisfy the constraint ${\bf w}_{2}^{\rm r}=4L_{9}^{\rm r}$. This has been
done in Ref.~\cite{BrP93} and more recently in Ref.~\cite{DG94}. The outcome is
that the $K_{S}$--branching ratio is rather sensitive to small variations
of the
octet--dominance constraint. As a result, ``indirect'' CP--violation
branching ratios comparable, if not bigger, than the ``direct'' prediction
cannot be excluded for the time being. Once more, we find the need to develop
good models of the low--energy effective action, if one wants to make further
progress. There is not much else that $\chi$PT can do here, except wait
for the experimentalists to measure the rate of $K_{S}\ra
\pi^{0}e^{+}e^{-}$; not an easy task!

%%%%%%%%%%%%%%%%%%%%%%
\subsubsection{\sl The two--photon exchange amplitude}

The transition $K_{2}\ra \pi^{0}\gamma^*\gamma^*\ra
\pi^{0}e^{+}e^{-}$ is
CP--allowed. The lowest non--trivial order calculation of this process in
$\chi$PT involves at least two loops. It is possible, however, to obtain a
lower bound to the
$2\gamma$--exchange rate from the calculation of the contribution to the
absorptive part of the amplitude $A(K_{1}^{0}\ra
\pi^{0}e^{+}e^{-})$ coming from the
$\gamma\gamma$--discontinuity.

The general kinematics decomposition of the subprocess $K_{2}^{0}\ra
\pi^{0}\gamma\gamma$ involves two invariant amplitudes. With
$$K_{2}(p)\to \pi^{0}(p') + \gamma(q_1) + \gamma(q_2),
\qquad p^2=M^2, \quad p'^2=M'^2, \quad q_1^2=q_2^2=0 ,$$

\noi
gauge invariance and Lorentz invariance restrict the form of the
transition amplitude to
$$\cM [K_{2}(p)\to \pi^{0}(p')  \gamma(q_1)  \gamma(q_2)]=
G_{9}\frac{\alpha_{\rm em}}{4\pi}\,
\epsilon_{\mu}(q_1)\epsilon_{\nu}(q_2)\,\left[A(y,z)\,
\left(  q_2^\mu q_1^\nu - q_1\cdot q_2 \, g^{\mu\nu}\right)\right. $$
\be
\label{eq:kpigga}
\left.  +  2 B(y,z)\,
\Bigl(p\cdot q_1 \, q_2^\mu p^\nu + p\cdot  q_2\, q_1^\nu p^\mu - p^\mu p^\nu
\, q_1\cdot q_2 - p\cdot q_1\, p\cdot q_2 \, g^{\mu\nu}\Bigr)\right]\, ,
\ee

\noi where
\be  y= {p\cdot(q_1-q_2)\over M^2} \, , \qquad {\rm and}\qquad z=
{(q_1+q_2)^2\over M^2}\, .
\ee

\noi Bose symmetry requires the amplitudes $A$ and $B$ to be symmetric
functions of $q_1$ and $q_2$, i.e.
\be
 A(-y,z)  =  A(y,z), \qquad\qquad B(-y,z) =  B(y,z)\,.
\ee

To lowest non--trivial order in
$\chi$PT, only the amplitude proportional to $A$  contributes; and in fact, to
that order, $A$ depends only on $z$ i.e., the invariant mass squared of the
$\gamma\gamma$--pair. None of the local terms of
$\cO (p^4)$ in the effective $\Delta S=1$ non--leptonic Lagrangian can
contribute to this decay. As a result, the full contribution at lowest
non--trivial order in the chiral expansion, comes only from the finite  chiral
one loop amplitude. The predicted $z$--distribution has a very characteristic
shape; being highly peacked at the higher end of the spectrum. This
prediction~\cite{EPR87} has been subsequently confirmed experimentally by the
CERN-NA31 collaboration~\cite{Betal92}. However the
$\cO (p^4)$ predicted branching ratio:
\be {\rm Br}(K_{L}\ra \pi^{0}e^{+}e^{-})= 6.8\times 10^{-7}\,,
\ee

\noi turns out to be smaller than the observed rates:
\be
\label{eq:expbrgg} {\rm Br}(K_{L}\ra \pi^{0}e^{+}e^{-})=\left\{ \ba{cc} (1.7\pm
0.2\pm0.2)\times 10^{-6} & \mbox{NA31~\cite{Betal92}}\,,\\ (1.86\pm
0.60\pm0.60)\times 10^{-6} &
\mbox{E731~\cite{Petal91}}\,.\ea\right.
\ee

\noi
It has been claimed, however, that phenomenologically expected higher order
effects can explain this discrepancy~\cite{CEP93,KH94}.

The term proportional to the $A$--amplitude in Eq.(\ref{eq:kpigga}) has a
tensor structure which, when inserted in the two--body
$\gamma\gamma$ phase space integral to get the corresponding
$K_{2}^{0}\ra \pi^{0}e^{+}e^{-}$ amplitude, requires a flip in the helicity of
the electron line for the transition to be allowed; a fact which brings in the
electron mass $m_{e}$ as an overall suppression factor in amplitude. This
amplitude structure leads to a branching ratio:
\be
\label{eq:brchs}  {\rm Br}(K_{2}^{0}\ra \pi^{0}\gamma\gamma\ra
\pi^{0}e^{+}e^{-})
\Big|_{A(\gamma\gamma)}\simeq 5\times 10^{-15},
\ee

\noi
too small, by several orders of magnitude  to be competitive with the
one--photon exchange,
CP--violating amplitude, contributing to
$K_{L}\ra
\pi^{0}e^{+}e^{-}$.

The interesting issue here is that, contrary to what happens with the
contribution from the $A(\gamma\gamma)$--amplitude which we have just
discussed, the tensor structure proportional to the
$B$--amplitude in Eq.(\ref{eq:kpigga}), when inserted in the two--photon phase
space integral of the process $K_{2}^{0}\ra \pi^{0}
\gamma^{*}\gamma^{*}\ra
\pi^0 e^{+}e^{-}$ has contributions which are allowed without requiring an
electron helicity flip; i.e., they are not
$m_{e}$--suppressed. Although the
$B(\gamma\gamma)$--amplitude first appears at $\cO(p^6 )$ in
$\chi$PT; the fact that it can induce a helicity allowed transition to
$K_{2}^{0}\ra \pi^{0} \gamma^{*}\gamma^{*}\ra
\pi^0 e^{+}e^{-}$, makes it ``the dominant amplitude'' for this process.

There have been several estimates in the literature~\cite{SE90} of the
$B$--amplitude in Eq.(\ref{eq:kpigga}), based on vector meson  dominance
($VMD$) models. Here, however, one has to be careful not to overestimate its
size and therefore spoil the observed shape of the invariant
$\gamma\gamma$--mass distribution in the
$K_{L}\ra
\pi^{0}e^{+}e^{-}$ decay. Many of the early models are in fact now ruled out by
the NA31 experiment; but they have been very useful to sharpen our views on the
r\^{o}le of vector mesons in
$\chi$PT in general, with the results~\cite{EGPR89}$^{,}$~\cite{EGLPR89} we
already discussed in {\it subsec.\,4.2}.

Within the framework of $\chi$PT, one can parametrize the local $\cO (p^6)$
contributions to $K_{L}\ra \pi^{0}\gamma\gamma$ induced by vector--exchanges,
(including vector--exchanges in direct weak transitions,) by an effective
vector coupling $a_{V}$, such that~\cite{EPR90}:
\be
\label{eq:bmodel} B=-2a_{V}G_{8}M_{K}^2 \frac{\alpha}{\pi}\,.
\ee

\noi In terms of this parametrization, the contribution to the $K_{2}^{0}\ra
\pi^0 e^{+}e^{-}$ amplitude is as follows:
\be
\label{eq:k2pizepem_ba}  A(K_{2}^{0}\ra
\pi^{0}e^{+}e^{-})\mid_{B(\gamma\gamma )}= iG_{8}\frac{\alpha^2 }{3\pi}
a_{V}
\bar{u}(k)\gamma\cdot p \frac{p\cdot (k-k^{'})}{M_{K}^2 }v(k^{'}),
\ee

\noi
which leads to a branching ratio~\cite{EPR90}:
\be
\label{eq:brcba}  {\rm Br}(K_{2}^{0}\ra \pi^{0}\gamma\gamma\ra
\pi^{0}e^{+}e^{-})
\Big|_{B(\gamma\gamma)}=4.4\times 10^{-12}\,a_{V}^{2}\,.
\ee

\noi
The NA31--collaboration~\cite{Betal92} has performed an analysis of
$K_{L}\ra
\pi^{0}e^{+}e^{-}$ events with an invariant $\gamma\gamma$--mass:
$M_{\gamma\gamma} < 240 {\rm MeV}$, using this parametrization, and obtained
the following limits:
\be
\label{eq:avexp} -0.32<a_{V}<0.19 \qquad (90\%\:{\rm C.L.})\,.
\ee

The rate in Eq.(\ref{eq:brcba}), and hence the result in Eq.(\ref{eq:avexp})
does not take into account, however, the non--polynomial structure of the
$B$--amplitude in Eq.(\ref{eq:kpigga}) due to chiral loops. A recent
analysis~\cite{CEP93} which tries to fold the phenomenology of both local and
non--polynomial effects in the
$B$--amplitude with the observed data from the NA31--experiment, results in a
branching ratio for $K_{L}\ra \pi^{0}e^{+}e^{-}$ from the
$\gamma\gamma$--discontinuity:
\be {\rm Br}(K_{L}\ra \pi^{0}e^{+}e^{-})\Big|_{2\gamma\,{\rm Abs.}}=
\left\{\ba{cc} 0.13\times 10^{-12},\qquad a_{V}=0\,,\\ 1.8\times
10^{-12},\qquad a_{V}=-0.9\,. \ea \right.
\ee

\noi
The value $a_{V}=-0.9$ is the one which reproduces the observed
$z$--spectrum in the decay rate of $K_{L}\ra \pi^{0}\gamma\gamma$, in the
presence of the non--polynamial ansatz for the $B$--amplitude which has been
used in Ref.~\cite{CEP93}. The two effects combined also raise the predicted
$K_{L}\ra\pi^{0}\gamma\gamma$ branching ratio to $1.6\times 10^{-6}$, in good
agreement with the experimental values in Eq.(\ref{eq:expbrgg}).

As a summary, the theoretical prospects for more refined predictions on the
decay $K_{L}\ra \pi^{0}e^{+}e^{-}$ are the following:

\begin{itemize}

\item The branching ratio expected from ``direct'' CP--violation is rather well
known. The uncertainty in Eq.(\ref{eq:dirCP}) will be reduced once the top
quark mass is better determined; and the $B_{K}$--factor is pinned down more
accurately -- either by theoretical improved calculations, or
phenomenologically-- .

\item The error in the branching ratio expected from the CP--conserving
transition induced via two intermediate photons, can be reduced with a combined
effort of more accurate measurements of the mode $K_{L}\ra
\pi^{0}\gamma\gamma$ on the one hand, and an improvement in the
phenomenological ansatz of the $B$--amplitude in the theoretical analysis on
the other.

\item The largest uncertainty, at the moment, comes from the ``indirect''
CP--violation transition. Here, the only way I can see to make progress is via
the development of good models of the QCD low--energy effective action; e.g.,
the extension of the succesful ENJL--model in the strong sector to
non--leptonic weak transitions. Parallel to this theoretical effort, there
should be, of course, some progress as well in obtaining more experimental
results in rare
$K$--decays, so as to have enough observables to test the models.

\end{itemize}

\section{Acknowledgements}
I am very grateful to the students of the TASI summer school for their
enthusiastic participation and the good questions they
asked. I hope that this written version of my lectures will encourage some of
them to work on a subject where:

\begin{center}

THERE ARE A LOT OF INTERESTING THINGS TO BE DONE BOTH FOR
THEORISTS AND EXPERIMENTALISTS!!

\end{center}

Thank you.

\section{References}

%%%%%%%%%%%%%%%%%%%%%%%%%%%%%%%%%%%%%%%%%%%%%%%

%%%%%%%%%%%%%%%%%%%%%%%%%%%%%%%%%%%%%%%%%%%%%%%

\end{document}